\documentclass[11pt]{article}
\pdfoutput=1
\usepackage{jheppub}
\usepackage[usenames,dvipsnames,table]{xcolor}
\usepackage{graphicx,amsmath,amssymb,amsthm,multirow,array,bm,bbm,esint}
\usepackage[mathscr]{eucal}
\usepackage[bbgreekl]{mathbbol}
\usepackage{epsf,amsfonts}
\usepackage{slashed}
\usepackage[numbers,sort&compress]{natbib}
\usepackage{minibox}
\usepackage{pdflscape}
\usepackage{comment}

\DeclareSymbolFontAlphabet{\mathbbm}{bbold}
\DeclareSymbolFontAlphabet{\mathbb}{AMSb}%

\usepackage{slashed}

\def \Tr{\mbox{Tr\,}}

\def \tr{\mbox{tr\,}}

\def\+{{+\!\!\!+}}		
\def\ppmm{{\overset{+\!\!\!+}{=}} }

\newcommand{\cA}{\mathcal{A} }
\newcommand{\cB}{\mathcal{B} }
\newcommand{\cC}{\mathcal{C} }
\newcommand{\cD}{\mathcal{D} }

\newcommand{\cH}{\mathcal{H} }

\newcommand{\cJ}{\mathcal{J} }
\newcommand{\cK}{\mathcal{K} }
\newcommand{\cL}{\mathcal{L} }
\newcommand{\cM}{\mathcal{M} }
\newcommand{\cN}{\mathcal{N} }
\newcommand{\cO}{\mathcal{O} }
\newcommand{\cP}{\mathcal{P} }

\newcommand{\bQ}{\mathbb{Q} }
\newcommand{\bP}{\mathbb{P} }   

\newcommand{\bW}{\mathbb{W} }

\newcommand{\bPi}{\Pi }    
\newcommand{\mpi}{\cK}    

\newcommand{\pz}{\mathbbm{p}}    

\newcommand{\momp}{p}    

\newcommand{\fpi}{\pi }    

\newcommand{\spi}{\pi }    

\newcommand{\tS}{\mathtt{S} }
\newcommand{\tK}{\mathtt{K} }

\newcommand{\wH}{\widetilde{H} }

\newcommand{\pb}{\text{\tiny PB} }
\newcommand{\dirac}{\text{\tiny D} }

\newcommand{\cdag}{ {\dag_\cJ} }    

\def \cT{\mathcal{T}}

\newcommand{\ttb}{$T\bar{T}$ }

\newcommand{\ie}{\emph{i.e.}, }
\newcommand{\eg}{\emph{e.g.}, }

\newcommand{\etc}{\emph{etc}.}

\title{$T\bar T$-deformed Fermionic Theories Revisited}

\author[\, a]{Kyung-Sun Lee}
\author[\, b]{\!, Piljin Yi}
\author[\, b,c]{\!, Junggi Yoon}
\affiliation[\,a]{School of Physics and Chemistry, Gwangju Institute of Science and Technology,\\
123 Cheomdan- gwagiro, Gwangju 61005, Korea}
\affiliation[\,b]{School of Physics, Korea Institute for Advanced Study\\
85 Hoegiro Dongdaemun-gu, Seoul 02455, Republic of Korea.}
\affiliation[\,c]{Asia Pacific Center for Theoretical Physics, Postech, Pohang 37673, Korea}
\emailAdd{kyungsun.cogito.lee@gmail.com}
\emailAdd{piljin@kias.re.kr}
\emailAdd{junggiyoon@gmail.com}

\preprint{ {\raggedleft \tt KIAS-P21013 \par}  }

\abstract{
We revisit $T\bar T$ deformations of $d=2$ theories with fermions
with a view toward the quantization. As a simple illustration, 
we compute the deformed Dirac bracket for a Majorana doublet and 
confirm the known eigenvalue flows perturbatively. We mostly consider 
those $T\bar T$ theories that can be reconstructed from string-like 
theories upon integrating out the worldsheet metric. After a quick 
overview of how this works when we add NSR-like or GS-like 
fermions, we obtain a known non-supersymmetric $T\bar T$ deformation of 
a $\cN=(1,1)$ theory from the latter, based on the Noether energy-momentum. 
This worldsheet reconstruction implies that the latter is actually 
a supersymmetric subsector of a $d=3$ GS-like model, implying hidden 
supercharges, which we do construct explicitly. This brings us to ask 
about different \ttb deformations, such as manifestly supersymmetric 
$T\bar T$ and also more generally via the symmetric energy-momentum. 
We show that, for theories with fermions, such choices often lead us 
to doubling of degrees of freedom, with potential unitarity issues.
We show that the extra sector develops a divergent gap in the 
``small deformation" limit and decouples in the infrared, although 
it remains uncertain in what sense these can be considered a deformation.

}

\begin{document}
\maketitle


\section{Summary}
\label{sec: introduction}

In recent years, a particular form of irrelevant deformation of field theories received much attention, namely the \ttb deformation~\cite{Zamolodchikov:2004ce}. A most surprising fact was that the exact energy eigenvalue flow of such deformations could be computed~\cite{Zamolodchikov:2004ce,Smirnov:2016lqw}. The eigenvalues are anticipated to flow as, provided that the spatial direction is compactified on a circle,
\begin{align}\label{energy}
    E(\lambda)\,\sim \,{1 \over 2\lambda} \left[\sqrt{1+ {4\lambda E(0)} +{4\lambda^2 P(0)^2}} - 1\right] \hspace{3mm} ,\hspace{9mm} P(\lambda)\,=\, P(0) \ ,
\end{align}
where $\lambda$ is the coefficient of the deformation. $E(0)$ and $P(0)$ are the energy and momentum eigenvalues of the theory before the deformation.\footnote{This assumes the absence of certain zero-modes, such as the Wilson line of a gauge theory, whose scaling with the spatial radius is exceptional.} The effect on $S$-matrices can also be kept track of relatively easily~\cite{Smirnov:2016lqw,Cavaglia:2016oda}, and the latter's equivalence to those of an undeformed theory coupled to dynamical gravity is known~\cite{Dubovsky:2017cnj,Dubovsky:2018bmo}, further suggesting another special nature of this set of theories.

Despite these well-known features, the first-principle quantization via the usual canonical approach remains beyond the reach as the deformed Lagrangians are heavily non-linear. In the simplest known example of a single real scalar, the deformed Lagrangian is of Nambu-Goto or of Born-Infeld type~\cite{Cavaglia:2016oda,Conti:2018jho}, with the kinetic terms enclosed inside a square root. The appearance of such a square-rooted Lagrangian implies the same for a similar structure for the Hamiltonian density,
\begin{align}\label{hamiltonian}
	\cH\,=\,&{1\over 2 \lambda} \left[ \sqrt{1+ 4\lambda \left({1\over 2} \spi^2 + {1\over 2} \phi'^2\right) + 4\lambda^2 (\spi \phi')^2 } -1 \right]\ ,
\end{align}
with the conjugate momentum $\pi$ of $\phi$, in an apparent similarity with the known eigenvalue flow. However, the two expressions are actually a world-apart as $\cH$ needs a further integration over the space while the eigenvalue flow tells us that actual Hamiltonian involves free Hamiltonian and momentum operators, rather than densities thereof, inside the square-root. The tantalizing similarity between Eq.~\eqref{energy} and Eq.~\eqref{hamiltonian} is thus even more vexing.

For fermions, the potential non-linearity of the Hamiltonian density truncates at a finite order, which also appears naively at odd with the above universal eigenvalue flow with a would-be infinite Taylor series in $\lambda$. The resolution is conceptually simpler for fermions, as it turns out.  Recall that quantization of usual Fermionic theories must start with a second-class constraint that relates the conjugate momenta linearly to the original fermion variable, 
\begin{align}
\pi_\psi\,\approx\,  {i\over 2} \psi\ ,
\end{align}
whereby the Dirac bracket \cite{Dirac} replaces the naive Poisson bracket. What happens with \ttb-deformed fermions is that this canonical structure is itself deformed, again up to some finite order in $\lambda$. Combining this polynomial-modified Dirac bracket and the polynomial-modified Hamiltonian sets up a perturbative eigenvalue problem that does not truncate in $\lambda$. For actual eigenvalues, one would prefer to rephrase the Hamiltonian in terms of simple harmonic oscillators, which would then involve an appropriate infinite Taylor series and reproduce the general eigenvalue flow formula. We have performed this to the first nontrivial order in Section~\ref{sec: hamiltonian analysis of ttb deformation}. 

For the scalar theory above, something different happens. Between the natural harmonic oscillators that come out of the canonical commutator,
\begin{align}
    [\pi,\phi]\,=\,-i\delta\ ,
\end{align}
and those that construct the free Hamiltonian and the translational Momentum inside the square root of~\eqref{energy}, we find a nontrivial canonical transformation. In other words, a non-linear
map between the two sets of oscillators such that both obey the common commutator algebra, yet one is an infinite Taylor series of the other. These preliminary investigations can be found in Section~\ref{sec: hamiltonian analysis of ttb deformation}, where the above perturbative computation is carried out up to the
second-order in $\lambda$ and confirms the known eigenvalue flow. Section~\ref{sec: hamiltonian analysis of ttb deformation} also reviews the Dirac quantization procedures in some detail, which will prove useful and important in later Sections. 

However, neither of these is efficient or extendable to higher-order, if interesting conceptually. As such, we also explore other potential venues. For the above free scalar theory, deformed by $T\bar{T}$, an indirect method is known~\cite{theisenTalk:2018,Baggio:2018rpv,Frolov:2019nrr,Frolov:2019xzi,Callebaut:2019omt,Jorjadze:2020ili} where the theory is realized as a unit winding number sector of a Polyakov action after integrating out a worldsheet metric, and the energy as the conserved conjugate momenta of the time-like target variable. 
It is natural to imagine that this method is straightforwardly generalized to Polyakov models with fermions, either Neveu-Schwarz-Ramond~(NSR) type or Green-Schwarz~(GS) type, and we explore such possibilities and find that some of such approaches do give the desired form of \ttb-deformed theories. We find, unfortunately, these allow only limited handles on the exact quantization, unlike the simplest scalar-only theory. Nevertheless, they offer structures of the \ttb deformation that would not have been easily available without such routes. The bulk of this note is dedicated to these explorations, as delineated in Sections~\ref{sec: string theory and ttbar deformed hamiltonian} and~\ref{sec: ttbar deformation and gs shifted background}.

In particular, we devote the bulk of Section~\ref{sec: ttbar deformation and gs shifted background} on \ttb deformation of the simplest $\cN=(1,1)$ theory, with the deformation operator being {\it not supersymmetry completed}. Despite this explicit supersymmetry breaking, the universal nature of the eigenvalue flow suggests preservation of the supersymmetry. This is also reinforced by the worldsheet reconstructions of the same theory, starting from a $3D$ GS-like model with spacetime supersymmetry. From both \ttb side and the worldsheet side, we construct and explore the supercharges of the deformed models. In particular, the worldsheet viewpoint gives a rather clear picture of why such supercharges exist, and we further explore the subsector of the Hilbert space that preserves the supersymmetry partially. This observation may lead to interesting topological observables for more complicated supersymmetric models and \ttb deformation thereof.

Along the way, we find a peculiar problem with the generic $T\bar{T}$ deformation of fermionic theories. Recall that the fermion quantization, with its at most single time derivative in the Lagrangian, starts with a second-class constraint that linearly relates the conjugate momentum to the configuration fermion as $ \pi_\psi\approx {i\over 2}\psi$ ``weakly'' \cite{Dirac}. Such a constraint is essential in all fermionic theory, free or interacting, for the usual counting of degrees of freedom, where one complex Grassmannian field carries the same number of independent canonical variables as one real scalar field. This counting holds for the initial undeformed theories, but does it for the deformed theories? The \ttb operator of free fermion theory containing up to two derivatives, and even worse, once deformation starts, the number of time derivatives can easily proliferate. The crucial second-class constraint can be lost depending on the precise nature of the energy-momentum tensor used to build \ttb.

As such, we find an important dividing feature among various \ttb deformations. Classified by precisely which energy-momentum tensor is used or the deformation, there appear to be two types of \ttb deformation of fermionic theories. In one type, this constraint remains a constraint only to be modified by non-linear corrections. The Dirac bracket is additively modified by the higher-order multi-fermion terms, as we noted already. The would-be second-class constraint itself is lifted entirely by acquiring a term with a time derivative in the other type. The latter means that the degrees of freedom are doubled than otherwise, so that the Hilbert space is much larger than that of the initial undeformed theory, rendering the terminology ``deformation" a misnomer. One fortunate aspect, though, is that these extra degrees of freedom are gapped amply, scaling inversely with the small $\lambda$ so that one may hope for a decoupling of the extra sector can be argued in the infrared limit. 

In the latter class, the Lagrangian involves terms with more than one time derivatives on fermions. Such theories are expected to be riddled with unitarity issues, generically, such as negative norm states or non-Hermitian Hamiltonian \cite{Henneaux:1992ig}. For a $d=1$ toy example and for some simplest $d=2$ examples, we see a glimpse of how the unitarity might be restored by replacing the naive inner product on the Hilbert space with the one consistent with the path integral. However, we performed the latter exercise only for specific limits in the case of $d=2$ examples, so it is not completely clear whether such ``deformation" leads to an inconsistent extension or to a different but sensible UV completion merely with twice the degrees of freedom. Either way, the choice of the energy-momentum tensor for \ttb deformation appears far more critical than naive expectations. 

Interestingly, for those simplest examples considered here, the \ttb deformations based on the Noether energy-momentum tensor do not generate such issues. With symmetric energy-momentum, we do find such problems even for the simplest possible theory. A further divide of this kind exists between this standard \ttb deformation, at least naively non-supersymmetric, of $\cN=(1,1)$ theory of a single scalar multiplet and the supersymmetry completed \ttb deformation thereof. It is unclear to us precisely what aspect of the Noether energy-momentum tensor allows such a nice behavior and whether the same would happen with a larger fermion content.

\section{Hamiltonian Analysis of \ttb Deformation}
\label{sec: hamiltonian analysis of ttb deformation}

The \ttb deformation features a universal form of the deformed spectrum in terms of undeformed energy and momentum~\cite{Smirnov:2016lqw,Cavaglia:2016oda}, 
\begin{align}
    E_n(L,\lambda)\,=\,&{L\over 2\lambda} \left[\sqrt{1+ {4\lambda \over L}E_n +{4\lambda^2 \over L^2}P_n^2}-1\right]\ ,\label{eq: deformed energy value}\\
    P_n(L,\lambda)\,=\,&P_n(L)\ .\label{eq: deformed momentum value}
\end{align}
%
%
It is tempting to claim that this universal form of the deformed spectrum also holds at the operator level. Namely, one can ask whether the Hamiltonian operator of the deformed theory can universally be written as 
\begin{align}
    H\,=\,{L\over 2\lambda} \left[\sqrt{1+ {4\lambda \over L}H_{(0)} + {4\lambda^2 \over L^2} P_{(0)}^2 }-1\right]\ ,\label{eq: deformed hamiltonian conjecture}
\end{align}
where $H_{(0)}$ and $P_{(0)}$ is Hamiltonian and momentum operator of the undeformed theory. We emphasize that $H_{(0)}$ and $P_{(0)}$ are not density operators. 
On the other hand, starting from the deformed Lagrangian density following the flow equation 
\begin{align}
    \partial_\lambda \cL \,=\, {1\over 2} \epsilon_{\mu\nu }\epsilon^{\rho\sigma}{T^\mu}_\rho {T^\nu}_\sigma\ ,
\end{align}
one can derive the Hamiltonian density of the deformed theory. However, it is highly non-trivial to see how the Hamiltonian from the deformed Lagrangian density is related to the universal form~\eqref{eq: deformed hamiltonian conjecture}. Also, the normal ordering of the deformed Hamiltonian~\eqref{eq: deformed hamiltonian conjecture} is not clear.\footnote{Considering the deformed energy spectrum, it is tempting to guess that the normal ordering chosen in the undeformed theory had better be applied to each $H_{(0)}$ and $P_{(0)}$ individually inside of the square-root.} Therefore, in this section, we will carry out the Hamiltonian analysis for the \ttb deformation of free theories. Then, we will give an explicit construction of the conjectured Hamiltonian~\eqref{eq: deformed hamiltonian conjecture} perturbatively by field redefinition.

\subsection{$T\bar{T}$ Deformation of Free Scalar Field}
\label{sec: perturbation scalar}

First, let us consider the \ttb deformation of the scalar field of which Lagrangian is given by
\begin{align}
	\cL\,=\,-{ 1\over 2 \lambda } \left[  \sqrt{1+ 2 \lambda  (-\dot{\phi}^2 +\phi'^2)} -1 \right]\ ,\label{eq: ttb deformed lagrangian of scalar field}
\end{align}
where $\dot{\phi}\,\equiv\, \partial_t \phi$ and $\phi'\,\equiv\, \partial_x \phi$. Using the conjugate momentum
\begin{align}
	\spi\,=\, {\dot{\phi}\over \sqrt{1+ 2\lambda(-\dot{\phi}^2+\phi'^2)}}\ ,
\end{align}
the Hamiltonian density can be written as follows,
\begin{align}
	\cH\,=\,&{1\over 2 \lambda} \left[ \sqrt{1+ 4\lambda \left({1\over 2} \spi^2 + {1\over 2} \phi'^2\right) + 4\lambda^2 (\spi \phi')^2 } -1 \right]\ ,\cr
	\,=\,&{1\over 2} \left( \spi^2 + \phi'^2\right) - \lambda \left({1\over 2}\spi^2 - {1\over 2}\phi'^2 \right)^2 +\cO(\lambda^2)\ ,\label{eq: deformed hamiltonian density of free scalar field}
\end{align}
where $\phi$ and $\spi$ satisfy the (equal-time) canonical commutation relations:
\begin{align}
	[\phi(x_1),\spi(x_2)]\,=\, i \delta(x_1-x_2)\;\;,\qquad [\phi(x_1),\phi(x_2)]\,=\,[\spi(x_1),\spi (x_2)]=0\ .
\end{align}
Note that the Hamiltonian~\eqref{eq: deformed hamiltonian density of free scalar field} is significantly different from Eq.~\eqref{eq: deformed hamiltonian conjecture} in that the Hamiltonian~\eqref{eq: deformed hamiltonian density of free scalar field} is the integral of a local density while the conjectured Hamiltonian~\eqref{eq: deformed hamiltonian conjecture} is a function of charge operators (\ie $H_{(0)}$ and $P_{(0)}$) so that it contains infinitely many integrals. This implies that the equivalence of two Hamiltonian is highly non-trivial, and one needs non-local canonical transformation\footnote{Note that the commutation relation is not changed under such a transformation for the case of a scalar field.} to reach the conjectured form~\eqref{eq: deformed hamiltonian conjecture} if exists.

It was shown~\cite{theisenTalk:2018,Jorjadze:2020ili} that the equivalence of the Hamiltonian~\eqref{eq: deformed hamiltonian density of free scalar field} and the conjectured Hamiltonian~\eqref{eq: deformed hamiltonian conjecture} can be proved by using Polyakov action (up to the issue of operator ordering and the equivalence between Polyakov action and Nambu-Goto action at quantum level). Although the equivalence of two Hamiltonians is proved, it is still non-trivial to find an explicit map between operators. The dynamical coordinate transformation~\cite{Conti:2018tca,Conti:2019dxg,Coleman:2019dvf} and the canonical map discussed in Ref.~\cite{Jorjadze:2020ili} could, in principle, give the answer--a map between operators. But, since they are an operator-dependent coordinate transformation, it would not be easy to utilize them at the quantum level from the point of view of 2D QFT. Hence it is worthwhile to find an explicit map between operators which transforms the Hamiltonian~\eqref{eq: deformed hamiltonian density of free scalar field} into the conjectured Hamiltonian~\eqref{eq: deformed hamiltonian conjecture}, at least, perturbatively in small $\lambda$. Such an operator map could be used in other quantum calculations (\eg correlation function \etc).

For this, we consider the Fourier expansion of $\phi$ and $\Pi$:
\begin{align}
	\phi(x)\,=\,&\sum_k  \phi_k e^{{2\pi i k x\over L}}\ , \\
	\Pi(x)\,=\,&{1\over L} \sum_k  \Pi_k e^{{2\pi i k x\over L}}\ , 
\end{align}
where $L$ denotes the circumference of $x$. Here, the normalization is chosen in a way that $\phi_k$ and $\Pi_k$ are dimensionless 
%
%
and the commutation relations of the modes are 
\begin{align}
	[\phi_k,\Pi_q]\,=\,i\delta_{k+q,0}\quad,\quad [\phi_k,\phi_q]\,=\,[\Pi_k,\Pi_q]=0\ .
\end{align}
It is useful to define oscillators $A_k$ $\bar{A}_k$ by
\begin{align}
	A_k\, \equiv\, & -\sqrt{\pi} i k \phi_k +{1\over 2\sqrt{\pi}} \Pi_k \ , \label{def: oscillators ak} \\
	\bar{A}_k\, \equiv\, & -\sqrt{\pi} i k \phi_{-k} +{1\over 2\sqrt{\pi}} \Pi_{-k} \ .\label{def: oscillators abark}
\end{align}
The commutation relations of $A_k$ and $\bar{A}_k$ are 
\begin{align}
	[A_k,A_q]\,=\,& k \delta_{k+q,0}\ ,\label{eq: com relation of a1}\\
	[\bar{A}_k,\bar{A}_q]\,=\,&k\delta_{k+q,0}\ ,\label{eq: com relation of a2}\\
	[A_k,\bar{A}_q]\,=\,&0\ .\label{eq: com relation of a3}
\end{align}
We will consider the small $\lambda/L^2$ expansion of $A_k$ and $\bar{A}_k$:
\begin{align}
	A_k\,=\,A^{(0)}_k + {\lambda\over L^2} A^{(1)}_k +\cdots\;\;,\qquad \bar{A}_k\,=\,\bar{A}^{(0)}_k + {\lambda\over L^2} \bar{A}^{(1)}_k +\cdots\ .
\end{align}
Because we have the free scalar field for $\lambda=0$, the leading operator $A_k^{(0)}$ and $\bar{A}_k^{(0)}$ should be the free oscillators $\alpha_k$ and $\bar{\alpha}_k$ of the free scalar field,
\begin{align}
	A^{(0)}_k\,=\, \alpha_k\quad,\quad \bar{A}^{(0)}_k\,=\, \bar{\alpha}_k\ ,
\end{align}
where the commutation relation of the oscillator $\alpha_k$ and $\bar{\alpha}_k$ is the same as that of $A$'s:
\begin{align}
	[\alpha_k,\alpha_q]\,=\,& k \delta_{k+q,0}\ ,\label{eq: alpha algebra 1}\\
	[\bar{\alpha}_k,\bar{\alpha}_q]\,=\,&k\delta_{k+q,0}\ ,\\
	[\alpha_k,\bar{\alpha}_q]\,=\,&0\ . \label{eq: alpha algebra 3}
\end{align}

Now, we claim that there exists a canonical transformation from $A_{q}\;,\; \bar{A}_{q}$ to $\alpha_k\;,\; \bar{\alpha}_k $ such that the Hamiltonian $\widetilde{H}[\alpha,\bar{\alpha}]\, \equiv\, H\big[A(\alpha,\bar{\alpha}),\bar{A}(\alpha,\bar{\alpha}) \big]$ is consistent with the energy spectrum. Namely, $\widetilde{H}[\alpha,\bar{\alpha}]$ is of form 
\begin{align}
	H[A,\bar{A}]\,=\,\wH [\alpha,\bar{\alpha}]= {L \over 2\lambda } \left[ \sqrt{1+ {4\lambda\over  L } (H_++H_-) + { 4\lambda^2 \over L^2} ( H_+ - H_-)^2 } - 1 \right]\ ,\label{eq: ham eq scalar sec 2}
\end{align}
where $H_\pm$ is defined by
\begin{align}
	H_+\,\equiv \, {\pi \over L }\sum_k  \alpha_{-k}\alpha_k \quad,\qquad H_-\, \equiv\, {\pi \over L } \sum_k  \bar{\alpha}_{-k} \bar{\alpha}_k \ .
\end{align}
Perturbatively, we aim at finding a transformation in which the Hamiltonian becomes
\begin{align}
	&\int dx\;\left[{1\over 2} \left( \spi^2 + \phi'^2\right) - \lambda \left({1\over 2}\spi^2 - {1\over 2}\phi'^2 \right)^2 +\cO(\lambda^2)\right] \cr
	=\,& H_+[\alpha] + H_-[\bar{\alpha}] - {4\lambda\over L} H_+[\alpha]  H_-[\bar{\alpha}]  +\cO(\lambda^2)\ .\label{eq: hamiltonian eq scalar}
\end{align}
To avoid the ordering ambiguity, we will first find the transformation classically. Up to the leading order $\cO(\lambda^0)$, we have
\begin{align}
	A_k\,=\,\alpha_k+\cO(\lambda)\quad,\qquad \bar{A}_k\,=\,\bar{\alpha}_k+\cO(\lambda) \ ,
\end{align}
and we trivially have
\begin{align}
	H[A,\bar{A}]\,=\,{\pi \over L}\sum_{k} ( A_{-k}A_k + \bar{A}_{-k}\bar{A}_k) +\cO(\lambda)
	\,=\, \wH[\alpha,\bar{\alpha}] +\cO(\lambda)\ .
\end{align}
We solve \eqref{eq: hamiltonian eq scalar} perturbatively up to order $\cO(\lambda^2)$ by expanding $A_k\, ,\; \bar{A}_k$ with respect to ${\lambda\over L^2}$:
\begin{align}
	A_k\,=\,&\alpha_k+ {\lambda\over L^2} A^{(1)}_k[\alpha,\bar{\alpha}]+ {\lambda^2 \over L^4} A^{(2)}_k[\alpha,\bar{\alpha}]+\cO(\lambda^3)\ ,\\
	\bar{A}_k\,=\,&\bar{\alpha}_k+ {\lambda\over L^2} \bar{A}^{(1)}_k[\alpha,\bar{\alpha}]+ {\lambda^2 \over L^4} \bar{A}^{(2)}_k[\alpha,\bar{\alpha}]+\cO(\lambda^3)\ .
\end{align}
Demanding that the transformation is canonical, we get a solution for the canonical transformation from $A_k, \bar{A}_k$ to $\alpha_k, \bar{\alpha}_k$:
\begin{align}
	A^{(1)}_k\,=\,&  2 \pi  \sum_{\substack{r,s\\ r+s\ne 0}}  {k\over r+s} \alpha_{k-r-s}\bar{\alpha}_{-r} \bar{\alpha}_{-s} \ , \label{eq: first order sol scalar 1} \\
	\bar{A}^{(1)}_k\,=\,&  2 \pi \sum_{\substack{r,s\\ r+s\ne 0}} {k\over r+s} \alpha_{-r} \alpha_{-s} \bar{\alpha}_{k-r-s} \ , \label{eq: first order sol scalar 2}
\end{align}
and
\begin{align}
	&A_k^{(2)}\,=\, 2\pi^2  k \sum_{\substack{r,s,u,v\\u+v\ne 0,r+s\ne 0}} {k-r-s-u-v\over (u+v)(r+s)}\alpha_{k-r-s -u -v} \bar{\alpha}_{-u} \bar{\alpha}_{-v} \bar{\alpha}_{-r}\bar{\alpha}_{-s} \cr
	&- 4 \pi  k  \sum_{\substack{u,v\\u+v\ne 0}} {1\over u+v} \alpha_{k-u-v} \bar{ \alpha}_{-u}\bar{ \alpha}_{-v} L(H_++H_-)  + 4\pi^2  k \sum_{\substack{r,s,u,v\\r+s\ne 0}} {1 \over r+s}  \alpha_{k-r-s-u-v} \alpha_{r}\alpha_{s}\bar{\alpha}_{-u}  \bar{\alpha}_{-v}\ ,\cr\\
	&\bar{A}_k^{(2)}\,=\, 2\pi^2  k \sum_{\substack{r,s,u,v\\u+v\ne 0,r+s\ne 0}} {k-r-s-u-v\over (u+v)(r+s)}\bar{\alpha}_{k-r-s -u -v}  \alpha_{-u}  \alpha_{-v} \alpha_{-r} \alpha_{-s} \cr
	&- 4 \pi  k  \sum_{\substack{u,v\\u+v\ne 0}} {1\over u+v} \bar{\alpha}_{k-u-v}  \alpha_{-u}  \alpha_{-v} L(H_+ + H_-)  + 4\pi^2  k \sum_{\substack{r,s,u,v\\r+s\ne 0}} {1 \over r+s}  \bar{\alpha}_{k-r-s-u-v} \bar{\alpha}_{r}\bar{\alpha}_{s}  \alpha_{-u}   \alpha_{-v}\ .
\end{align}
See Appendix~\ref{app: perturbation scalar field} for details. Indeed, one can check that under this solution, the Hamiltonian $H[A,\bar{A}]$ and the momentum $P$ defined by
\begin{align}
	P[A,\bar{A}]\,\equiv\,\int dx \; \spi \phi'\,=\, - {\pi \over L}\sum_k [A_{-k}A_{k}- \bar{A}_{-k}\bar{A}_k]\ ,
\end{align}
are mapped to the expected result for the \ttb deformed spectrum,
\begin{align}
	H[A,\bar{A}]\,=\,&H_++ H_- -4{\lambda\over L} H_+ H_- + {8\lambda^2\over L^2} H_+H_-(H_++H_-)+\cO(\lambda^3)\ ,\\
	P[A,\bar{A}]\,=\,&-H_+ + H_- + \cO(\lambda^3)\ .
\end{align}
So far, our analysis remained classical to avoid the normal ordering issue in finding the solution. We confirm our solution up to order $\cO(\lambda)$ at the quantum level. We find that the normal ordering $:\;\;:$ with respect to the free oscillators $\alpha\,,\, \bar{\alpha}$ can still be used up to order $\cO(\lambda)$, and we take the normal ordering with respect to the free oscillators $\alpha$'s and $\bar{\alpha}$'s for $H, P$ and the solution $A_k^{(1)},\bar{A}_k^{(1)}$ up to order $\cO(\lambda)$,
\begin{align}
	H\,=\,&  {2\pi \over 2L } :\left( A_{-k}A_k + \bar{A}_{-k} \bar{A}_k \right) : - {4\pi^2 \lambda\over L^3 } \sum_{k,q,r} : A_k \bar{A}_{-q} A_{-k-q-r} \bar{A}_{-r}: + \cO(\lambda^2)\ ,\cr
	P\,=\,& - {\pi \over L}\sum_k : [A_{-k}A_{k}- \bar{A}_{-k}\bar{A}_k]:\ ,
\end{align}
\begin{align}
	A^{(1)}_k\,=\,&  2 \pi  \sum_{\substack{r,s\\ r+s\ne 0}}  {k\over r+s} \alpha_{k-r-s}\;:\bar{\alpha}_{-r} \bar{\alpha}_{-s}:\ , \label{eq: quantum first order sol scalar 1} \\
	\bar{A}^{(1)}_k\,=\,&  2 \pi \sum_{\substack{r,s\\ r+s\ne 0}} {k\over r+s} :\alpha_{-r} \alpha_{-s}:\;  \bar{\alpha}_{k-r-s} \ . \label{eq: quantum first order sol scalar 2}
\end{align}
After repeating the same calculations with careful operator ordering, we have
\begin{align}
	H\,=\,& :H_+: + :H_-: -{4\lambda \over L} :H_+: :H_-: + \cO(\lambda^2)\ ,\cr
	P\,=\,& -:H_+: + :H_-: +\cO(\lambda^2)\ .
\end{align}
%

\subsection{\ttb Deformation of Free Fermion}
\label{sec: perturbation of fermion}

Now, we consider the \ttb deformation of the free Majorana fermion of which the undeformed Lagrangian density is given by
\begin{equation}
	\cL_0= i\psi_+ \partial_= \psi_+ + i\psi_- \partial_\+ \psi_- \ .\label{eq: lagrangian of free fermion}
\end{equation}
Here, we denote the light-cone derivative by $\partial_\ppmm\equiv{1\over 2}(\partial_0\pm \partial_1)$. The energy-momentum tensor of a fermionic model is given by
\begin{align}
	{T^=}_\+\,=\,& {\overleftarrow{\delta} \cL \over \overleftarrow{\delta} \partial_= \psi_+ }\partial_\+ \psi_+ + {\overleftarrow{\delta} \cL \over \overleftarrow{\delta} \partial_= \psi_- }\partial_\+ \psi_- \ , \\
	{T^\+}_=\,=\,& {\overleftarrow{\delta} \cL \over \overleftarrow{\delta} \partial_\+  \psi_+ } \partial_= \psi_+ + {\overleftarrow{\delta} \cL \over \overleftarrow{\delta} \partial_\+ \psi_- }\partial_= \psi_-\ , \
	{T^=}_= \,=\,&{\overleftarrow{\delta} \cL \over \overleftarrow{\delta} \partial_= \psi_+ } \partial_= \psi_+ +  {\overleftarrow{\delta} \cL \over \overleftarrow{\delta}  \partial_= \psi_- } \partial_= \psi_-  -\mathcal{L}\ ,\\
	{T^\+}_\+ \,=\,&{\overleftarrow{\delta} \cL \over \overleftarrow{\delta} \partial_\+ \psi_+ }\partial_\+ \psi_+ +  {\overleftarrow{\delta} \cL \over \overleftarrow{\delta} \partial_\+ \psi_- }\partial_\+ \psi_-  -\cL\ ,
\end{align}
where ${\overrightarrow{\delta}  \over \overrightarrow{\delta}\psi}$ and ${\overleftarrow{\delta}  \over \overleftarrow{\delta}\psi} $ denotes the left and right functional derivative with respect to a Grassmannian variable defined by
\begin{align}
    \delta X\, =\, \delta \psi {\overrightarrow{\delta} X  \over \overrightarrow{\delta}\psi}\, =\, {\overleftarrow{\delta} X \over \overleftarrow{\delta}\psi} \delta \psi\ .
\end{align}
Note that the resulting energy-momentum tensor is not symmetric. We will solve the flow equation perturbatively for the $T\bar{T}$ deformation with non-symmetric energy-momentum tensor,
\begin{align}
	&\partial_\lambda \cL\,=\, - \left({T^\+}_\+{T^=}_= - {T^\+}_= {T^=}_\+ \right)\ ,
\end{align}
by expanding the Lagrangian with respect to $\lambda$
\begin{equation}
	\cL\,=\,\sum_{n=0}^\infty\;  \lambda^n \cL_n\ ,
\end{equation}
where $\cL_0$ is the Lagrangian of the free fermion in Eq.~\eqref{eq: lagrangian of free fermion}.
%
%
The (leading) energy-momentum tensor from $\cL_0$ is 
\begin{align}
	{T_{(0)}^=}_\+\,=\,& i\psi_+\partial_\+ \psi_+\ ,\\
	{T_{(0)}^\+}_=  \,=\,& i\psi_-\partial_= \psi_-\ ,\\
	{T_{(0)}^=}_= \,=\,&  - i \psi_-\partial_\+ \psi_-\ ,\\
	{T_{(0)}^\+}_\+ \,=\, &- i \psi_+\partial_=\psi_+\ ,
\end{align}
%
%
%
which determines the first correction $\cL_1$,
\begin{align}
	\cL_1\,=\, \psi_+\partial_=\psi_+\psi_-\partial_\+ \psi_-  -  \psi_+\partial_\+ \psi_+\psi_-\partial_= \psi_- \ .
\end{align}
One can confirm that $\cL_0+\lambda \cL_1$ is indeed the exact solution of the flow equation because of Fermi statistics. Hence, the \ttb deformed Lagrangian of the free fermion is
\begin{align}
	\cL\,=\,& i\psi_+ \partial_= \psi_+ + i\psi_- \partial_\+ \psi_-  + \lambda  \left( - \psi_+\partial_\+ \psi_+\psi_-\partial_= \psi_-  + \psi_+\partial_=\psi_+\psi_-\partial_\+ \psi_- \right)\ .
\end{align}
To calculate Hamiltonian, it is useful to write the Lagrangian in the Cartesian coordinates:
\begin{align}
	\cL\,=\,
	{i\over 2}\psi_+  \dot{\psi}_+ + {i\over 2} \psi_- \dot{\psi}_- - {i\over 2}\psi_+  \psi'_+ + {i\over 2} \psi_- \psi'_-  + {\lambda\over 2}  \left( - \psi_+  \psi'_+ \psi_-\dot{\psi}_-  + \psi_+  \dot{\psi}_+ \psi_-\psi'_- \right)\ ,
\end{align}
where $\dot{\psi}\,\equiv\, \partial_t \psi$ and $\psi'\,\equiv\, \partial_x \psi$. The conjugate momentum $\fpi_\pm$ of the fermion $\psi_\pm$ is calculated by right functional derivative,
\begin{align}
	\fpi_+\,=\, &{\overleftarrow{\delta} \cL \over \overleftarrow{\delta} \dot{\psi}_+}={i\over 2}\psi_+ + {\lambda\over 2}\psi_+ \psi_- \psi'_-\ ,\\
	\fpi_-\,=\,& {\overleftarrow{\delta} \cL \over \overleftarrow{\delta} \dot{\psi}_-}={i\over 2} \psi_- - {\lambda \over 2} \psi_+\psi'_+ \psi_-\ .
\end{align}
The right-hand side does not contain $\dot{\psi}_\pm$, and they form the second-class constraints,
\begin{align}
	\cC_1\,\equiv\, & \fpi_+ -{i\over 2}\psi_+ -  {\lambda\over 2}\psi_+ \psi_- \psi'_-\ ,\\
	\cC_2\,\equiv \,& \fpi_-  -{i\over 2}\psi_- + {\lambda \over 2} \psi_+\psi'_+ \psi_-\ .
\end{align}
We define the Poisson bracket of $F$ and $G$, consistent with any Grassmannian variables $F$ and $G$, by
\begin{align}
	\{F,G\}_\pb\,\equiv\, \sum_{\alpha=\pm }\int dx\; \left[{\overleftarrow{\delta} F \over \overleftarrow{\delta} \psi_\alpha (x) }{\overrightarrow{\delta}G \over \overrightarrow{\delta} \fpi_\alpha(x) }+ {\overleftarrow{\delta} F \over \overleftarrow{\delta} \fpi_\alpha (x) }{\overrightarrow{\delta}G \over \overrightarrow{\delta} \psi_\alpha(x) }\right]\ .\label{def: fermion poisson bracket}
\end{align}
For the Dirac bracket, we need to evaluate the Poisson brackets of the constraints,
\begin{align}
    \cM(i,x_1;j,x_2)\,\equiv\, \{\cC_i(x_1),\cC_j(x_2)\}_\pb\ ,
\end{align}
where each component of the matrix $\cM$ is given by
\begin{align}
	\{\cC_1(x_1),\cC_1(x_2)\}_\pb\,=\,&-i\left[1 -  i\lambda \psi_- (x_1)\psi'_-(x_1) \right]\delta(x_1-x_2)\ ,\\
	\{\cC_2(x_1),\cC_2(x_2)\}_\pb\,=\,& -i \left[1 +  i \lambda  \psi_+(x_1)\psi'_+(x_1) \right]\delta(x_1-x_2)\ ,\\
	\{\cC_1(x_1),\cC_2(x_2)\}_\pb
	\,=\,& \lambda   (\psi'_+ \psi_- +\psi_+ \psi'_-)\delta(x_1-x_2)\ .
\end{align}
%
%
%
%
%
%
%
Using the matrix $\cM$, one can evaluate the Dirac bracket,
\begin{align}
    &\{F(x_1),G(x_2)\}_\dirac \,\equiv\,\{F(x_1),G(x_2)\}_\pb \cr
    &\hspace{10mm}-  \sum_{i,j=1,2}\int dx_3 dx_4 \; \{F(x_1),\cC_i(x_3)\}_\pb \cM^{-1}(i,x_3;j,x_4)\{\cC_j(x_4),G(x_2)\}_\pb\ .
\end{align}
The Dirac brackets of $\psi_\pm$ are found to be
\begin{align}
	i\{\psi_+(x_1),\psi_+(x_2)\}_\dirac\,=\,& ( 1+  \lambda   S_-  + 2\lambda^2 S_+S_- )\delta(x_1-x_2)\ ,\label{eq: dirac bracket 1} \\
	i\{\psi_-(x_1),\psi_-(x_2)\}_\dirac\,=\,&(1-  \lambda  S_+  + 2 \lambda^2 S_+S_- )\delta(x_1-x_2)\ ,\label{eq: dirac bracket 2}\\
	i\{\psi_+(x_1),\psi_-(x_2)\}_\dirac\,=\,& - i \lambda  (\psi'_+\psi_- + \psi_+ \psi'_- )\delta(x_1-x_2)\ ,\label{eq: dirac bracket 3}
\end{align}
where it is useful to introduce Hermitian operators $S_\pm$ defined by
\begin{equation}
	S_\pm\,\equiv\, i \psi_\pm \psi'_\pm\ .
\end{equation}
%
%
%
%
Also, the Dirac brackets of $\fpi_+$ and $\psi_\pm$ are
\begin{align}
	\{ \fpi_+(x_1), \psi_+(x_2)\}_\dirac
	\,=\, &{1\over 2} (1- \lambda^2 S_+S_-) \delta(x_1-x_2)\ ,\\
	\{\fpi_+(x_1),\psi_-(x_2)\}_\dirac
	\,=\,&-{i \lambda\over 2}\psi'_+\psi_-\delta(x_1- x_2)\ ,
\end{align}
and similar for $\fpi_-$. With constraint, the Hamiltonian (density) is given by
\begin{align}
	&\cH\,\equiv \,\fpi_+ \dot{\psi}_+ + \fpi_- \dot{\psi}_-  -\mathcal{L}\cr
	\,=\,&\left(\fpi_+ -{i\over 2} \psi_+ - {\lambda\over 2} \psi_+ \psi_- \psi'_- \right) \dot{\psi}_+ + \left(\fpi_- - {i\over 2} \psi_- + {\lambda \over 2} \psi_+\psi'_+ \psi_- \right) \dot{\psi}_-   + {i\over 2}\psi_+ \psi'_+- {i\over 2}\psi_- \psi'_-\cr
	\,=\,&{i\over 2}\psi_+ \psi'_+- {i\over 2}\psi_- \psi'_-\ .\label{eq: ham density of free Majorana fermion}
\end{align}
%
%
%
At first glance, it does not seem possible to get the conjectured form of Hamiltonian~\eqref{eq: deformed hamiltonian conjecture} from the deformed Hamiltonian density~\eqref{eq: ham density of free Majorana fermion} because the Hamiltonian density~\eqref{eq: ham density of free Majorana fermion} is of the same form as that of the undeformed model. However, unlike the scalar field case, the deformed fermions satisfy non-trivial Dirac brackets~\eqref{eq: dirac bracket 1}$\sim$\eqref{eq: dirac bracket 3}, which allows us to match two Hamiltonians. To see this, we first consider the mode expansion of the fermions (with circumference $L$):
\begin{align}
	\psi_\pm(x)\, =\, {1 \over\sqrt{ L} } \sum_k \psi_{\pm,k} e^{2\pi i k x \over L} \ ,
\end{align}
where $k\in \mathbb{Z}+1/2$ for anti-periodic boundary condition and $k\in \mathbb{Z}$ for periodic boundary condition. It is also useful to Fourier expand $S_\pm\, ,\;  S_+S_-$ and $i(\psi'_+\psi_- + \psi_+ \psi'_-)$:
\begin{align}
	S_\pm(x)\,=\,&{1\over L^2} \sum_p S_{\pm, p} e^{2\pi i p x\over L}\ ,\\
	(S_+S_-)(x)\,=\,&{1\over L^4} \sum_p (S_+S_-)_{p} e^{2\pi i p x\over L}\ ,\\
	i(\psi'_+\psi_- + \psi_+ \psi'_-)(x)\,=\,&{1\over L^2} \sum_p K_{p} e^{2\pi i p x\over L}\ .
\end{align}	
The Dirac brackets of the modes $\psi_{\pm,k}$ can be written as
\begin{align}
	i\{\psi_{+, k},\psi_{+,q}\}_\dirac\,=\,& \delta_{k+q,0} + { \lambda\over L^2} S_{-,k+q} +  {2\lambda^2 \over L^4}(S_+S_-)_{k+q}\ , \label{eq: dirac bracket mode 1}\\
	i\{\psi_{-, k},\psi_{-,q}\}_\dirac\,=\,& \delta_{k+q,0} - {  \lambda\over L^2} S_{+,k+q} + {2 \lambda^2 \over L^4}(S_+S_-)_{k+q} \ , \label{eq: dirac bracket mode 2}\\
	i\{\psi_{+, k},\psi_{-,q}\}_\dirac\,=\,&-{  \lambda \over L^2} K_{k+q} \ .\label{eq: dirac bracket mode 3}
\end{align}
This nonstandard Dirac bracket is at the heart of how the Hamiltonian, which apparently has no  $\lambda$ at all when written in terms of $\psi$'s, can generate a Taylor series with infinitely many terms. For actual evaluation of the eigenvalues of Hamiltonian and momentum, we will introduce a new set of operators that obey the usual fermi harmonic oscillator algebra,
\begin{align}
	i\{b_k,b_q\}_\dirac\,=\,\delta_{k+q,0}\;\;,\;\; i \{\bar{b}_k,\bar{b}_q\}_\dirac \, =\,  \delta_{k+q,0}\;\;,\;\; i\{b_k,\bar{b}_q\}_\dirac\,=\,0\ .\label{eq: dirac bracket of b}
\end{align}
and we reconstruct $\psi$ by
\begin{align}
	\psi_{+,k}\,=\, b_k + {\lambda\over L^2} \psi^{(1)}_{+,k}[b,\bar{b}] +\cdots\;\;,\qquad \psi_{-,k}\,=\,\bar{b}_k + {\lambda\over L^2} \bar{\psi}^{(1)}_{-,k}[b,\bar{b}] +\cdots\ , \label{eq: expansion of psi}
\end{align}
such that the Dirac bracket of $\psi$'s is properly incorporated.

We will further demand that the map between $\psi_{+,k}\;,\; \psi_{-,k}$ and $b_q\;,\; \bar{b}_q $ is such that the Hamiltonian $\widetilde{H}[b_k,\bar{b}_k]\, \equiv\, H\big[\psi_{+}(b,\bar{b}),\psi_{-}(b,\bar{b}) \big]$ is consistent with the energy spectrum. Namely, $\widetilde{H}[b,\bar{b}]$ is of form 
\begin{align}
	H[\psi_{+},\psi_{-}]\,=\,\wH [b,\bar{b}]\,=\, {L \over 2\lambda } \left[ \sqrt{1+ {4\lambda\over  L } (H_++H_-) + { 4\lambda^2 \over L^2} ( H_+ - H_-)^2 } - 1 \right]\ ,\label{eq: fermion ttb hamiltonian equation sec 2}
\end{align}
where $H_\pm[b,\bar{b}] $ is defined by
\begin{align}
	H_+\,\equiv\, -{\pi \over L} \sum_k k b_{-k} b_k \qquad, \qquad H_- \,\equiv\,  {\pi \over L} \sum_k k \bar{b}_{-k} \bar{b}_k\ .
\end{align}
If these steps are achieved, we will have effectively shown that the first-principle quantization of \ttb deformed Majorana fermion does reproduce the anticipated eigenvalue flow. 
We find a map between $\psi$'s and $b$'s perturbatively up to order $\cO(\lambda)$ by solving the equation~\eqref{eq: fermion ttb hamiltonian equation sec 2} and by demanding that non-trivial Dirac brackets of $\psi$'s is realized by $b$ and $\bar{b}$. To avoid the ordering ambiguity, one may first search for the map classically. At the quantum level, one needs to take care of operator ordering. Especially, the \ttb deformation of the fermion has additional ambiguity of operator ordering in the Dirac bracket. That is, when we promote the Dirac bracket~\eqref{eq: dirac bracket mode 1}$\sim$\eqref{eq: dirac bracket mode 3} to the anti-commutation relation, the operators on the right-hand side of the Dirac brackets have ordering ambiguity. One may choose a prescription for the operator ordering, and one can test it with certain criteria, such as the Jacobi identity of the Dirac bracket. But, although this could rule out inconsistent one, it is not clear which criteria is a sufficient condition for a consistent prescription for the ordering.

One can see that the Jacobi identity of Dirac brackets of $\psi_{\pm,k}$ holds up to order $\cO(\lambda)$ irrespective of any choice of ordering. Thus, it is natural to choose the normal ordering prescription with respect to the free fermi oscillators $b$ and $\bar{b}$ up to order $\cO(\lambda)$. Hence, we obtain a solution,
\begin{align}
	\psi_{+,k}^{(1)}\,=\,&2\pi \sum_{\substack{r,s\\r+s\ne 0}} {(k-r-s)s\over r+s}b_{k-r-s} : \bar{b}_r \bar{b}_s : - \pi    b_k \sum_r  r  : \bar{b}_{-r} \bar{b}_r: \ , \\
	\psi_{-,k}^{(1)}\,=\,&-2\pi \sum_{\substack{r,s\\r+s\ne 0}}{(k-r-s)s\over r+s} :b_r b_s: \bar{b}_{k-r-s} +  \pi  \sum_r r  :b_{-r} b_r: \bar{b}_k\ .
\end{align}
We confirm that the map gives the expected result at the quantum level up to order $\cO(\lambda)$,
\begin{align}
	H^{\text{\tiny qu}}\,=\,& :H_+: + :H_-: - {4 \lambda \over L^2}:H_+:  :H_-: + \cO(\lambda^2)\ ,\\
	P^{\text{\tiny qu}}\,=\,& :H_+: - :H_-: + \cO(\lambda^2)\ .
\end{align}
%

\section{\ttb Eigenvalue Flows via Worldsheet Theories}
\label{sec: string theory and ttbar deformed hamiltonian}

We have seen that the direct canonical quantization of the simplest \ttb deformed theories 
reproduce the leading Taylor expansions of the anticipated eigenvalue flow \eqref{eq: deformed energy value}. 
However, it would be difficult to go beyond order $\cO(\lambda^2)$ with this perturbative method. 
On the other hand, it was demonstrated by Refs.~\cite{theisenTalk:2018,Callebaut:2019omt,Jorjadze:2020ili} 
that the worldsheet Polyakov action for flat 3D target space with a unit ``winding number" and in 
the light-cone gauge reproduces the \ttb deformed Hamiltonian for massless scalars. With a view toward 
fermions and  supersymmetric theories, we wish to extend the same approach to worldsheet theories  
with fermions, either Neveu-Schwarz-Ramond~(NSR) type or Green-Schwarz (GS) type, included. 
Unlike the scalar-only case, which we review first, we encounter various difficulties in the presence of fermions, 
so this approach does not offer a good handle for the full quantum spectrum. Nevertheless,
there are lessons to be learned, which we will later make use of for the canonical analysis of a \ttb-deformed
supersymmetric theory.

\subsection{Via Polyakov Worldsheet in Lightcone: A Review}
\label{sec: polyakonv action}

We begin with the review of Ref.~\cite{Jorjadze:2020ili} for the \ttb deformed spectrum from the Polyakov action with flat target space metric,
\begin{align}
    S\,=\,- {1\over 2\lambda } \int d\tau d\sigma\; {1\over 2} \sqrt{-h} h^{\alpha\beta }\partial_\alpha X^\mu \partial_\beta X_\mu +  {1\over 2\lambda}\int d\tau d\sigma \;{1\over 2}B_{\mu\nu}\epsilon^{\alpha\beta}\partial_\alpha X^\mu \partial_\beta X^\nu\ , \label{eq: polyakov action theisen}
\end{align}
where ${1\over 2\lambda}$ denotes the string tension. Here, we add\footnote{One may consider the Polyakov action without the $B$-field as in Ref.~\cite{Jorjadze:2020ili}. In that case, one needs a constant shift to identify the string energy with \ttb deformed energy.} the constant Kalb-Ramond $B$-field~\cite{Hashimoto:2019wct,Callebaut:2019omt} given by
\begin{align}
    B_{\mu\nu}\,=\,\begin{pmatrix}
    0 & 1 & 0\\
    -1 & 0 & 0\\
    0 & 0 & 0\\
    \end{pmatrix}\qquad, \hspace{12mm} (\mu,\nu\,=\,+,-,2)\ .
\end{align}
One can rewrite the Polyakov action in the first order form by introducing the conjugate momentum $p_\mu$,
\begin{align}
    S\,=\,\int d\tau \int d\sigma\, \big[\momp_\mu \dot{x}^\mu  +\lambda_1\cC_1+\lambda_2\cC_2 \big]\ ,\label{eq: polyakov first order form}
\end{align}
where the Lagrange multiplier is given by $\lambda_1\,=\,\frac{\gamma^{\tau\sigma}}{\gamma^{\tau\tau}}$ and $\lambda_2\,=\,\frac1{2\gamma^{\tau\tau}}$ with $\gamma^{\alpha\beta}\,=\, \sqrt{-h}h^{\alpha\beta}$. In addition, the Virasoro constraints $\cC_1$ and $\cC_2$ are\footnote{In this note, the light-cone target coordinates are defined by $X^+\,=\,{1\over 2} (X^1+X^0)\; ,\; X^-\,=\,X^1-X^0$.}
\begin{align}
    \cC_1\,\equiv\, \momp_\mu X^{\mu\prime}\;\;,\quad \cC_2\,\equiv\,4\lambda \bigg(\momp_+ - {1\over 2 \lambda} X^{-\prime}\bigg)\bigg( \momp_- + {1\over 2\lambda} X^{+\prime}\bigg)+2\lambda (\momp_2)^2+\frac1{2\lambda} X'_\mu X^{\mu\prime}\ .
\end{align}
Note that the shift in the momentum comes from the $B$-field term. To make contact with the \ttb deformation, a subsector in which the target coordinate $X^1$ has winding number $1$ was considered in Refs.~\cite{theisenTalk:2018,Callebaut:2019omt,Jorjadze:2020ili}. For this, one can choose the following gauge condition,
\begin{align}
    X^+\,= \, \bigg(2\lambda \pz_- +{L\over 4\pi} \bigg)\tau + {L\over 4\pi }\sigma\label{eq: theisen gauge boson}\ ,
\end{align}
where $L$ is the circumference of the target coordinate $X^1$ and $\pz_\mu$ is the zero-mode of the conjugate momentum $\momp_\mu$:
\begin{align}
    \pz_\mu \,\equiv \, {1\over 2\pi }\int_0^{2\pi} d\sigma\; \momp_\mu\ .
\end{align}
Compared with the \ttb deformation, the target coordinate $X^2$, which corresponds to the scalar field in the \ttb deformation, is rescaled by $\lambda$ to make $\phi$ and $\spi$ dimensionless as follows,
\begin{equation}
    X^2\,=\, \sqrt{2\lambda}\phi\;\;,\qquad \momp_2\,=\,\frac\spi{\sqrt{2\lambda}}\ .\label{eq: rescaling of x2}
\end{equation}
In this light-cone gauge~\eqref{eq: theisen gauge boson}, the Virasoro constraints can be written as
\begin{align}
    \cC_1\,=\,&{L\over 4\pi }\momp_++\pz_- X^{-\prime}+\cP_{b,0}=0\label{eq: Theisen constraint1}\ ,\\
    \cC_2\,=\,& 4\lambda \pz_- \momp_+ -2\pz_- X^{-\prime} +{L\over 2\pi} \momp_+  +2 \cH_{b,0}=0\label{eq: Theisen constraint2}\ ,
\end{align}
where $\cH_{b,0}$ and  $\cP_{b,0}$ is the Hamiltonian and momentum density of the 2D free scalar field
\begin{align}
    \cH_{b,0}\,=\,& {1\over 2}\spi^2 + {1\over 2} \phi'^2\ ,\\
    \cP_{b,0}\,=\,& \spi \phi'\,.
\end{align}
We integrate the constraints $\cC_1$ and $\cC_2$ in Eqs.~\eqref{eq: Theisen constraint1} and \eqref{eq: Theisen constraint2} over $\sigma\in [0,2\pi ]$. Also, it was demanded~\cite{theisenTalk:2018,Callebaut:2019omt,Jorjadze:2020ili} that $X^0$ coordinate has zero winding number. Together with Eq.~\eqref{eq: theisen gauge boson}, this implies that $X^-$ has $1$ winding number, and we have
\begin{align}
    \int_0^{2\pi}  d\sigma \; \partial_\sigma X^- \,=\, L\ .
\end{align}
Therefore, the integrated Virasoro constraints become
\begin{align}
    &\frac12\pz_+ + \pz_-+{1\over 2\pi }P_{b,0}\,=\,0\ , \label{eq: polakov zero-mode constraint 1}\\
    &4\pi\lambda  \pz_+ \pz_-  + L\big( {1\over 2}\pz_+ -\pz_- \big)  + {L\over 2\pi} H_{b,0}\,=\,0\ , \label{eq: polakov zero-mode constraint 2}
\end{align}
where we defined the charge $H_{b,0}$ and $P_{b,0}$ (of length dimension $-1$) by
\begin{align}
    H_{b,0}\,\equiv \, {2\pi \over L} \int_0^{2\pi }d\sigma\; \cH_{b,0}\ ,\\
    P_{b,0}\,\equiv \, {2\pi \over L} \int_0^{2\pi }d\sigma\; \cP_{b,0}\ .
\end{align}
A solution of those equations is
\begin{align}
    \pz_+\,=\,&- {P_{b,0}\over 2\pi } +{L\over 4\pi \lambda} - {L \over 4\pi \lambda } \sqrt{1+{4\lambda\over L} H_{b,0}+{4\lambda^2\over L^2} P_{b,0}^2 }\ ,\\
    \pz_-\,=\,&- {P_{b,0}\over 4\pi }- {L\over 8\pi \lambda} + {L \over 8\pi \lambda } \sqrt{1+{4\lambda\over L} H_{b,0}+{4\lambda^2\over L^2} P_{b,0}^2 }\ .
\end{align}
The string ``energy''\footnote{More precisely, the target time coordinate $X^0$ is identified with the worldsheet time $\tau$ in a way that $\dot{X}^0\,>\,0$ \cite{Jorjadze:2020ili}. Hence, $p^0$ is positive, and we identify it with energy.} $E_{str}$ related to the charge generating the translation along $X^0$ is
\begin{equation}
    E_{str} \,\equiv\, \int_0^{2\pi} d\sigma \; \momp^0 \,= \, -2\pi \left({1\over 2}\pz_+-\pz_-\right)={L\over 2\lambda}\bigg[\sqrt{1+{4\lambda\over L} H_{b,0}+{4\lambda^2\over L^2} P_{b,0}^2}- 1\bigg] \ . \label{eq: theisen string energy}
\end{equation}
This agrees with the \ttb deformed energy~\eqref{eq: deformed hamiltonian conjecture}.

\subsection{NSR-like Extension}
\label{sec: rns superstring}

In Section~\ref{sec: perturbation of fermion}, we have shown perturbatively up to $\cO(\lambda)$ that the \ttb deformed spectrum~\eqref{eq: deformed hamiltonian conjecture} can be obtained for the case of the \ttb deformation of fermion. Now, we will generalize the Jorjadze-Theisen’s method~\cite{theisenTalk:2018,Jorjadze:2020ili} to derive the \ttb deformed Hamiltonian for fermion. To incorporate fermion, we first utilize the NSR-like action for 3D flat target space with the three pairs of worldsheet Majorana fermions~$\psi_\pm^\mu$\;\; $(\mu=\pm, 2)$,
\begin{align}
	S_{\text{\tiny NSR}}\,=\,&\frac1{2\lambda}\int d\tau d\sigma \;\left(2\partial_\+ X^\mu\partial_=X_\mu+i\Psi_+^\mu\partial_=\Psi_{+\mu}+i\Psi_-^\mu\partial_\+\Psi_{-\mu} \right) \cr
	&+ {1\over 2\lambda}\int d\tau d\sigma\; {1\over 2} B_{\mu\nu}\epsilon^{\alpha\beta}\partial_\alpha X^\mu \partial_\beta X^\nu \ ,\label{3DRNSaction}
\end{align}
where the Regge slope $2\lambda$ will become the \ttb deformation parameter and the flat target space metric\footnote{The light-cone target coordinates are defined by $X^+\,=\,\frac12(X^1+X^0)\; ,\; X^-\,=\,X^1-X^0$ and $\Psi^+_\pm\,=\,\frac12(\Psi^1_\pm+\Psi^0_\pm)\; ,\; \Psi_\pm^-\,=\,\Psi_\pm^1-\Psi_\pm^0$. On the other hand, the worldsheet light-cone is defined by $x^\pm\,=\,\tau\pm \sigma$. Hence, we have $\partial_\+\,=\,{1\over 2}(\partial_0 +\partial_1)$ and $\partial_=\,=\,{1\over 2}(\partial_0 -\partial_1)$.} is given by
\begin{align}
    ds^2\,=\, 2dX^+ dX^-+(dX^2)^2\ .
\end{align} 
We also include the coupling to the constant $B$-field,
\begin{align}
    B_{\mu\nu}=\begin{pmatrix}
    0 & 1 & 0\\
    -1 & 0 & 0\\
    0 & 0 & 0\\
    \end{pmatrix}\qquad, \hspace{12mm} (\mu,\nu\,=\,+,-,2)\ .\label{eq: b field nsr}
\end{align}
The Virasoro constraints of the NSR-like model in Eq.~\eqref{3DRNSaction} are
\begin{align}
	T_{\+\+}\,=\,&\partial_\+ X^\mu \partial_\+ X_\mu+\frac i2\Psi_+^\mu\partial_\+\Psi_{+\mu}=0\ ,\label{RNSVirasoro1}\\
	T_{==}\,=\,&\partial_= X^\mu \partial_= X_\mu+\frac i2\Psi_-^\mu\partial_=\Psi_{-\mu}=0\ .\label{RNSVirasoro2}
\end{align}
As in the Polyakov case, demanding that $X^1$ have winding number $1$, we choose the following light-cone gauge:
\begin{equation}
	X^+\,=\, \bigg(2\lambda \pz_- + {L\over 4\pi }\bigg)\tau+{L\over 4\pi}\sigma \; ,\quad \Psi^+_\pm\,=\,0\ .\label{RNSgauge}
\end{equation}
Also, we rescale $X^2$ and $\Psi_\pm^2$ by $2\lambda$ to make them dimensionless,
\begin{equation}
	X^2\,=\, \sqrt{2\lambda}\phi\ ,\quad \Psi^2_\pm\,=\,\sqrt{2\lambda}\psi_\pm\ .
\end{equation}
With the light-cone gauge~\eqref{RNSgauge}, the Virasoro constraints~\eqref{RNSVirasoro1} and \eqref{RNSVirasoro2} can be written as
\begin{align}
	T_{\+\+}\,=\,&\bigg(\lambda \pz_-	+ {L\over 4\pi}\bigg)(\dot X^-+X^{-\prime})+\lambda(\cH_{b,0}+\cP_{b,0})+\lambda(\cH_{f,0}+\cP_{f,0})=0\ ,\\[6pt]
	T_{==}\,=\,&\lambda \pz_-	(\dot X^--X^{-\prime})+\lambda(\cH_{b,0}-\cP_{b,0})+\lambda(\cH_{f,0}-\cP_{f,0})=0\ ,
\end{align}
where $\cH_{(b,f),0}$, $\cP_{(b,f),0}$ are defined by
\begin{alignat}{4}
    \cH_{b,0}\,\equiv \,& {1\over 2}\pi^2 + {1\over 2} \phi'^2 \;\;,\qquad&\cP_{b,0}\,\equiv \,&  \pi \phi'\ , \\
    \cH_{f,0}\,\equiv\,& {i\over 2}\psi_+\psi'_+ - {i\over 2}\psi_-\psi'_-  \;\;,\qquad &\cP_{f,0}\,\equiv\,&  {i\over 2}\psi_+\psi'_+ + {i\over 2}\psi_-\psi'_-  \ ,
\end{alignat}
which correspond to the Hamiltonian and momentum density operators of the undeformed model from the point of \ttb deformation. By integrating the constraints over $\sigma$, we have
\begin{align}
	\int_0^{2\pi} d\sigma\; T_{\+\+}\,=\, &4\pi\lambda\pz_+ \bigg(\lambda \pz_-	+{L\over 4\pi}\bigg)+\frac{L\lambda}{2\pi}(H_{b,0}+H_{f,0}+P_{b,0}+P_{f,0})=0\ ,\label{eq: nsr zero-mode constraint 1}\\
	\int_0^{2\pi} d\sigma\; T_{==}\,=\, &\lambda \pz_-\left(4\pi\lambda\pz_+-2L\right)+\frac{L\lambda}{2\pi}(H_{b,0}+H_{f,0}-P_{b,0}-P_{f,0})=0\ . \label{eq: nsr zero-mode constraint 2}
\end{align}
Here we used the fact that there exists winding number $1$ along $X^1$. In addition, we defined the operators $H_{b,0}\,,\;P_{b,0}\,,\;H_{f,0}$ and $P_{f,0}$ of length dimension $-1$ by 
\begin{alignat}{4}
    H_{b,0}\,\equiv \,& {2\pi \over L} \int d\sigma \;\cH_{b,0}\;\;,\qquad P_{b,0}\,\equiv \,&{2\pi \over L} \int d\sigma \;\cP_{b,0}\ , \\
    H_{f,0}\,\equiv\,& {2\pi \over L}\int d\sigma\; \cH_{f,0}\;\;,\qquad P_{f,0}\,\equiv\,& {2\pi \over L}\int d\sigma\; \cP_{f,0}\ .
\end{alignat}
One of solutions of the constraints~\eqref{eq: nsr zero-mode constraint 1} and \eqref{eq: nsr zero-mode constraint 2} is 
\begin{align}
	\pz_+\,=\,&-{P_{0}\over 2\pi}+{L\over 4\pi \lambda}-{L\over 4\pi\lambda}\sqrt{1+{4\lambda\over L} H_{0}+{4\lambda^2\over L^2} P_{0}^2 }\ ,\\
	\pz_-\,=\,&-{P_{0}\over 4\pi}-{L\over 8\pi \lambda}+{L\over 8\pi\lambda}\sqrt{1+{4\lambda\over L} H_{0}+{4\lambda^2\over L^2} P_{0}^2 }\ .
\end{align}
The string ``energy'' $p^0$ related to the translation in the $X^0$ direction is
\begin{equation}
	E_{str}\equiv\int_0^{2\pi} d\sigma\,\momp^0=-2\pi\left({1\over2}\pz_+-\pz_-\right)={L \over 2\lambda} \bigg[\sqrt{1+ {4\lambda \over L}H_{0} + {4\lambda^2 \over L^2} P_{0}^2 } -1 \bigg]\ , \label{eq: deformed hamiltonian nsrlike string}
\end{equation}
which reproduces the \ttb deformed spectrum~\eqref{eq: deformed hamiltonian conjecture}.

As in Polyakov action, it is natural to ask which Lagrangian we will get if we integrate out the worldsheet metric of NSR-like action. For this, let us consider NSR-like action for 3D flat target space with constant background $B$-field:
\begin{align}
	S_{\text{\tiny NSR}}\,=\,& {1\over 2\lambda}\int d\tau d\sigma \sqrt{-h} \;\left( - {1\over 2}h^{\alpha\beta}\partial_\alpha X^\mu\partial_\beta X_\mu+{i\over 4} \bar{\Psi}^\mu \gamma^\alpha \nabla_\alpha \Psi_{\mu} - {i\over 4} \nabla_\alpha \bar{\Psi}^\mu \gamma^\alpha  \Psi_{\mu} \right)\cr
	&+{1\over 2\lambda}\int d\tau d\sigma \; {1\over 2} B_{\mu\nu} \epsilon^{\alpha \beta} \partial_\alpha X^\mu \partial_\beta X^\nu \ ,\label{eq: nsrlike string action}
\end{align}
where the constant $B$-field is given in Eq.~\eqref{eq: b field nsr}. Moreover, the action of the covariant derivative is defined by
\begin{align}
    \nabla_\alpha \Psi = \partial_\alpha \Psi -  \omega_\alpha \sigma_3 \Psi\;\;,\quad \nabla_\alpha \bar{\Psi} = \partial_\alpha \bar{\Psi} +  \omega_\alpha  \Psi \sigma_3\ ,
\end{align}
where we defined $\omega_\alpha^{ab}\,\equiv\,\epsilon^{ab}\omega_\alpha$. Due to the spin connection $\omega_\alpha$, one has to vary the action with respect to the zweibein $e^\alpha_a$. Since the fermion part in the action~\eqref{eq: nsrlike string action} is anti-symmetrized by integrating by part, its dependence on the zweibein becomes extremely simple~\cite{Bonelli:2018kik}, and one can easily obtain the equation of motion for $e^\alpha_a$ or $h_{\alpha\beta}$:
\begin{align}
	-{4\lambda \over \sqrt{-h} } e^{\alpha}_a e^{\beta}_b {\delta S_{\text{\tiny NSR}}\over \delta h^{\alpha \beta}}\,=\,e^{\alpha}_a e^{\beta}_b \left[\rho_{\alpha \beta} - {1\over 2} h_{\alpha \beta} h^{\sigma \delta} \rho_{\sigma \delta}\right]\,=\,0\ , \label{eq: nsr ws eom}
\end{align}
where we define $\rho_{\alpha \beta}$ by
\begin{align}
	\rho_{\alpha \beta}\,\equiv\,\partial_\alpha X^\mu \partial_\beta X_\mu - {i\over 2}\bar{\Psi}\gamma_\alpha \partial_\beta \Psi - {i\over 2}\bar{\Psi}\gamma_\beta \partial_\alpha \Psi \ .
\end{align}
Note that in terms of $\rho_{\alpha \beta}$ and $h_{\alpha\beta}$ the NSR-like action~\eqref{eq: nsrlike string action} can be written as
\begin{align}
    S_{\text{\tiny NSR}}\,=\, -{1\over 4\lambda} \int d\tau d\sigma \sqrt{-h} h^{\alpha \beta} \rho_{\alpha\beta}+{1\over 2\lambda}\int d\tau d\sigma\;{1\over 2} B_{\mu\nu} \epsilon^{\alpha \beta} \partial_\alpha X^\mu \partial_\beta X^\nu\ .
\end{align}
As in derivation of the Nambu-Goto action from Polyakov action, the solution of the equation~\eqref{eq: nsr ws eom} is $h_{\alpha\beta}= f(\tau,\sigma) \rho_{\alpha \beta}$ where $f(\tau,\sigma)$ is a function which is irrelevant in the final result. From the solution, we have
\begin{align}
    S_{\text{\tiny NSR}}\,=\, -{1\over 2\lambda} \int d\tau d\sigma \sqrt{-\rho} +{1\over 2\lambda}\int d\tau d\sigma\; {1\over 2}  B_{\mu\nu} \epsilon^{\alpha \beta} \partial_\alpha X^\mu \partial_\beta X^\nu\ ,\label{eq: nsrlike action ngtype}
\end{align}
where we defined $\rho\,\equiv\,\det\rho_{\alpha\beta}$. We find that extra degrees of freedom, especially, related to fermion appear in this action. To see this in a simple way, it is enough to truncate all fields except for $\Psi^2$. \eg
\begin{align}
	\Psi^2\,=\,\sqrt{2\lambda} \begin{pmatrix}
	\psi_+\\
	\psi_-\\
	\end{pmatrix}\quad \mbox{and} \quad  X^2\,=\,0 \ .
\end{align}
By this truncation, together with the static gauge
\begin{align}
	X^0\,=\,{L\over 2\pi}\tau\;\;,\quad X^1\,=\,{L\over 2\pi }\sigma\ ,\label{eq: static gauge theisen}
\end{align}
the action~\eqref{eq: nsrlike action ngtype} is reduced to an action of the fermion $\psi_\pm$,
\begin{align}
	S^{\text{\tiny tr}}_{\text{\tiny NSR}} \,=\,& \int d\tau d\sigma \left[ {i\over 2} ( \psi_+\dot{\psi}_+ +\psi_-\dot{\psi}_- - \psi_+\psi'_+  + \psi_-\psi'_-  ) \right]\cr
	- &\Lambda\int d\tau d\sigma \left[ \psi_+ \dot{\psi}_+\psi_- \dot{\psi}_-  - \psi_+ \dot{\psi}_+\psi_- \psi'_- + \psi_+ \psi'_+\psi_- \dot{\psi}_- - \psi_+ \psi'_+\psi_- \psi'_- \right]\ ,\label{eq: truncated nsr action}
\end{align}
where we define the dimensionless parameter $\Lambda\,\equiv\, {4\pi^2\lambda\over L^2}$. 

Usual fermion theories have second-class constraints that impose a relation between the fermion and the conjugate momentum, leading to a Dirac bracket. However, the action~\eqref{eq: truncated nsr action} does not have such constraints because of the term which is quadratic in the time derivatives of fermions, \ie $\psi_+ \dot{\psi}_+\psi_- \dot{\psi}_-$. This implies that extra degrees of freedom which the constraints would have removed are now coupled like the symplectic fermion in Refs.~\cite{LeClair:2007iy,Robinson:2009xm}. Furthermore, those extra degrees of freedom often imply a unitarity issue.
We will come back to and address such subtleties in Section~\ref{sec: instability}.

\subsection{GS-like Extension, or a Failure Thereof}
\label{sec: gs superstring theory and hamiltonian}

There is another way to incorporate fermions--Green-Schwarz-like action. This section considers the Green-Schwarz~(GS)-like action to find a relation to \ttb deformation. Let us consider the $\cN=2$ GS-like action for 3D target space given by
\begin{align}
    S_{\text{\tiny GS}}\,=\,{1\over 2\lambda} \int d\tau d\sigma \; \mathcal{L}_{\text{\tiny GS}}\ ,
\end{align}
with
\begin{align}
	\mathcal{L}_{\text{\tiny GS}}\,=\,&-\frac12\gamma^{\alpha\beta}\bPi^\mu_\alpha\bPi^\nu_\beta G_{\mu\nu} -i\epsilon^{\alpha\beta}\partial_\alpha X^\mu (\bar\Psi_+\Gamma^\nu\partial_\beta\Psi_+-\bar\Psi_-\Gamma^\nu\partial_\beta\Psi_-)G_{\mu\nu}\cr
	&-\epsilon^{\alpha\beta}(\bar\Psi_+\Gamma^\mu\partial_\alpha\Psi_+)(\bar\Psi_-\Gamma^\nu\partial_\beta\Psi_-)G_{\mu\nu}\ ,\label{eq: 3DGSaction for flat}\\[6pt]
	\bPi^\mu_\alpha\,\equiv \,& \partial_\alpha X^\mu+i\bar\Psi_+\Gamma^\mu\partial_\alpha\Psi_++i\bar\Psi_-\Gamma^\mu\partial_\alpha\Psi_-\ ,
\end{align}
where $2\lambda$ corresponds to the Regge slope parameter which will become the \ttb deformation parameter. Moreover, we defined $\epsilon^{\tau \sigma}=1$ and $\gamma^{\alpha\beta}\,\equiv\, \sqrt{-h}h^{\alpha\beta}$ with worldsheet metric $h^{\alpha\beta}$. Also, note that $\Psi_\pm$ denotes the two-component Majorana spinors for 3D target space with $\bar\Psi_\pm\,=\,\Psi^T\Gamma^0$.  We consider a flat target space\footnote{Again, note that the light-cone coordinates for target space is defined by $X^+\,\equiv {1\over 2} (X^1+X^0) $ and $X^-=X^1-X^0$.} 
\begin{align}
    ds^2\,=\, 2dX^+ dX^-+(dX^2)^2\ .
\end{align}
As in the bosonic string, we are interested in the background where the target coordinate $X^1$ has winding number $1$ (and the winding number of $X^0$ is $0$). Hence, we take the following light-cone gauge,
\begin{align}
    X^+\,=\, 2\lambda \pz_- \tau + {L\over 4\pi }\sigma\ , \label{eq: lightcone gauge gs theisen}
\end{align}
where $\pz_\mu$ is the zero-mode of the conjugate momentum $\momp_\mu$:
\begin{align}
    \pz_\mu \,\equiv \, {1\over 2\pi }\int_0^{2\pi} d\sigma\; \momp_\mu\ .
\end{align}
We will solve the (zero-mode) Virasoro constraints to find the relation to the \ttb deformed Hamiltonian as in Polyakov action. The Virasoro constraint can be read off when we write the GS-like Lagrangian in the first-order form,
\begin{align}
\mathcal L\,=\,&\mpi_\mu\bPi_\tau^\mu+\frac{\gamma^{\tau\sigma}}{\gamma^{\tau\tau}}\cC_1+\frac1{2\gamma^{\tau\tau}}\cC_2-i\epsilon^{\alpha\beta}\partial_\alpha X^\mu (\bar\Psi_+\Gamma^\nu\partial_\beta\Psi_+-\bar\Psi_-\Gamma^\nu\partial_\beta\Psi_-)G_{\mu\nu} \cr
	&\qquad-\epsilon^{\alpha\beta}(\bar\Psi_+\Gamma^\mu\partial_\alpha\Psi_+)(\bar\Psi_-\Gamma^\nu\partial_\beta\Psi_-)G_{\mu\nu}\ ,\cr
=\,&\momp_\mu\dot X^\mu+\fpi_+ \dot\psi_++\fpi_-\dot\psi_-+\frac{\gamma^{\tau\sigma}}{\gamma^{\tau\tau}}\cC_1+\frac1{2\gamma^{\tau\tau}}\cC_2\,,\label{eq: 1st order gs lag theisen}
\end{align}
where we define
\begin{equation}
\mpi_\mu\,\equiv\,\frac{\delta \mathcal S}{\delta\bPi^\mu_\tau}\;\; ,\quad \momp_\mu\,\equiv\,\frac{\delta\mathcal S}{\delta \dot X^\mu}\;\; ,\quad \fpi_\pm\,\equiv\,\frac{\delta\mathcal S}{\delta \dot \psi_\pm}\ , \label{eq: gs momenta in theisen}
\end{equation}
and the Virasoro constraints $\cC_1$ and $\cC_2$ are given by
\begin{align}
\cC_1=\mpi_\mu\bPi_\sigma^\mu\;\;,\qquad \cC_2=2\lambda G^{\mu\nu}\mpi_\mu\mpi_\nu+ {1\over 2\lambda} G_{\mu\nu} \bPi^\mu_\sigma \bPi^\nu_\sigma 	\ . \label{eq: virasoro constraint gs theisen}
\end{align}

The GS-like action has additional local fermionic symmetry~\cite{Green:1983wt,Green:1983sg,Mezincescu:2011nh}. For example, the spinor $\Psi_\pm$ is transformed as 
\begin{align}
    \delta_\kappa \Psi_\pm\,=\, \Gamma_\mu\bigg(\momp^\mu\pm {1\over 2\lambda}\bPi^\mu_\sigma \bigg)\kappa_\pm\ .\label{eq: gs kappa symmetry psi}
\end{align}
By fixing the $\kappa$ symmetry, one can get correct propagating fermi degrees of freedom. For the background that we are interested in, we find that a proper gauge condition for $\kappa$ symmetry is $\Gamma^\pm \Psi_\pm\,=\,0$ instead of the usual one $\Gamma^+\Psi_\pm\,=\,0$. As in the previous section, for the relation to \ttb deformation, we need to choose a background where $X^1$ has winding number $1$ while $X^0$ has no winding mode. In this case, we should check whether the gauge condition for $\kappa$ symmetry is proper or not. That is, for a given general configuration of $\Psi_\pm$, one should ask whether there is a $\kappa$ gauge transformation to the gauge condition that we chose. In particular, this should also be possible perturbatively around the background (\ie zero-modes without the oscillators). Hence, turning the oscillators off, the $\kappa$ transformation of the spinor~\eqref{eq: gs kappa symmetry psi} becomes
\begin{align}
    \delta_\kappa \Psi_\pm
    \,=\, \begin{pmatrix}
    0 &  \pz_+ \pm {L\over 4\pi \lambda}\\
    2\pz_- \pm {L\over 4\pi \lambda} & 0\\
    \end{pmatrix}\kappa_\pm  \ ,\label{eq: kappa sym zero-mode gs}
\end{align}
and $\pz_\pm$ are subjected to the following zero-mode Virasoro constraints
with the fluctuations turned off:
\begin{align}
    {L\over 2\pi }\bigg({1\over 2}\pz_+ + \pz_-\bigg)\,=\,0\;\; ,\qquad \pz_+ \pz_-+{L^2\over 32\pi^2 \lambda^2}=0\ .    \label{eq: zero-mode constraint gs no fluc}
\end{align}
These constraints make the rank of the matrix~\eqref{eq: kappa sym zero-mode gs} $1$ so that we have two Majorana fermions at the end of the gauge fixing. Note that the terms related $L$ in Eqs.~\eqref{eq: kappa sym zero-mode gs} and \eqref{eq: zero-mode constraint gs no fluc} come from the winding mode along $X^1$, which does not appear in the usual light-cone quantization $X^+=\tau$. Indeed, in the usual light-cone quantization (\eg $L=0$), it is easy to see that one could choose $\Gamma^+\Psi_\pm\,=\,0$ because a solution of the constraints \eqref{eq: zero-mode constraint gs no fluc} is $\pz_+=0\;,\;\; \pz_-=0$ and the $\kappa$ transformation becomes 
\begin{align}
    \delta_\kappa \Psi_\pm \,=\, \begin{pmatrix}
    0\\
    2\pz_- \kappa_\pm^1\\
    \end{pmatrix}\quad,\hspace{12mm}\mbox{where}\quad \kappa_\pm\,=\, \begin{pmatrix}
    \kappa_\pm^1\\
    \kappa_\pm^2\\
    \end{pmatrix}\ .
\end{align}
However, the solution of Eq.~\eqref{eq: zero-mode constraint gs no fluc} for the usual light-cone quantization is different from the solution that we are interested in. For the \ttb deformed Hamiltonian, the background solution of the Virasoro constraints is
\begin{align}
    \pz_- \,=\,-{1\over 2} \pz_+ \,=\, \pm {L\over 8\pi \lambda} \ ,\label{eq: zero-mode sol no fluc}
\end{align}
which is not smoothly connected to that of the usual light-cone quantization in the limit $L\rightarrow 0$. With the solution~\eqref{eq: zero-mode sol no fluc}, the $\kappa$ gauge transformation can be written as
\begin{alignat}{5}
     \delta \Psi_+ \,=\,& \begin{pmatrix}
     0 & 0 \\
     {L\over 2\pi \lambda} & 0 \\
     \end{pmatrix}\kappa_+\;\;,\qquad    &  \delta \Psi_- \,=\, &\begin{pmatrix}
     0 & -{L\over 2\pi \lambda} \\
     0 & 0 \\
     \end{pmatrix}\kappa_- \\
     \mbox{or}\quad \delta \Psi_+ \,=\,& \begin{pmatrix}
     0 & {L\over 2\pi \lambda} \\
     0 & 0 \\
     \end{pmatrix}\kappa_+\;\;,\qquad   &   \delta \Psi_- \,=\,& \begin{pmatrix}
     0 & 0 \\
     -{L\over 2\pi \lambda} & 0 \\
     \end{pmatrix}\kappa_-
\end{alignat}
depending on the sign of $\pz_-$. Therefore, when considering fluctuations around the background perturbatively, we can choose either $\Gamma^\pm\Psi_\pm\;=0$ or $\Gamma^\mp\Psi_\pm\;=0$, but it is not possible to choose that gauge condition $\Gamma^+\Psi_\pm\,=\,0$ when $X^1$ has a winding number (and $X^0$ does not). 

Furthermore, the problem of the gauge choice $\Gamma^+ \Psi_\pm\,=\,0$ for our background is also revealed at the level of the GS-like action. As in the usual quantization of string theory, one can choose a flat worldsheet metric using the reparametrization and Weyl invariance. For the background $\pz_-=\pm {L\over 8\pi \lambda}$, one of the kinetic terms of the fermions vanishes\footnote{Here, we also consider small fluctuation around the background.}, and the corresponding fermions become singular.

Hence, we choose the gauge condition for the $\kappa$ symmetry, consistent with the background,
\begin{align}\label{kappa-projection}
    \Gamma^\pm\Psi_\pm\,=\,0\ ,\hspace{12mm} \mbox{for}\quad \pz_-\,>\,0
\end{align}
and one can write the spinor as
\begin{align}
    \Psi_+\,=\,{1\over 2\sqrt{\xi_+}}\begin{pmatrix}
    \psi_+\\
    0
    \end{pmatrix}\;\;,\quad  \Psi_-\,=\,{1\over 2\sqrt{\xi_-}}\begin{pmatrix}
    0\\
    \psi_-
    \end{pmatrix}\ ,\label{eq: kappa sym condition theisen}
\end{align}
where we choose $\xi_\pm\,\equiv\, {L\over 4\pi \lambda}$.

Now, with our gauge choices, $\bPi^\mu_\sigma $ and $\mpi_\mu$ in Eq.~\eqref{eq: gs momenta in theisen} are found to be
\begin{align}
	\bPi^\mu_\sigma=\begin{pmatrix}
		{L\over 4\pi } - {\pi i \lambda \over L} \psi_- \psi_-' \\
		X^{-\prime} +  {2\pi i \lambda \over L} \psi_+ \psi_+' \\
		\sqrt{2\lambda}\phi'
	\end{pmatrix}\;\; ,\qquad \mpi_\mu=\begin{pmatrix}
	\momp_+ + {\pi i  \over L} \psi_+ \psi_+' \\
	\momp_- + {\pi i  \over 2 L} \psi_- \psi_-' \\
	{1\over \sqrt{2\lambda}}\spi\end{pmatrix}\;\; ,\hspace{8mm} (\mu=+,-,2)
\end{align}
where we also rescaled $X^2$ and $\momp_2$ by $\sqrt{2\lambda}$:
\begin{equation}
    X^2\,=\, \sqrt{2\lambda}\phi\;\;,\qquad \momp_2\,=\,{1\over \sqrt{2\lambda}}\spi\ .
\end{equation}
Unlike the Polyakov or NSR-like cases, the Virasoro constraints contain cross terms between 
oscillators in the light-cone sector, 
\begin{align}
    \cC_1\,=\,& - {\pi i \lambda \over L} \psi_- \psi_-' \momp_+ +  {\pi i  \over 2 L} \psi_- \psi_-' X^{-\prime} + \cdots\\
    \cC_2\,=\,& { 2\pi i \lambda \over  L} \psi_- \psi_-' \momp_+ -  {\pi i  \over  L} \psi_- \psi_-' X^{-\prime} +\cdots
\end{align}
Hence, the integrated constraints do not give simple algebraic equations for zero-modes as in Eqs.~\eqref{eq: polakov zero-mode constraint 1} and \eqref{eq: polakov zero-mode constraint 2}. With the gauge conditions \eqref{eq: lightcone gauge gs theisen} and \eqref{eq: kappa sym condition theisen}, we fail to reproduce the anticipated Hamiltonian~\eqref{eq: deformed hamiltonian nsrlike string} from the GS-like action. 

The failure can be traced to how the $\kappa$ symmetry was fixed by demanding residual degree of fermionic freedom to obey Eq.~\eqref{kappa-projection}, an avoidable consequence of the gauge choice made here. Even if we had demanded a more familiar one $\Gamma^+\Psi_\pm\,=\,0$ as in the standard light-cone approach, an invalid choice anyway, the reduced action becomes degenerate in that some of the necessary quadratic kinetic terms are killed. 

This complication with the $\kappa$ gauge fixing disappears for $\cN=1$ GS-like model, equivalent 
to removing  one of the spinors from the $\cN=2$ GS-like action. For example, truncating the spinor $\Psi_-$, 
we may choose the gauge condition $\Gamma^+\Psi_+\,=\,0$ (for $\pz_-\,>\,0$). Then, we have
\begin{align}
    \Psi_+ \,= \, {1\over 2\sqrt{\xi}_+}\begin{pmatrix}
    \psi_+\\
    0\\
    \end{pmatrix}\ .\label{eq: n1 kappa gauge condition theisen}
\end{align}
Now, with the light-cone gauge~\eqref{eq: lightcone gauge gs theisen} and the $\kappa$ gauge condition~\eqref{eq: n1 kappa gauge condition theisen}, $\mpi_\mu$ and $\bPi^\mu_\sigma$ defined in Eq.~\eqref{eq: gs momenta in theisen} is found to be
\begin{align}
    \bPi^\mu_\sigma\,=\, \begin{pmatrix}
    {X^+}'\\
    {X^-}'+  \widetilde{\cP}_{f,0}\\
    \sqrt{2\lambda}\phi'\\
    \end{pmatrix}\;\;,\qquad \mpi_\mu\,=\,\begin{pmatrix}
    \momp_++ {1\over 2 \lambda}\widetilde{\cH}_{f,0}\\
    \momp_-\\
    {1\over\sqrt{2\lambda}}\pi\\
    \end{pmatrix}\ ,
\end{align}
where we defined 
\begin{align}
    \widetilde{\cH}_{f,0}\,\equiv\, {i\over 2\xi_+}\psi_+\psi'_+\;\;,\qquad \widetilde{\cP}_{f,0}\,\equiv\, {i\over 2\xi_+}\psi_+\psi'_+ \ .
\end{align}
Using the choice\footnote{We determine the coefficient $\xi_+$ in a way that in the light-cone gauge with flat worldsheet metric, the $\cN=1$ GS-like action becomes
\begin{align}
    S^{\text{\tiny l.c.}}_{\text{\tiny $\cN=1$ GS}}= \int d\tau d\sigma \bigg[-{1\over 4\lambda}\partial_\alpha X^2 \partial^\alpha X^2 + i\psi_+\partial_=\psi_+  \bigg]\ .
\end{align}
} of $\xi_+ \,=\, \pz_- + {L\over 8\pi \lambda} $, the zero-mode of the constraints $\cC_1$ and $\cC_2$ in Eq.~\eqref{eq: virasoro constraint gs theisen} becomes
\begin{align}
    &{1\over 2}\pz_+ + \pz_-  + {1\over 2\pi }P_0\,=\,0 \label{eq: zero-mode constraint 3}\ ,\\
    &{4\pi \lambda\over L} \pz_+ \pz_- +{L\over 8\pi \lambda}+{1\over 2\pi }H_0\,=\,0\label{eq: zero-mode constraint 4}\ ,
\end{align}
where the total Hamiltonian and momentum of the undeformed theory is 
\begin{align}
    H_0\,\equiv\,& H_{b,0}+H_{f,0}\;\;,\qquad P_0\,\equiv\, P_{b,0}+P_{f,0} \ ,
\end{align}
and 
\begin{alignat}{5}
    H_{b,0}\,\equiv \,& {2\pi \over L} \int d\sigma \;\left[ {1\over 2}\pi^2 + {1\over 2} \phi'^2 \right] \;\;,\qquad &P_{b,0}\,\equiv \,&{2\pi \over L} \int d\sigma \; \pi \phi'\ , \\  
    H_{f,0}\,\equiv\,& {2\pi \over L}\int d\sigma\;  {i\over 2}\psi_+\psi'_+ \;\;,\qquad &P_{f,0}\,\equiv\,& {2\pi \over L}\int d\sigma\;   {i\over 2}\psi_+\psi'_+ \ .
\end{alignat}
Hence, one can immediately get the energy corresponding to the translation of the target coordinate $X^0$,
\begin{equation}
    E_{str} \,\equiv\, \int_0^{2\pi} d\sigma \; \momp^0 \,= \, -2\pi \bigg({1\over 2}\pz_+-\pz_-\bigg)=\frac{L}{2\lambda}\sqrt{1+{4\lambda\over L} H_{0}+{4\lambda^2\over L^2} P_{0}^2}\ . \label{eq: gs string energy theisen}
\end{equation}
Modulo an additive constant, this is again the \ttb deformed spectrum of the free $\cN=(1,0)$ SUSY model. 

One must remember, however, this last model by itself would have a worldsheet diffeomorphism anomaly unless
embedded into a larger anomaly-free set-up, either by adding canceling chiral fields or by introducing an inflow 
mechanism. We will not pursue the latter possibility in this note, although it may lead to an interesting variation
on this general theme of \ttb vs. worldsheet. 

%

\section{\ttb Deformation and $N=(1,1)$ Supersymmetry}
\label{sec: ttbar deformation and gs shifted background}

The relation between  \ttb deformation and a Polyakov worldsheet theory was first pointed 
out in Ref.~\cite{Cavaglia:2016oda} of which the \ttb deformed Lagrangian of the free scalar field 
is a Nambu-Goto Lagrangian. A Polyakov action was designed to produce Nambu-Goto action after 
integrating out the worldsheet metric, so the connection is obvious, at least for the massless 
scalar theories. 

In the previous section, we reviewed how even the \ttb deformed quantum spectrum can 
be sometimes read off via the worldsheet analysis, say, by taking a particular gauge 
and concentrating on a winding sector. We have also seen that such a clean result 
is no longer immediate when fermions are involved. For instance, when we start 
with a GS-like worldsheet theory, the zero-mode part of the conjugate momentum of 
$X^0$, the would-be Hamiltonian on the \ttb side, cannot be cleanly factored out 
from oscillators in the worldsheet diffeomorphism constraints, rendering the 
the mechanism employed in Ref.~\cite{Jorjadze:2020ili} unusable.

Nevertheless, this alternate approach to the \ttb deformed theory offers us  more 
handles than otherwise, so in this section,  we will further pursue this relationship
with an emphasis on the canonical structures and the symmetry algebra. 
In particular, we wish to concentrate on a $\cN=(1,1)$ supersymmetric theory, for which 
GS-like worldsheet models with their manifest spacetime supersymmetry is a natural place to start. 

\subsection{Two Distinct Constructions via GS-like Worldsheet Theory}
\label{sec: review sfondrini}

This dictionary between \ttb deformation and worldsheet models is not unique.
With $3D$ GS-like models we have already encountered, one can take at least two different 
gauge choices that reproduce the same \ttb deformation of the simplest $\cN=(1,1)$ supersymmetric 
theory with a single massless scalar supermultiplet. Either way, the deformation resorts 
to the ordinary \ttb operator rather than its supersymmetry completion, yet the worldsheet 
interpretation tells us that $N=(1,1)$ supersymmetry would be actually intact throughout 
the deformation. We will investigate the canonical and symmetry structures from 
the original \ttb perspective and the alternate worldsheet perspective.

\subsubsection*{Static Gauge with a Light-Cone Target}
\label{sec: gslike action theisen}

In Section~\ref{sec: gs superstring theory and hamiltonian}, we have already studied the GS-like action in a light-cone gauge. We choose the static gauge,
\begin{align}
    X^0\,=\,{L\over 2\pi } \tau\;\;,\quad X^1\,=\, {L\over 2\pi }\sigma\ , \label{eq: static gauge of gslike action sec 4}
\end{align}
and we integrate out the worldsheet metric in the GS-like Lagrangian~\eqref{eq: 3DGSaction for flat},
%
%
%
%
%
%
\begin{align}
    S_{\text{\tiny GS}}\,=\,& {1\over 2\lambda }\int d\tau d\sigma\; \bigg[-\sqrt{-\det \widetilde{G}} -i\epsilon^{\alpha\beta}\partial_\alpha X^\mu (\bar\Psi_+\Gamma^\nu\partial_\beta\Psi_+-\bar\Psi_-\Gamma^\nu\partial_\beta\Psi_-)G_{\mu\nu}\cr
    &\hspace{60mm}-\epsilon^{\alpha\beta}(\bar\Psi_+\Gamma^\mu\partial_\alpha\Psi_+)(\bar\Psi_-\Gamma^\nu\partial_\beta\Psi_-)G_{\mu\nu}\bigg]\ , \label{eq: gslike action flat metric 0}
\end{align}
where $\widetilde{G}$ is the induced metric given by
\begin{align}
    \widetilde{G}_{\alpha\beta}\,\equiv \, \bPi^\mu_\alpha \bPi^\nu_\beta G_{\mu\nu}\ .
\end{align}
As we have discussed in Section~\ref{sec: gs superstring theory and hamiltonian}, we need to choose the following $\kappa$ gauge condition for the background~\eqref{eq: static gauge of gslike action sec 4},
\begin{align}
    \Gamma^\pm\Psi_\pm\,=\,0\ , 
\end{align}
and we have
\begin{align}
    \Psi_+\,=\,{1\over 2\sqrt{\xi_+}}\begin{pmatrix}
    \psi_+\\
    0
    \end{pmatrix}\;\;,\quad  \Psi_-\,=\,{1\over 2\sqrt{\xi_-}}\begin{pmatrix}
    0\\
    \psi_-
    \end{pmatrix}\ ,\label{eq: kappa sym condition theisen sec 4}
\end{align}
where we choose $\xi_\pm\,\equiv\, {L\over 4\pi \lambda}$.

In the usual light-cone gauge, we fix the $\kappa$ gauge symmetry by $\Gamma^+\Psi_\pm\,=\,0$ so that the GS-like action with flat worldsheet metric becomes free-field action of scalar and fermion with suitable normalization. On the other hand, our gauge condition~\eqref{eq: kappa sym condition theisen sec 4} gives quartic interactions of fermions in the same calculation. But, in this gauge, after integrating out the worldsheet metric, the GS-like action becomes free-field action of scalar and fermion (up to divergent constant) in the limit where the string length goes to zero compared to the circumference $L$ of the target coordinate $X^1$, \ie
\begin{align}
   {4\pi^2\lambda\over L^2}\quad\longrightarrow \quad 0\ .
\end{align}
It is convenient to define the dimensionless (\ttb deformation) parameter $\Lambda$ by
\begin{align}
     \Lambda\,\equiv\, {4\pi^2\lambda\over L^2}\ .
\end{align}
With our gauge choice~\eqref{eq: kappa sym condition theisen sec 4} for $\kappa$ symmetry together with the static gauge~\eqref{eq: static gauge of gslike action sec 4}, the GS-like action~\eqref{eq: gslike action flat metric 0} is written as
\begin{align}
    S_{\text{\tiny GS}}^{\text{\tiny s.t.}}\,=\,\int d\tau d\sigma\; \bigg[ -{1\over 2\Lambda}\sqrt{\cA}+ \cB \bigg]\label{eq: static gauge gs action}\ ,
\end{align}
where $\cA$ and $\cB$ is given by
\begin{align}
    \cA\,=\,& 1-2\Lambda(4\partial_\+\phi\partial_=\phi + i\psi_+\partial_=\psi_++i\psi_-\partial_\+\psi_-)\cr
    &-2\Lambda^2(\psi_+\partial_=\psi_+ \psi_-\partial_\+\psi_--\psi_+\partial_\+\psi_+\psi_-\partial_=\psi_-) -\Lambda^2(\psi_+\partial_=\psi_++\psi_-\partial_\+\psi_-)^2\cr
    &-8i\Lambda^2\bigg[(\partial_\+\phi)^2\psi_-\partial_=\psi_-+(\partial_=\phi)^2\psi_+\partial_\+\psi_+-\partial_\+\phi\partial_=\phi(\psi_+\partial_=\psi_++\psi_-\partial_\+\psi_-)\bigg]\cr
    &-8\Lambda^3(\partial_\+\phi\psi_+\partial_=\psi_+-\partial_=\phi\psi_+\partial_\+\psi_+)(\partial_\+\phi\psi_-\partial_=\psi_--\partial_=\phi\psi_-\partial_\+\psi_-)\ ,
\end{align}
and
\begin{align}
    \cB\,=\, {i\over2}(\psi_+\partial_=\psi_++\psi_-\partial_\+\psi_-)+{\Lambda\over2}(\psi_+\partial_=\psi_+\psi_-\partial_\+\psi_--\psi_+\partial_\+\psi_+\psi_-\partial_=\psi_-)\ ,
\end{align}
where we used the rescaling of $X^2$ by $\sqrt{2\lambda}$ as before:
\begin{equation}
    X^2\,=\, \sqrt{2\lambda}\phi \ .
\end{equation}
By expanding the square-root in Eq.~\eqref{eq: static gauge gs action} with respect to the fermions, one can compare\footnote{For concrete comparison, one needs to rescale the fermion and coordinates $(t,x)$ by the length ${L\over 2\pi }$ in the \ttb deformation side to make them dimensionless.} the result with the \ttb deformed Lagrangian of the free $\cN=(1,1)$ SUSY model~\eqref{eq: susy lag} which we will discuss in detail in the next section. One can find that they differ by the constant ${1\over 2\Lambda}$, 
\begin{align}
    \cL_{T\bar{T}}\,=\, \cL_{\text{\tiny GS}}^{\text{\tiny s.t.}} +{1\over 2\Lambda}\ .\label{eq: gs shift energy}
\end{align}
%
%
%

\subsubsection*{With a Shifted Light-Cone Target}
\label{sec: gslike action frolov}

There is another way to construct \ttb deformed Lagrangian from the GS-like worldsheet theory. It was shown in Ref.~\cite{Baggio:2018rpv,Frolov:2019nrr,Frolov:2019xzi} that the Lagrangian density of the \ttb deformed theory can be obtained from the GS-like action as well as Polyakov action in the uniform light-cone gauge. We review the approach of Refs.~\cite{Baggio:2018rpv,Frolov:2019nrr,Frolov:2019xzi}, in particular, for the case of the $\cN=2$ GS-like action of 3D target space in the uniform light-cone gauge to reproduce the Hamiltonian density of the \ttb deformation of 2D free $\cN=(1,1)$ SUSY model.

The 3D $\cN=2$ GS-like Lagrangian with WZ term is given by
\begin{align}
	\mathcal{L}_{\text{\tiny GS}}\,=\,&-\frac12\gamma^{\alpha\beta}\bPi^\mu_\alpha\bPi^\nu_\beta G_{\mu\nu} -i\epsilon^{\alpha\beta}\partial_\alpha X^\mu (\bar\Psi_+\Gamma^\nu\partial_\beta\Psi_+-\bar\Psi_-\Gamma^\nu\partial_\beta\Psi_-)G_{\mu\nu}\cr
	&-\epsilon^{\alpha\beta}(\bar\Psi_+\Gamma^\mu\partial_\alpha\Psi_+)(\bar\Psi_-\Gamma^\nu\partial_\beta\Psi_-)G_{\mu\nu}\ ,\label{3DGSaction}\\[6pt]
	\bPi^\mu_\alpha\,\equiv \,& \partial_\alpha X^\mu+i\bar\Psi_+\Gamma^\mu\partial_\alpha\Psi_++i\bar\Psi_-\Gamma^\mu\partial_\alpha\Psi_-\ .
\end{align}
We introduce a new light-cone target coordinate $X^\pm$ \cite{Baggio:2018rpv,Frolov:2019nrr,Frolov:2019xzi} defined by
\begin{align}
    X^+\,\equiv\, &\left({1\over 2} -\Lambda \right) X^1+ \left({1\over 2} + \Lambda \right)X^0 \ , \label{def: light-cone1}\\
    X^-\, \equiv \,&X^1-X^0\ .\label{def: light-cone2}
\end{align}
In addition, we will consider the flat target space. But, due to the new definition of the light-cone~\eqref{def: light-cone1}, the target space metric\footnote{In this review for the GS-like model, we take $2\pi \ell_s^2=1$ for simplicity. Accordingly, we use dimensionless parameter $\Lambda={\lambda \over 2\pi \ell_s^2}$ where $\lambda$ is the \ttb deformation parameter. In Section~\ref{sec: revisit ttbar deformation of gs superstring}, we will retrieve the string length $\ell_s$ to see the detailed relation.} denoted by $G_{\mu\nu}$ becomes
\begin{align}
    ds^2\,=\, G_{\mu\nu} X^\mu X^\nu \,=\, 2\Lambda (dX^-)^2+ 2 dX^+ dX^-+(dX^2)^2\ .\label{new target light cone}
\end{align}

To fix the $\kappa$ symmetry of the GS-like action, we introduce projectors~\cite{Baggio:2018rpv}
\begin{align}
    \Upsilon^\pm \,=\, {1\over 2}(\Gamma^1\pm \Gamma^0)\ ,\label{def: projector}
\end{align}
which satisfies
\begin{align}
    \Upsilon^\pm \Upsilon^\pm\,=\,0\;,\quad \Upsilon^+ \Upsilon^- + \Upsilon^- \Upsilon^+=1\;,\quad (\Upsilon^\pm)^T\,=\, \Upsilon^\mp\,.
\end{align}
Note that the projector $\Upsilon^\pm$ is different from the gamma matrix $\Gamma^\pm$ of in light-cone coordinates because of the new definition of the light-cone~\eqref{def: light-cone1} and \eqref{def: light-cone2}. The target space gamma matrices in the light-cone direction are given by
\begin{align}
    \Gamma^+ \,=\, \left({1\over 2} - \Lambda \right)\Gamma^1 + \left({1\over 2} + \Lambda \right) \Gamma^0= \Upsilon^+ - 2\Lambda \Upsilon^-\quad,\qquad \Gamma^-\,=\,  \Gamma^1 + \Gamma^0= 2\Upsilon^- \ ,
\end{align}
%
%
where $\Gamma^\mu$ is chosen to be
\begin{align}
    \Gamma^0\,=\, i\sigma_2 \;,\quad \Gamma^1\,=\,\sigma_1\;,\quad \Gamma^2\,=\,\sigma_3\,.
\end{align}
We fix the $\kappa$ gauge by the condition\footnote{We confirm that this choice of gauge condition is valid by a similar analysis as in Section~\ref{sec: gs superstring theory and hamiltonian} with the light-cone gauge~\eqref{eq: gs lightcone gauge review}.}
\begin{align}
    \Upsilon^+ \Psi_\pm\,=\,0\ .\label{eq: gauge condition for kappa sym}
\end{align}
%
%
%
This condition projects out half of the Majorana spinor,
\begin{equation}
	\Psi_\pm=\frac12\begin{pmatrix}\psi_\pm\\0\end{pmatrix}\ . \label{eq: spinor gauge fixed}
\end{equation}
%
%
Then, one can write the 3D GS-like Lagrangian in the first-order form as follows,
\begin{align}
\mathcal L\,=\,&\mpi_\mu\bPi_\tau^\mu+\frac{\gamma^{\tau\sigma}}{\gamma^{\tau\tau}}\cC_1+\frac1{2\gamma^{\tau\tau}}\cC_2-i\epsilon^{\alpha\beta}\partial_\alpha X^\mu (\bar\Psi_+\Gamma^\nu\partial_\beta\Psi_+-\bar\Psi_-\Gamma^\nu\partial_\beta\Psi_-)G_{\mu\nu} \cr
	&\qquad-\epsilon^{\alpha\beta}(\bar\Psi_+\Gamma^\mu\partial_\alpha\Psi_+)(\bar\Psi_-\Gamma^\nu\partial_\beta\Psi_-)G_{\mu\nu}\ ,\cr
=\,&\momp_\mu\dot X^\mu+\fpi_+ \dot\psi_++\fpi_-\dot\psi_-+\frac{\gamma^{\tau\sigma}}{\gamma^{\tau\tau}}\cC_1+\frac1{2\gamma^{\tau\tau}}\cC_2\,,\label{eq: 1st order gs lag}
\end{align}
where $\dot{\psi}_\pm\,\equiv\, \partial_\tau \psi_\pm $ and $\psi'_\pm\,\equiv \, \partial_\sigma \psi_\pm$ \etc Also, we define
\begin{equation}
\mpi_\mu\,\equiv\,\frac{\delta \mathcal S}{\delta\bPi^\mu_\tau}\;\; ,\quad \momp_\mu\,\equiv\,\frac{\delta\mathcal S}{\delta \dot X^\mu}\;\; ,\quad \fpi_\pm\,\equiv\,\frac{\delta\mathcal S}{\delta \dot \psi_\pm}\ ,
\end{equation}
and the Virasoro constraints $\cC_1$ and $\cC_2$ are given by
\begin{align}
\cC_1=\mpi_\mu\bPi_\sigma^\mu\;\;,\qquad \cC_2=G^{\mu\nu}\mpi_\mu\mpi_\nu+G_{\mu\nu} \bPi^\mu_\sigma \bPi^\nu_\sigma 	\ .
\end{align}
%
%
By choosing the uniform light-cone gauge
\begin{equation}
    X^+=\tau\; ,\quad \momp_-=1\ , \label{eq: gs lightcone gauge review}
\end{equation}
we have
\begin{align}
	\bPi^\mu_\sigma=\begin{pmatrix}
		-\Lambda \cP_{f,0}\\
		X^{-\prime}+\cP_{f,0}\\
		\phi'
	\end{pmatrix}\;\; ,\qquad \mpi_\mu=\begin{pmatrix}
	\momp_++\cH_{f,0}\\
	1+\Lambda\cH_{f,0}\\
	\spi\end{pmatrix}\ ,
\end{align}
for $\mu=+,-,2$. To make contact with \ttb deformation, we identified $X^2$ with $\phi$, and accordingly we also identified its conjugate momentum, \ie $\mpi_2\, \equiv\,\spi$. Also, $\cH_{(b,f),0}, \cP_{(b,f),0}$ are the Hamiltonian and momentum density of the free scalar and fermion,
\begin{alignat}{3}
	&\cH_{f,0}\,\equiv \,\frac i2\psi_+\psi'_+-\frac i2\psi_-\psi'_-\;\;,\qquad &&\cP_{f,0}\,\equiv \,\frac i2\psi_+\psi'_++\frac i2\psi_-\psi'_-\ ,\\
	&\cH_{b,0}\,\equiv \,\frac12(\spi^2+\phi'^2)\;\;, &&\cP_{b,0}\,\equiv \, \spi\phi'\ .
\end{alignat}
Using them, the Virasoro constraints $\cC_1,\cC_2$ can be written as
\begin{align}
\cC_1\,=\,&\mpi_+\bPi^+_\sigma+\mpi_-\bPi^-_\sigma+\mpi_\mu\bPi^\mu_\sigma\cr
\,=\,&-\left(\momp_++\cH_{f,0}\right)\Lambda\cP_{f,0}+(1+\Lambda\cH_{f,0})\bPi^-_\sigma+\cP_{b,0}=0\ ,\label{vir1}\\[6pt]
\cC_2\,=\,&2\mpi_-\mpi_+-2\Lambda \pi_+^2+2\Lambda (\bPi^-_\sigma)^2+2\bPi^+_\sigma\bPi^-_\sigma+2\cH_{b,0}\cr
\,=\,&2\left(1+\Lambda \cH_{f,0}\right)\left(\momp_++\cH_{f,0}\right)-2\Lambda\left(\momp_++\cH_{f,0}\right)^2 \cr
&+2\Lambda (\bPi^-_\sigma)^2-2\Lambda\cP_{f,0}\bPi^-_\sigma+2\cH_{b,0}=0\label{vir2}\ .
\end{align}
We solve Virasoro constraints $\cC_1$ and $\cC_2$ for $\momp_+$ and $\bPi^-_\sigma$. In particular, one can express $\momp_+$ in terms of $\cH_{(b,f),0}$ and $\cP_{(b,f),0}$. Then, the Lagrangian of the GS-like model can be written as
\begin{align}
\mathcal L\,=\,&\momp_\mu\dot X^\mu+\fpi_+ \dot\psi_++\fpi_-\dot\psi_-
=\,\spi\dot\phi +\fpi_+ \dot\psi_++\fpi_-\dot\psi_--(-\momp_+) \ ,
\end{align}
where the total derivative term $\dot X^-$ is ignored. From the first-order Lagrangian, one can read off the Hamiltonian density of the 3D $\cN=2$ GS-like action,
\begin{align}
\cH&\,=\,-\momp_+\\
&\,=\,   {1\over 2\Lambda }\left[\sqrt{1+4\Lambda \cH_{b,0} +4\Lambda^2\left( \cP_{b,0}\right)^2}-1 \right]\cr
	&\qquad+ {i\over 4} \left({1-4\Lambda^2 \left( \cP_{b,0}\right)^2\over \sqrt{1+4\Lambda \cH_{b,0} +4\Lambda^2\left( \cP_{b,0}\right)^2}}+1 \right)(\psi_+\psi'_+-\psi_-\psi'_-) \nonumber\\
    &\qquad + {i\Lambda \cP_{b,0}\over 2}(\psi_+\psi'_+ + \psi_-\psi'_-)- {2\left(\left(\cH_{b,0}\right)^2-\left( \cP_{b,0}\right)^2\right)\Lambda^3 \over \left(1+4\Lambda \cH_{b,0} +4\Lambda^2\left( \cP_{b,0}\right)^2\right)^{3\over 2} }(\psi_+\psi'_+\psi_-\psi'_-)\ .
\end{align}
This agrees with the \ttb deformed Hamiltonian density of free $\cN=(1,1)$ SUSY model, as will be explicitly shown next.

\subsection{Does $\cN=(1,1)$ SUSY Survive Ordinary \ttb ?}
\label{sec: hamiltonian analysis for susy model}

We have seen that a \ttb deformation of  a single real scalar and 
a single Majorana doublet can be obtained from a GS-like worldsheet theory in two 
different backgrounds and the accompanying gauge choices. One important fact is
that in neither approaches, the spacetime supersymmetry of the GS-like action
was broken. As we will see in this section, however, the common end result is
that of non-supersymmetric \ttb deformation of $N=(1,1)$ theory of a single scalar 
supermultiplet. 

On the other hand, the generality of the eigenvalue flow, which does not 
distinguish the bosonic or the fermionic nature of the state, does naturally 
suggest that the supersymmetry would not be explicitly broken by \ttb and 
one can say that the above analysis realizes this anticipation. We 
will devote this subsection and the next to an exploration of this supersymmetry 
structure, from both the \ttb side and the worldsheet side. 

We start from the free $N=(1,1)$ theory of a single massless scalar supermultiplet,
\begin{align}
    \cL_0\,=\, 2\partial_\+\phi\partial_=\phi+i\psi_+\partial_=\psi_++i\psi_-\partial_\+\psi_-\ ,
\end{align}
and solve the flow equation of the non-SUSY \ttb deformation
\begin{align}
    \partial_\lambda \cL\, =\, {1\over 2} \epsilon_{\mu\nu }\epsilon^{\rho\sigma}{T^\mu}_\rho {T^\nu}_\sigma\ .
\end{align}
%
%
We emphasize that we used the energy-momentum tensor calculated by the Noether procedure, which differs from the usual symmetric energy-momentum tensor. This simple deformation is referred to as non-SUSY \ttb deformation since the $\cN=(1,1)$ supersymmetry is not manifestly respected by the deformation, unlike the manifestly supersymmetric versions, where the \ttb operator is supersymmetry-completed, as in Refs.~\cite{Baggio:2018rpv,Chang:2018dge}. 

One can directly solve this flow equation by taking the following ansatz,
\begin{align}
    \cL\,=\,& f_0[\chi]+ f_1[\chi]S_{\+,=}+f_2[\chi]S_{=,\+} + f_3[\chi](\partial_=\phi)^2 S_{\+,\+}+ f_4[\chi](\partial_\+ \phi)^2 S_{=,=}+ f_5[\chi]S_{\+,\+}S_{=,=} \cr 
    & + f_6[\chi]S_{\+,=}S_{=,\+} + f_7[\chi] (\partial_\+\phi)^2 S_{\+,=} S_{=,=} + f_8[\chi](\partial_= \phi)^2 S_{\+,\+}S_{=,\+} \ ,
\end{align}
where we defined
\begin{align}
	\chi \, \equiv\, &  -  4  \lambda \partial_\+\phi \partial_=\phi \ ,\\
	S_{\+\,,\,\mu}\,\equiv\, & i \psi_+ \partial_\mu \psi_+\ ,\\
	S_{=\,,\,\mu}\,\equiv\, & i \psi_- \partial_\mu \psi_-\ .
\end{align}
Note that higher-order terms in $S_{\mu,\nu}$ is truncated because of fermi-statistics. With initial condition $\cL[\lambda=0]=\cL_0$, the flow equation determines $f_i[\chi]$ $i=0,1,\cdots, 8$, and we have
\begin{align}
	\mathcal{L}\,=\,&-{1\over 2\lambda} \left[ \sqrt{1 + 2\chi } -1 \right] + { 1 +\chi + \sqrt{1  +2 \chi } \over 2 \sqrt{1+ 2\chi  } } (S_{\+,=}+S_{=,\+}) \cr
	&+{2\lambda \over \sqrt{1+ 2 \chi }} [ (\partial_=\phi)^2 S_{\+,\+}+ (\partial_\+ \phi)^2 S_{=,=} ] \cr
	& + \lambda{ 1+ \chi -\chi^2 + ( 1+ 2\chi )^{3\over 2}  \over 2 ( 1+ 2\chi )^{3\over 2}}S_{\+,\+}S_{=,=} - \lambda{ 1 + 3 \chi + \chi^2 + ( 1+ 2\chi )^{3\over 2}  \over 2 ( 1+ 2\chi )^{3\over 2}}S_{\+,=}S_{=,\+} \cr
	& - {2\lambda^2 \chi \over  ( 1+ 2\chi )^{3\over 2} } [ (\partial_\+\phi)^2 S_{\+,=} S_{=,=} + (\partial_= \phi)^2 S_{\+,\+}S_{=,\+} ] \ .\label{eq: susy lag}
\end{align}
This Lagrangian has appeared elsewhere~\cite{Baggio:2018rpv,Frolov:2019nrr,Coleman:2019dvf}.

For Hamiltonian analysis, we first calculate the conjugate momentum of $\phi$ and $\psi_\pm$: 
\begin{align}
    \spi\,\equiv&\, {\partial \cL \over \partial \dot{\phi} }\ ,\label{def: scalar momentum pi}\\
    \fpi_\pm \,\equiv& \, {\partial \cL\over \partial \dot{\psi}_\pm }\ .\label{def: fermi momentum pi}
\end{align}
Note that the right-hand side of Eq.~\eqref{def: fermi momentum pi} contains $\dot{\psi}_\pm$, unlike the free fermion. We invert Eq.~\eqref{def: scalar momentum pi} to express $\dot{\phi}$ in terms of others:
\begin{align}
	\dot{\phi}\,=\,&{\spi \sqrt{1+ 2 \lambda\phi'^2   } \over \sqrt{1 + 2\lambda \spi^2}}  -{ \lambda \spi (1+ \lambda \spi^2 +3 \lambda \phi'^2 + 4 \lambda^2 \spi^2 \phi'^2) \over \sqrt{1+2\lambda \phi'^2} ( 1+2\lambda \spi^2 )^{3\over 2}  }  (S_{\+,\+} + S_{=,=}) \cr
	&+ \lambda \phi'  (S_{\+,\+} - S_{=,=})  - { \lambda^2 \spi (\spi^2-\phi'^2)\over \sqrt{1+2\lambda \phi'^2} ( 1+2\lambda \spi^2 )^{3\over 2}  } [S_{\+,=} +S_{=,\+} ]\cr
	&+ { \lambda^3 \spi (\spi^2-\phi'^2) (2+ \lambda \spi^2 +3 \lambda \phi'^2) \over (1+ 2\lambda \phi'^2)^{3\over 2} ( 1+ 2\lambda \spi^2)^{5\over 2} } (S_{\+,\+}-S_{\+,=})(S_{=,\+}  - S_{=,=} )\cr
	& - { \lambda^3 \spi (\spi^2-\phi'^2)\over \sqrt{1+2\lambda \phi'^2} ( 1+2\lambda \spi^2 )^{3\over 2}  } [S_{\+,\+}S_{=,=}-S_{\+,=}S_{=,\+}] \ .\label{eq: phi dot}
\end{align}
%
%
%
When inserting Eq.~\eqref{eq: phi dot} into Eq.~\eqref{def: fermi momentum pi}, one can see that the right-hand side of Eq.~\eqref{def: fermi momentum pi} has no $\dot{\psi}_\pm$ dependence, and they become the second-class constraints $\cC_1$ and $\cC_2$,
\begin{align}
	\cC_1=&\fpi_+ - {i\over 4} \psi_+ \left( 1- 2\lambda \spi \phi' +\sqrt{(1+2\lambda \spi^2)( 1 + 2\lambda \phi'^2 )} \right) \cr
	&- \lambda \left[ { 1+ \lambda (\spi^2 + \phi'^2) \over 4 \sqrt{(1+2\lambda \spi^2)( 1 + 2\lambda \phi'^2 )} } +{1\over 4} \right]\psi_+ \psi_-\psi'_-\ ,\label{eq: susy const1}\\
	\cC_2=&\fpi_-  - {i\over 4} \psi_- \left( 1+ 2\lambda \spi \phi' +\sqrt{(1+2\lambda \spi^2)( 1 + 2\lambda \phi'^2 )} \right) \cr
	& + \lambda \left[ { 1+ \lambda (\spi^2 + \phi'^2) \over 4 \sqrt{(1+2\lambda \spi^2)( 1 + 2\lambda \phi'^2 )} } +{1\over 4} \right]\psi_+ \psi'_+\psi_-\ .\label{eq: susy const2}
\end{align}
Hence, in this \ttb deformed Lagrangian, the dimension of the Hilbert space is not changed under the deformation. 

From the Noether procedure, Hamiltonian and momentum density is given by
\begin{align}
    \cH\, = \, & {1\over 2\lambda }\left[\sqrt{1+4\lambda \cH_{b,0} +4\lambda^2 \cP_{b,0}^2}-1 \right]+ {1\over 4} \left({1-4\lambda^2 \cP_{b,0}^2\over \sqrt{1+ 4\lambda \cH_{b,0} +4\lambda^2 \cP_{b,0}^2}}+1 \right)(S_\+-S_=) \cr
    & + {\lambda \cP_{b,0}\over 2}(S_\+ + S_=)+ {2(\cH_{b,0}^2-\cP_{b,0}^2)\lambda^3 \over (1+ 4\lambda \cH_{b,0} + 4\lambda^2 \cP_{b,0}^2)^{3\over 2} }S_\+S_= \ ,\label{eq: ham op susy}\\
    \cP\,=\, & \cP_{b,0} +{1\over 4}\left[1+\sqrt{1+ 4\lambda \cH_{b,0} +4\lambda^2 \cP_{b,0}^2}\right](S_\++ S_=) -{\lambda\cP_{b,0} \over 2}(S_\+-S_=) \ ,\label{eq: momentum op susy}
\end{align}
where $\cH_{b,0}$ and $\cP_{b,0}$ denotes the undeformed Hamiltonian and momentum density of the scalar field, respectively,
\begin{align}
    \cH_{b,0}\,\equiv \,& {1\over 2} \spi^2 +{1\over 2} \phi'^2\ , \\
    \cP_{b,0}\, \equiv \,& \spi \phi'\ ,
\end{align}
and $S_{\+}$ and $S_=$ is defined by
\begin{align}
	S_{\+}\, \equiv\, i \psi_+\psi'_+\;\;,\quad S_{=}\,\equiv \, i \psi_-\psi'_-\ .
\end{align}
Note that the constraints~\eqref{eq: susy const1} and \eqref{eq: susy const2} can be written as
\begin{align}
    \cC_1\,=\,& \fpi_+ -{i\over 2}\psi_+ - {i\over 2}\lambda (\cH -\cP) \psi_+\ , \label{eq: susy const p1}\\
    \cC_2\,=\,& \fpi_- -{i\over 2}\psi_- - {i\over 2}\lambda (\cH +\cP) \psi_-\ .\label{eq: susy const p2}
\end{align}

Using the Poisson bracket
\begin{align}
	\{F,G\}_\pb\,\equiv\, & \int dz\; \left[{\partial F \over \partial \phi_\nu(z) }{\partial G \over \partial \spi(z) }- {\partial F \over \partial \spi(z) }{\partial G \over \partial \phi(z) }\right]  \cr
	&+\sum_{\alpha=\pm }\int dz\; \left[{F \overleftarrow{\partial}\over \overleftarrow{\partial} \psi_\alpha(z) }{\overrightarrow{\partial}G \over \overrightarrow{\partial} \fpi_\alpha(z) }+ {F \overleftarrow{\partial}\over \overleftarrow{\partial} \fpi_\alpha(z) }{\overrightarrow{\partial}G \over \overleftarrow{\partial} \psi_\alpha(z) }\right]\ ,\label{def: susy poisson bracket}
\end{align}
%
%
we can evaluate the Poisson bracket of the constraints
\begin{align}
    \cM(i,x;j,y)\,\equiv\, \{\cC_i(x),\cC_j(y)\}_\pb\ ,
\end{align}
where $\cM(i,x;j,y)$ are found to be
\begin{align}
	&\cM(1,x;1,y)\cr
	\,=\,&\left[-{i\over 2} (1- 2\lambda\spi\phi' +\sqrt{(1+2\lambda \spi^2)(1+2\lambda \phi'^2)} )+ \lambda^2\left( -\spi \phi' + { \spi^2+\phi'^2 +4\lambda \spi^2 \phi'^2 \over 2\sqrt{(1+2\lambda \spi^2)(1+2\lambda \phi'^2)} } \right) \psi_+\psi'_+\right.\cr
	& \left.- {\lambda\over 2} \left( 1+  {1+ \lambda(\spi^2 + \phi'^2) \over \sqrt{(1+2\lambda \spi^2)(1+2\lambda \phi'^2)}} \right) \psi_-\psi'_-  - {i \lambda^4 (\spi^2-\phi'^2)^2 \over 2 \left[(1+2\lambda \spi^2)(1+2\lambda \phi'^2)\right]^{3\over 2} }\psi_+\psi'_+ \psi_-\psi'_-\right]\delta(x-y)\ ,\cr\\
	&\cM(2,x;2,y)\cr
	\,=\,&\left[-{i\over 2} (1+ 2\lambda\spi\phi' +\sqrt{(1+2\lambda \spi^2)(1+2\lambda \phi'^2)} )- \lambda^2\left( \spi \phi' + { \spi^2+\phi'^2 +4\lambda \spi^2 \phi'^2 \over 2\sqrt{(1+2\lambda \spi^2)(1+2\lambda \phi'^2)} } \right) \psi_-\psi'_-\right.\cr
	& \left.+ {\lambda\over 2} \left( 1+  {1+ \lambda(\spi^2 + \phi'^2) \over \sqrt{(1+2\lambda \spi^2)(1+2\lambda \phi'^2)}} \right) \psi_+\psi'_+  - {i \lambda^4 (\spi^2-\phi'^2)^2 \over 2 \left[(1+2\lambda \spi^2)(1+2\lambda \phi'^2)\right]^{3\over 2} }\psi_+\psi'_+ \psi_-\psi'_-\right]\delta(x-y)\\
	&\cM(1,x;2,y)= {\lambda\over 2} \left( 1+  {1+ \lambda(\spi^2 + \phi'^2) \over \sqrt{(1+2\lambda \spi^2)(1+2\lambda \phi'^2)}} \right) (\psi_+\psi'_- + \psi'_+\psi_- )\delta(x-y)\ .
\end{align}
From the matrix $\cM(i,x;j,y)$, one can calculate the Dirac bracket:
\begin{align}
    &\{F(x),G(y)\}_\dirac\cr
    \equiv\,&\{F(x),G(y)\}_\pb -  \sum_{i,j=1,2}\int dzdw\; \{F(x),\cC_i(z)\}_\pb \cM^{-1}(i,z;j,w)\{\cC_j(w),G(y)\}_\pb\ .
\end{align}
%

We found that the Dirac brackets among the scalar field and its conjugate momentum are the same as their Poisson brackets,
\begin{align}
	\{\phi(x),\spi(y)\}_\dirac\,=\,& \delta(x-y)\ ,\\
	\{\phi(x),\phi(y)\}_\dirac\,=\,& \{\spi(x),\spi(y)\}_\dirac\,=\,0\ .
\end{align}
The Dirac brackets among fermions are found to be
\begin{align}
	&i\{\psi_+(x),\psi_+(y)\}_\dirac\,=\,{ 2\lambda \spi \phi' -1+ \sqrt{(1+2\lambda \spi^2)(1+2 \lambda \phi'^2)} \over \lambda (\spi+\phi')^2}\delta(x-y)\cr
	&+{-1-\lambda(\spi+\phi')^2+ \sqrt{(1+2 \lambda \spi^2)(1+2\lambda \phi'^2)} \over (\spi+\phi')^2 \sqrt{(1+2\lambda \spi^2)(1+2\lambda \phi'^2)} }S_\+ \delta(x-y)\cr
	&+ \left[ { 2\lambda \spi \phi'\over  (\spi+\phi')^2}  + { \lambda(\spi^2+\phi'^2 +4\lambda \spi^2 \phi'^2) \over   (\spi+\phi')^2  \sqrt{(1+2\lambda \spi^2)(1+2\lambda \phi'^2)}  }  \right]S_= \delta(x-y)\cr
	&+ { \lambda^2\over (1+2\lambda \spi^2)(1+2\lambda \phi'^2) }\left[ 1+2 \lambda \spi \phi'  + {1+ \lambda\left[ (\spi+\phi')^2+ 4\lambda \spi^2\phi'^2 \right]\over \sqrt{(1+ 2\lambda \spi^2)(1+2\lambda \phi'^2)}} \right]S_\+ S_=\delta(x-y)\ ,\cr\\
	&i\{\psi_-(x),\psi_-(y)\}_\dirac\,=\,{ -2\lambda \spi \phi' -1+ \sqrt{(1+2\lambda \spi^2)(1+2 \lambda \phi'^2)} \over \lambda (\spi-\phi')^2}\delta(x-y)\cr
	&+{1+\lambda(\spi+\phi')^2- \sqrt{(1+2 \lambda \spi^2)(1+2\lambda \phi'^2)} \over (\spi - \phi')^2 \sqrt{(1+2\lambda \spi^2)(1+2\lambda \phi'^2)} }S_= \delta(x-y)\cr
	&+ \left[ { 2\lambda \spi \phi'\over  (\spi-\phi')^2}  - { \lambda(\spi^2+\phi'^2 +4\lambda \spi^2 \phi'^2) \over   (\spi - \phi')^2  \sqrt{(1+2\lambda \spi^2)(1+2\lambda \phi'^2)}  }  \right]S_\+ \delta(x-y)\cr
	&+ { \lambda^2\over (1+2\lambda \spi^2)(1+2\lambda \phi'^2) }\left[ 1-2 \lambda \spi \phi'  + {1+ \lambda\left[ (\spi - \phi')^2+ 4\lambda \spi^2\phi'^2 \right]\over \sqrt{ (1+ 2\lambda \spi^2) (1+2\lambda \phi'^2)}} \right]S_\+ S_=\delta(x-y)\ ,\cr\\
	&i\{\psi_+(x),\psi_-(x)\}_\dirac= - {i \lambda \over \sqrt{ (1+2\lambda \spi^2)(1+2\lambda \phi'^2) }}(\psi_+\psi_- )' \delta(x-y)\ .
\end{align}
Also, the Dirac brackets between the fermion and the scalar field are 
\begin{align}
	&\{\phi(x),\psi_+(y)\}_\dirac\,=\,{-2\lambda \spi^2 -1 + \sqrt{(1+2\lambda \spi^2)(1+2\lambda \phi'^2)}\over 2 (1+2\lambda \spi^2)(\spi +\phi')}\psi_+\delta(x-y) \cr
	&\hspace{20mm} - {i\lambda^2 \spi\over 2(1+2\lambda \spi^2))}\left[ 1+ {1+2\lambda \spi \phi' \over \sqrt{ (1+2\lambda \spi^2)(1+2\lambda \phi'^2) }} \right]\psi_+ \psi_- \psi'_- \delta(x-y)\ ,\\
	&\{\phi(x),\psi_-(y)\}_\dirac\,=\, {-2\lambda \spi^2 -1 + \sqrt{(1+2\lambda \spi^2)(1+2\lambda \phi'^2)}\over 2 (1+2\lambda \spi^2)(\spi - \phi')}\psi_-\delta(x-y) \cr
	&\hspace{20mm}+ {i\lambda^2 \spi\over 2(1+2\lambda \spi^2))}\left[ 1+ {1-2\lambda \spi \phi' \over \sqrt{ (1+2\lambda \spi^2)(1+2\lambda \phi'^2) }} \right]\psi_+ \psi'_+ \psi_-\delta(x-y)\ ,
\end{align}
\begin{align}
	&\{\spi(x),\psi_+(y)\}_\dirac\,=\,{ 2\lambda \phi'^2+1-\sqrt{(1+2\lambda \spi^2)(1+2\lambda \phi'^2)} \over 2 (1+2\lambda \phi'^2)(\spi+\phi')} \psi_+ \delta(x-y) \partial_x\cr
	&\hspace{20mm}+{i\lambda^2 \phi' \over 2(1+2\lambda \phi'^2)} \left[1+{1+2\lambda \spi \phi' \over \sqrt{ (1+2\lambda \spi^2)(1+2\lambda \phi'^2) } }\right]\psi_+\psi_-\psi'_- \delta(x-y) \partial_x\ ,\\
	&\{\spi(x),\psi_-(y)\}_\dirac \,=\,{ -2\lambda \phi'^2-1+\sqrt{(1+2\lambda \spi^2)(1+2\lambda \phi'^2)} \over 2 (1+2\lambda \phi'^2)(\spi - \phi')} \psi_- \delta(x-y) \partial_x\cr
	&\hspace{20mm}- {i\lambda^2 \phi' \over 2(1+2\lambda \phi'^2)} \left[1+{1 - 2\lambda \spi \phi' \over \sqrt{ (1+2\lambda \spi^2)(1+2\lambda \phi'^2) } }\right]\psi_+\psi'_+\psi_- \delta(x-y) \partial_x\ .
\end{align}

It is interesting to ask whether this deformed model has supersymmetry or not. At first glance, it is not guaranteed that the non-SUSY deformation preserves supersymmetry. But, the deformed spectrum follows the universal formula~\eqref{eq: deformed energy value}, and the Bose-Fermi degeneracy of the undeformed theory will be preserved under the non-SUSY deformation. This implies a supercharge operator with Hamiltonian and momentum and maps between those degenerate states, and indeed the deformed model will be supersymmetric. In Ref.~\cite{Coleman:2019dvf}, the supersymmetry of this model was confirmed perturbatively up to order $\cO(\lambda^2)$. Still, it is not clear whether this supercharge operator can be written as an integration of a local supercharge density or not.

Using the Dirac bracket, we find that two supercharges of the deformed model are
\begin{align}
    Q^1_+\, \equiv \, &\int dx\; \psi_+(\spi+\phi')\ , \label{eq: supercharge ttb susy 1}\\
    Q^1_-\, \equiv \, &\int dx\; \psi_-(\spi-\phi')\ .\label{eq: supercharge ttb susy 2}
\end{align}
This expression is identical to that of the undeformed theory. But, it is important to note that the conjugate momentum $\spi$ is not equal to $\dot{\phi}$. First, we confirm that $Q^1_+$ anti-commutes with $Q^1_-$,
\begin{align}
    \{Q^1_+,Q^1_-\}_\dirac\,=\,0\ .\label{eq: sq algebra1}
\end{align}
Furthermore, the Hamiltonian and the momentum from Eqs.~\eqref{eq: ham op susy} and \eqref{eq: momentum op susy} can be expressed in terms of Dirac brackets of the supercharges,
\begin{align}
    H\, =\, &  \int dx\; \cH\, =\, {i\over 4}\{Q^1_+,Q^1_+\}_\dirac+{i\over 4}\{Q^1_-,Q^1_-\}_\dirac\ ,\label{eq: sq algebra2}\\
    P\, =\, &  \int dx\; \cP \, =\, {i\over 4} \{Q^1_+,Q^1_+\}_\dirac-{i\over 4}\{Q^1_-,Q^1_-\}_\dirac\ .\label{eq: sq algebra3}
\end{align}
From the Jacobi identity of the Dirac bracket, one can easily deduce that $Q^1_\pm$ commutes with Hamiltonian and momentum operator,
\begin{align}
    \{Q^1_\pm,H\}_\dirac\,=\, \{Q^1_\pm,P\}_\dirac\,=\, 0\ .
\end{align}
Therefore, this explicitly proves that the non-SUSY \ttb deformation of $\cN=(1,1)$ SUSY model has supersymmetry (at least, classically). One can also evaluate the SUSY transformation of $\phi$ and $\psi_\pm$:
\begin{align}
    &\delta_+ \phi(x)\,\equiv\,\{Q^1_+,\phi(x)\}_\dirac \cr
    =\,&-{1\over 2} \left(1+\sqrt{1+2\lambda \phi'^2\over 1+2\lambda \spi^2}\right)\psi_+ + {i\lambda^2 \spi (\spi+\phi')\over 2(1+2\lambda\spi^2) }\left( 1+{1+2\lambda \spi \phi'\over \sqrt{(1+2\lambda \spi^2)(1+2\lambda \phi'^2)}} \right) \psi_+\psi_-\psi'_-\ ,\cr \\
    &\delta_- \phi(x)\,\equiv\, \{Q^1_-,\phi(x)\}_\dirac\cr
    =\,&-{1\over 2} \left(1+\sqrt{1+2\lambda \phi'^2\over 1+2\lambda \spi^2}\right)\psi_- - {i\lambda^2 \spi (\spi - \phi')\over 2(1+2\lambda\spi^2) }\left( 1+{1-2\lambda \spi \phi'\over \sqrt{(1+2\lambda \spi^2)(1+2\lambda \phi'^2)}} \right) \psi_+\psi'_+\psi_- \ .
\end{align}
\begin{align}
    &\delta_+ \psi_+\,\equiv \, \{Q^1_+,\psi_+\}_\dirac\cr
    =\,&i{1-2\lambda \spi \phi' -\sqrt{(1+2\lambda \spi^2)(1+2\lambda \phi'^2)} \over \lambda (\spi +\phi')}\cr
    &+ \lambda \left({2\spi \phi' \over \spi +\phi'}+ {\phi'^2 +\spi^2 +4 \lambda \spi^2 \phi'^2 \over (\spi + \phi') \sqrt{(1+2\lambda \spi^2)(1+2\lambda \phi'^2) } } \right) \psi_-\psi'_-\cr
    &+i{\lambda^2 (\spi + \phi')(1+2\lambda \spi \phi' + \sqrt{(1+2\lambda \spi^2)(1+2\lambda \phi'^2)})\over 2 (1+2\lambda \spi^2)(1+2\lambda \phi'^2)}\psi_+\psi'_+\psi_- \psi'_-\ ,\\
    &\delta_- \psi_+\, \equiv\, \{Q^1_-,\psi_+\}_\dirac=-{\lambda(\spi - \phi') \over \sqrt{(1+2\lambda \spi^2)(1+2\lambda \phi'^2)}}\psi'_+\psi_-\ ,\\
    &\delta_- \psi_- \,\equiv\,  \{Q^1_-,\psi_-\}_\dirac\cr
    =\,&i{1+2\lambda \spi \phi' -\sqrt{(1+2\lambda \spi^2)(1+2\lambda \phi'^2)} \over \lambda (\spi - \phi')}\cr
    &+ \lambda \left({2\spi \phi' \over \spi - \phi'} - {\phi'^2 +\spi^2 +4 \lambda \spi^2 \phi'^2 \over (\spi - \phi') \sqrt{(1+2\lambda \spi^2)(1+2\lambda \phi'^2) } } \right) \psi_+\psi'_+\cr
    &+i{\lambda^2 (\spi - \phi')(1-2\lambda \spi \phi' - \sqrt{(1+2\lambda \spi^2)(1+2\lambda \phi'^2)})\over 2 (1+2\lambda \spi^2)(1+2\lambda \phi'^2)}\psi_+\psi'_+\psi_- \psi'_- \ ,\\
    &\delta_+ \psi_- \,\equiv \, \{Q^1_+,\psi_-\}_\dirac=-{\lambda(\spi + \phi') \over \sqrt{(1+2\lambda \spi^2)(1+2\lambda \phi'^2)}}\psi_+\psi'_-\ .
\end{align}

The undeformed free $\cN=(1,1)$ SUSY model has additional global symmetry in shifting $\phi$ and $\psi_\pm$ by $a$ and $\eta_\pm$:
\begin{align}
    \phi \quad &\longrightarrow \quad \phi + a\ ,\\
    \psi_\pm\quad&\longrightarrow \quad \psi_\pm + \eta_\pm\ ,
\end{align}
where $a$ and $\eta_\pm$ is Grassmannian even and odd constant, respectively. This symmetry is generated by
\begin{align}
    \bP^2\, \equiv&\, {2\pi \over L}\int dx \; \spi\ ,\\
    Q^2_\pm\,\equiv& \,- {8\pi i\over L} \int dx\; \fpi_\pm\ .\label{def: fermi global charge}
\end{align}
Here, we emphasize that $\fpi_\pm$ is related to the $\psi_\pm\,,\; \phi$ and $\spi$ via the constraints~\eqref{eq: susy const p1} and \eqref{eq: susy const p2}, which is not equal to ${i\over 2}\psi_\pm$ for $\lambda\ne 0$. We confirm that this symmetry is also preserved under the \ttb deformation of $\cN=(1,1)$ SUSY model. Namely, they commute with Hamiltonian and momentum operator,
\begin{align}
    \{Q^2_\pm,H\}_\dirac\,=\,\{\bP^2,H\}_\dirac\,=\,\{Q^2_\pm,P\}_\dirac\,=\,\{\bP^2,P\}_\dirac\,=\,0\ .\label{eq: dirac brackets of h p}
\end{align}
And they satisfy the following algebra,
\begin{align}
    \{Q^2_\pm,Q^2_\pm\}_\dirac\, =\, &- {16\pi^2i\over L} - {16\pi^2i \lambda\over L^2} (H\mp P)\ ,\label{eq: sq algebra4}\\
    \{Q^2_+,Q^2_-\}_\dirac\, =\, &0\ ,\label{eq: sq algebra5}\\
    \{Q^1_+,Q^2_+\}_\dirac\, =\, &- 2i\left(\bP^2+ {4\pi^2\over L^2}\bW^2\right)\ ,\label{eq: sq algebra6} \\
    \{Q^1_-,Q^2_-\}_\dirac\,= \,&-2i \left(\bP^2-{4\pi^2\over L^2}\bW^2\right)\ ,\label{eq: sq algebra7}\\
    \{Q^1_\pm,Q^2_\mp\}_\dirac\,= \,&0\ ,\label{eq: sq algebra8}
\end{align}
where $\bW^2$ is the winding number of the scalar field $\phi$ if it is compactified:
\begin{align}
    \bW^2\,\equiv\, {L\over 2\pi }\oint dx\; \phi' \ .
\end{align}

\subsection{Lessons from the GS-like Worldsheet}
\label{sec: revisit ttbar deformation of gs superstring}

We saw the ordinary, i.e., non-supersymmetric \ttb deformation of the $N=(1,1)$ theory admit an $N=(1,1)$ supersymmetry. We constructed the superalgebra by exploring the canonical structure of the deformed theory. As shown earlier, the same theory can be obtained from integrating out the worldsheet metric of a GS-like theory with $D=3$ spacetime supersymmetry, where the presence of 
the supersymmetry is a little more transparent; The procedure of integrating out the worldsheet metric identifies part of 
spacetime with the worldsheet, such that the spacetime supersymmetry descends to that on the \ttb side. 
Now armed with the canonical analysis of the \ttb side, we wish to revisit the latter for the purpose of 
relating the symmetry algebra of the latter after the gauge fixing to that of  \ttb side. This should 
illuminate further on precise nature of the supersymmetry  in question.

We will follow the second approach to the GS-like action for the supersymmetry with shifted light-cone metric~\eqref{new target light cone} in Section~\ref{sec: review sfondrini}. Before moving on to the second approach, we shortly comment on the supersymmetry from the first approach to the $\cN=2$ supercharges of GS-like model with the static gauge~\eqref{eq: static gauge of gslike action sec 4}. We find that two of $\cN=2$ supercharges and their algebra exactly agree\footnote{For this we use the relation $P_{T\bar{T}}\,=\,-\bP^1$ deduced from the level-matching condition at $\lambda=0$. Recall that the momentum $P$ is invariant under the \ttb deformation.} with the supersymmetry of the \ttb deformed $\cN=(1,1)$ model in Section~\ref{sec: hamiltonian analysis for susy model}. The other two supercharges in $\cN=2$ SUSY are supposed to be related to the fermi global charges~\eqref{def: fermi global charge} in \ttb deformation. However, those two charges and their algebra do not seem to be simply expressed in terms of the fermi global charges.

Hence, in this section, we will focus on the second approach in Section~\ref{sec: review sfondrini}. Recall that the Lagrangian of the non-SUSY \ttb deformation agrees with that of GS-like action in uniform light-cone gauge with the shifted light-cone metric,
\begin{align}
    ds^2= 2\Lambda (dX^-)^2 + 2 dX^+ dX^- + (dX^i)^2\ ,\label{eq: target space metric}
\end{align}
where $\Lambda$ is a dimensionless parameter which we will clarify the relation to the \ttb deformation parameter $\lambda$ soon. We retrieve the string length scale ${1\over 2\pi \ell_s^2} = {1\over 2\pi \alpha'}$ to see the explicit relation to the length scale in the \ttb deformation~(\ie circumference $L$ of the cylinder). The length scale $L$ of the \ttb deformation is identified with the length scale of the string $\ell_s$ as follows,
\begin{align}
    {L\over 2\pi }\,= \, \sqrt{2\pi} \ell_s\ .
\end{align}
Accordingly, we relate the dimensionless parameter $\Lambda$ to the \ttb deformation parameter $\lambda$ by
\begin{align}
\Lambda\,\equiv {\lambda \over 2\pi \ell_s^2}\ .
\end{align}
Also, to compare the results explicitly, it is convenient to rescale the fields and variables by ${L\over 2\pi}$ to make them dimensionless in the \ttb deformation side. \eg
\begin{align}
    t\;,\;\;x\,\longrightarrow\, {L\over 2\pi } \tau\;,\;\; {L\over 2\pi } \sigma\;\;,\quad \lambda\, \longrightarrow  \, \left({L\over 2\pi}\right)^2 \Lambda\;\;,\quad \psi_\pm \,\longrightarrow\, \sqrt{2\pi \over L}\psi_\pm \;\;\;\;\etc
\end{align}
Then, the charge density and charge operators in the \ttb deformation are also rescaled properly. For example,
\begin{align}
    \cH\,=&\,{1\over 2\Lambda }\left(\sqrt{1+ 4\Lambda \cH_{b,0}+4\Lambda^2 \cP^2_{b,0}}-1\right)+\cdots  \ ,\\
    \cP\,=&\, \cP_{b,0}+{1\over 4}\left[ 1+ \sqrt{1+ 4\Lambda \cH_{b,0}+4\Lambda^2 \cP^2_{b,0}} \right](S_\++ S_=) \ ,\\
    H\,=&\,{2\pi \over L} \int_0^{2\pi} d\sigma \; \cH \ , \\
    P\,=&\, {2\pi \over L}\int_0^{2\pi} d\sigma\;\cP\ ,\label{eq: rescaled momentum}\\
    Q_\pm^1\,=&\, \sqrt{2\pi \over L }\int_0^{2\pi} d\sigma \; \psi_+ (\spi\pm\phi')\ ,\\
    Q_\pm^2\,=&\,-4i \sqrt{2\pi \over L}\int_0^{2\pi} d\sigma \; \fpi_\pm\qquad \etc
\end{align}
Here, we made the density operators $\cH$ and $\cP$ dimensionless while we keep the dimension of the charge operators. Also, we defined $\cH_{b,0}\,\equiv\, {1\over 2}\spi^2 + {1\over 2} \phi'^2$ and $\cP_{0,b}\,=\,\spi \phi'$.

The 3D GS-like action has 3D super Poincare symmetry, and we will show how this 3D super Poincare algebra is related to the algebra found in the previous section. We calculate the supercharges of the 3D $\cN=2$ GS-like model given in Ref.~\cite{Mezincescu:2011nh}\footnote{The index $a=1,2$ of supercharge $Q^\alpha_a$ in Ref.~\cite{Mezincescu:2011nh} corresponds to $-$ and $+$ in this paper, respectively.}
\begin{align}
    \bQ_\pm^\alpha= 2\oint d\sigma\; \left[ \left(\momp^\mu \pm  {1\over 2\pi \ell_s^2}\bPi_\sigma^\mu\right)(\Gamma_\mu\Psi_\pm)^\alpha \pm {2i\over 2\pi \ell_s^2} (\overline{\Psi}_\pm\Psi_\pm) \partial_\sigma \Psi^\alpha_\pm  \right]\hspace{8mm} (\alpha=1,2)\ ,
\end{align}
where $\momp^\mu$ is conjugate momentum of $X^\mu$. Due to the gauge condition $\Gamma^+\Psi_\pm=0$ fixing the $\kappa$ symmetry, the Majorana spinor $\Psi_\pm$ is written as follows,
\begin{align}
    \Psi_\pm\,=\, {1\over 2\sqrt{p_-}} \begin{pmatrix}
        \psi_\pm\\
        0\\
    \end{pmatrix}\ .
\end{align}
To make contact with (dimensionless) scalar field $\phi$ and its conjugate momentum $\spi$ in the \ttb deformation, we also rescale the transverse target coordinate $X^2$ and its conjugate momentum $p^2$ as follows,
\begin{align}
    X^2\,\equiv \, \sqrt{2\pi} \ell_s \phi\;\;,\quad p^2\,=\, {1\over \sqrt{2\pi } \ell_s} \spi\ .
\end{align}
Before obtaining the supercharges, we revisit the uniform light-cone quantization. In order to identify target coordinate $X^+$ with worldsheet time $\tau$,
\begin{align}
    X^+\,=\, \sqrt{2\pi } \ell_s \tau\ ,
\end{align}
the target coordinate $X^+$ should be either time-like or null. From Eq.~\eqref{eq: target space metric}, one can see that this is possible\footnote{For example, consider Eddington-Finkelstein metric or Vaidya metric.} when
\begin{equation}
\Lambda\;\geqq\; 0\ .
\end{equation}
Therefore, the relation between GS-like model and \ttb deformation in this work is valid for $\Lambda\;\geqq\; 0$. Following DLCQ~\cite{Susskind:1997cw}, we compactify null coordinate $X^-$. Then, the momentum zero-mode $p_-$ is quantized,\footnote{Consider the Wilson-Sommerfeld quantization for the symplectic one-form,
\begin{equation}
    \oint d\sigma \oint  p_- dX^-\,=\, 2\pi n\ .
\end{equation}} \ie
\begin{align}
    p_-\;=\;{2\pi n\over 2\pi R }\ ,
\end{align}
where $R$ is the circumference of $X^-$ and the range of the worldsheet spatial coordinate $\sigma$ belongs to $[0,2\pi]$. Recall that we chose the condition\footnote{In Section~\ref{sec: review sfondrini}, we worked out in the unit with $2\pi \ell_s^2=1$.}
\begin{align}
    p_-\,=\, {1\over \sqrt{2\pi} \ell_s}\ ,
\end{align}
for uniform light-cone gauge. This is possible when the circumference $R$ of the target coordinate $X^-$ becomes
\begin{align}
    R\,=\, \sqrt{2\pi }\ell_s n\ .\label{eq: circumference R}
\end{align}
Furthermore, due to the compactification of $X^-$ coordinate, we have a topological charge corresponding to winding mode along $X^-$, which is quantized, 
\begin{align}
    \bW^-\,\equiv \, \oint d\sigma \; \partial_\sigma X^-\,=\, - m R\,=\, - \sqrt{2\pi } \ell_s m n \ ,
\end{align}
where we used Eq.~\eqref{eq: circumference R}. The level matching condition of the string theory relates this topological charge to the momentum operator $P$~\eqref{eq: rescaled momentum} of the \ttb deformed theory, 
\begin{align}
    P\,=\,- {2\pi \over L} p_- \bW^-\,=\, {2\pi m R p_-\over L} \,=\, {2\pi n m \over L}\ .
\end{align}
Therefore, for $n=1$ or $m=1$, this reproduces the quantization of the momentum operator $P$ in the \ttb deformation. In addition to $\bW^-$, we could have additional topological charge $\bW^2$ if $X^2\,=\,\sqrt{2\pi}\ell_s\phi$ is also compactified.

With this gauge condition, the supercharge can be written as
\begin{align}
    \bQ_\pm\,=\, \oint d\sigma\;{1\over\sqrt{p_-}}\left[  \left( 2p_- - 2\Lambda p_+  \pm  {2\Lambda\over 2\pi \ell_s^2 } \partial_\sigma X^- \right) \begin{pmatrix}
        0\\
        \psi_\pm\\
    \end{pmatrix} + (p^2 \pm {1\over 2\pi \ell_s^2}\bPi^2_1) \begin{pmatrix}
        \psi_\pm\\
        0 \\
    \end{pmatrix} \right] \ .
\end{align}
%
%
%
One can connect the operators in string theory (on the left-hand side) with those in the \ttb deformed theory\footnote{Recall that we make the density operator $\cH$ and $\cP$ dimensionless, and the charge operator $H$ and $P$ has the length dimension $-1$.} (on the right-hand side) as follow
\begin{alignat}{4}
    p_+\,=&\, -{2\pi \over L}\cH\quad,\hspace{15mm} &\bP_+\,&=\,- H\ ,\\
    p_-\,=&\, {2\pi \over L}\quad,\hspace{15mm}  &\bP^2\, &= \,{2\pi \over L} \int d\sigma\; \spi\ ,\\
    p^2\,=&\,{2\pi \over L}\spi\quad,\hspace{15mm} &\bW^- \,&= \,  - {L^2\over 4\pi^2} P\ ,\\
    \bPi^2_1\,=&\, {L\over 2\pi} \phi'\quad,\hspace{15mm}  &\bW^2 \,&= \, {L\over 2\pi }\oint d\sigma\; \partial_\sigma \phi\ ,\\
    \partial_\sigma X^-\,= &\, -{L\over 2\pi} \cP \ .
\end{alignat}
Then, the supercharge of the $\cN=2$ super Poincare algebra can be written as
\begin{align}
    \bQ_\pm\,=\, \sqrt{2\pi \over L}\oint d\sigma\; \left[\begin{pmatrix}
        0\\
        -4i \fpi_\pm\\
    \end{pmatrix} +  \begin{pmatrix}
        \psi_\pm (\spi \pm \phi') \\
        0 \\
    \end{pmatrix}\right]\ ,
\end{align}
where we used the constraints~\eqref{eq: susy const p1} and \eqref{eq: susy const p2} for $\fpi_\pm$ which is conjugate momentum of $\psi_\pm$: 
\begin{equation}
    \fpi_\pm\,=\,{i\over 2}[1 + \Lambda (\cH\mp \cP) ]\psi_\pm\ .
\end{equation}
Hence, we confirm that this is exactly what we had in the previous section (with suitable rescaling as explained):
\begin{align}
    \bQ^\alpha_a \, =&\, Q^\alpha_a \hspace{9mm} (\alpha=1,2\;,\;\; a=\pm)\ .
\end{align}
Then, the algebra found in Eqs.~\eqref{eq: sq algebra1}$\sim$\eqref{eq: sq algebra3} and Eqs.~\eqref{eq: sq algebra4}$\sim$\eqref{eq: sq algebra8} can be summarized by
\begin{align}
    \{\bQ_a^\alpha, \bQ_b^\beta \}_\dirac\,=\,&-2i \delta_{ab}(\Gamma^\mu C)^{\alpha \beta} \bP_\mu - {2i\over 2\pi \ell_s^2} \sigma^3_{ab} A^{\alpha \beta }\ ,\label{eq: susy algebra central extension}
\end{align}
where $\Gamma^\mu C \bP_\mu$ and the topological charge $A^{\alpha \beta}$ is given by
\begin{align}
    \Gamma^\mu C \bP_\mu\,=\,\begin{pmatrix}
        -\bP^- & \bP^2\\
        \bP^2 & 2\bP^+ +2\Lambda \bP^-\\
    \end{pmatrix} \,=\,\begin{pmatrix}
        -\bP_+ & \bP^2\\
        \bP^2 & 2\bP_- - 2\Lambda \bP_+\\
    \end{pmatrix} \,=\,\begin{pmatrix}
        H & \bP^2\\
        \bP^2 & {8\pi^2\over L} + 2\Lambda H\\
    \end{pmatrix}\ , \label{eq: gamma C in algebra}
\end{align}
\begin{align}
    A\,=\, \Gamma_\mu C \oint d\sigma \; \partial_\sigma X^\mu\,=\, \begin{pmatrix}
        -\bW^- & \bW^2\\
        \bW^2 & 2\Lambda \bW^-\\
    \end{pmatrix}\,=\, {L^2\over 4\pi^2}\begin{pmatrix}
        P & {4\pi^2\over L^2}\bW^2\\
        {4\pi^2\over L^2}\bW^2 & -2 \Lambda P\\
    \end{pmatrix}\ .\label{eq: topological central charge}
\end{align}
%
%
%
%
%
%
We can write them explicitly as
\begin{align}
     \{\bQ_+^\alpha, \bQ_+^\beta \}_\dirac\,=\, -2i\begin{pmatrix}
        H + P & \bP^2 + {4\pi^2 \over L^2}\bW^2\\
        \bP^2+ {4\pi^2 \over L^2}\bW^2 & {8\pi^2\over L} + 2\Lambda (H-P) \\
    \end{pmatrix} \ , \\
     \{\bQ_-^\alpha, \bQ_-^\beta \}_\dirac\,=\,-2i\begin{pmatrix}
        H - P & \bP^2 - {4\pi^2 \over L^2}\bW^2\\
        \bP^2- {4\pi^2 \over L^2}\bW^2 & {8\pi^2\over L} + 2\Lambda (H+P) \\
    \end{pmatrix} \ .
\end{align}
This is exactly the supersymmetry algebra of $\cN=2$ super Poincare with topological charge~\cite{deAzcarraga:1989mza}. From the point of view of the 3D GS-like model, the topological charge in the supersymmetry algebra is originated from the WZ term of the GS-like model~\cite{deAzcarraga:1989mza}.

Recall the $\lambda$ dependence of the algebra of the fermi global charges~\eqref{eq: sq algebra4}. In string theory, the deviation of the light-cone coordinate $X^+$  from the usual one is responsible for this $\lambda$ dependence. On the other hand, from the point of view of \ttb deformation, this is an example of the symmetry algebra deformed by \ttb deformation.

Recall that the operators $H,\;P,\; \bP^2$ and $\bW^2$ commute.\footnote{See Eq.~\eqref{eq: dirac brackets of h p}. Also, from Eqs.~\eqref{eq: sq algebra6} and \eqref{eq: sq algebra7}, one can also deduce that $\bW^2$ commutes with others by Jacobi identity if $\phi$ is compact.} Hence, at quantum level one can consider an eigenstate of those operators
\begin{align}
    |E,P,p_m,p_w\rangle\ ,
 \end{align}
where $E,\;P,\; p_m$ and $p_w$ is the eigenvalue of $H,\;P,\;\bP^2$ and $\bW^2$, respectively. In general, we have 16 degeneracy in this eigenspace. To see this, let us consider degenerate states by acting $Q_+^1$ and $Q_+^2$ on $|E,P,p_m,p_w\rangle$. Note that $|E,P,p_m,p_w\rangle\;,\;\;Q_+^1Q_+^2|E,P,p_m,p_w\rangle\;,\;\; Q_+^2Q_+^1|E,P,p_m,p_w\rangle$ are not linearly independent because of the anti-commutation relation of $Q_+^1$ and $Q_+^2$. Hence, we have 4 states instead of 5 states,
\begin{align}
    &|E,P,p_m,p_w\rangle\;\;,\quad \bQ^1_+ \bQ_+^2 |E,P,p_m,p_w\rangle \ ,\\
    &\bQ_+^1 |E,P,p_m,p_w\rangle\;\;,\quad \bQ_+^2 |E,P,p_m,p_w\rangle \ .
\end{align}
Together with the action of $Q_-^1$ and $Q_-^2$, we have 16 degenerate states in general. This agrees with the usual free $\cN=(1,1)$ SUSY model. In free $\cN=(1,1)$ SUSY model ($\lambda=0$), we have 4 degeneracy by the $\cN=(1,1)$ supercharges $Q_\pm^1$. Furthermore, the fermion zero-modes\footnote{In $\lambda=0$, $\fpi_\pm= {i\over 2} \psi_\pm$.} $\bQ_\pm^2$ give 4 degenerate ground states. In total, we also have 16 degenerate states in general.

In $\cN=(1,1)$ supersymmetry of \ttb deformation, the positive definiteness\footnote{Recall that the Dirac bracket becomes (anti-)commutation relation by $i\{\;,\;\}_\dirac\;\;\longrightarrow\;\; [\;,\;]_\pm$. Moreover, there could be additional contributions from the normal ordering. Here, we include the (regularized) shifts from the normal ordering in the eigenvalues such as $E$ and $P$ if exists.} of the supersymmetric algebra,
\begin{align}
[\bQ_\pm^1,\bQ_\pm^1]_+\,=\, 2(H\pm P)\ ,
\end{align}
gives 
\begin{align}
    E\pm P \geqq 0\ .
\end{align}
This can be saturated by a state satisfying
\begin{align}
    E\,=\, |P|\label{eq: bps condition 1}\ ,
\end{align}
and this state preserve the half of the $\cN=(1,1)$ supersymemtry -- either $\cN=(1,0)$ or $\cN=(0,1)$ supersymmetry depending on the sign of $P$. The condition~\eqref{eq: bps condition 1} is protected under \ttb deformation~\cite{Datta:2018thy}. That is, let us consider a state in the undeformed theory ($\lambda=0$) satisfying
\begin{align}
    E^{(0)}\,=\, |P^{(0)}|\ ,
\end{align}
where $E^{(0)}$ and $P^{(0)}$ is the undeformed energy and momentum. Under the \ttb deformation, the momentum is invariant $P\,=\,P^{(0)}$, and the deformed energy~\eqref{eq: deformed energy value} becomes
\begin{align}
    E\,=\,{L \over 2\lambda}\left[\sqrt{1+ 4{\lambda\over L} |P| + 4{\lambda^2\over L^2} |P|^2} -1\right]\, =\, |P|\ .
\end{align}
Therefore, the deformed state also satisfies the condition~\eqref{eq: bps condition 1}. Without momentum and winding zero-modes of $\phi$, there always exists a state with the condition~\eqref{eq: bps condition 1} for each momentum $P$. On the other hand, there is no state with non-vanishing momentum zero-mode or winding zero-mode (if $X^2=\phi$ is compactified) which is annihilated by $\bQ^1_\pm$. One might try to find a
special cases\footnote{For example, non-compact $X^2=\phi$ with a special momentum zero-mode or compact $X^2=\phi$ with self-dual radius can make the energy of form ${2\pi n\over L}$ ($n\in\mathbb{Z}$).} satisfying the condition~\eqref{eq: bps condition 1} with non-vanishing zero-modes. Soon we will prove that no physical state with non-vanishing zero-mode satisfies the condition~\eqref{eq: bps condition 1} from the point of view of $\cN=2$ supersymmetry of GS-like model.

From the point of view of $\cN=(1,1)$ SUSY model, Eq.~\eqref{eq: susy algebra central extension} is the algebra among $\cN=(1,1)$ supercharges $\bQ_\pm^1$ and fermi global charges $\bQ_\pm^2$. Thus, it might look odd to consider its ``BPS'' bound. Nevertheless, one can still demand the positiveness of the anti-commutation relation~\eqref{eq: susy algebra central extension} whether it is a supercharge or a fermi global charge. Also, from the perspective of $\cN=2$ supersymmetry of GS-like model, it is natural to discuss the ``BPS'' bound. From Eq.~\eqref{eq: susy algebra central extension}, the ``BPS'' bound reads
\begin{align}
    & L(E\pm P)+ {\Lambda\over 4\pi^2} L^2 (E^2-P^2) - {1\over 2} \left( {L\over 2\pi }p_m \pm {2\pi \over L }p_w\right)^2\,\geqq\, 0\ ,\label{eq: bps condition1}\\
    &8\pi^2+2\Lambda L(E\mp P)+ L(E\pm P) \,\geqq\, 0\ .\label{eq: bps condition2}
\end{align}
where $p_m$ and $p_w$ denotes the eigenvalue of $\bP^2$ and $\bW^2$ corresponding to the momentum and winding zero-mode of $X^2={L\over 2\pi }\phi$. First, note that the BPS bound~\eqref{eq: bps condition2} will not be saturated because $E\pm P \geqq 0$, which leads to the partially broken rigid supersymmetry~(PBRS) by topological WZ term~\cite{deAzcarraga:1989mza}. From the point of view of free $\cN=(1,1)$ SUSY model, $\bQ_\pm^2\,\sim\,\int d\sigma \pi_\pm\,\sim\, \int d\sigma \pi_\pm $ is already ``broken'' (\ie $\langle\text{vac} | \bQ_+^2\bQ_+^2 | \text{vac}\rangle\,\ne \,0$). And we have never called it ``supersymmetry'' in the $\cN=(1,1)$ SUSY context.

The BPS bound~\eqref{eq: bps condition1} implies that when $p_m\ne0$ or $p_w\ne 0$, the condition $E=|P|$ violates the BPS bound~\eqref{eq: bps condition1}. This immediately prove that there is no physical state with non-vanishing zero-modes such that $E=|P|$.

From the point of view of $\cN=2$ SUSY of the GS-like model, one can consider a ``BPS'' state with non-vanishing zero-modes which saturates the BPS bound~\eqref{eq: bps condition1}. When one of the inequality in Eq.~\eqref{eq: bps condition1} is saturated, the other in Eq.~\eqref{eq: bps condition1} gives the inequality between $P$ and $p_m p_w$. And then, one can write the ``BPS'' condition with $\big|P-{1\over L}p_m p_w\big|$ instead of $\pm \big(P-{1\over L}p_mp_w\big)$,
\begin{align}
    E\,=\,{L\over 2\lambda}\left[\sqrt{1 + {4\lambda \over L }\bigg|P-{1\over L}p_mp_w\bigg| +{2\lambda \over L^2}\left( {L^2\over 4\pi^2 }p_m^2 + {4\pi^2 \over L^2 }p_w^2\right) + {4\lambda^2\over L^2}P^2 }-1\right]\ ,\label{eq: bps condition 2}
\end{align}
where we used $\Lambda={4\pi^2 \over L^2}\lambda$. Note that this condition returns to the condition~\eqref{eq: bps condition 1} for vanishing zero-modes $p_m=p_w=0$. This ``BPS'' condition is protected under the \ttb deformation. For $\lambda=0$, the ``BPS'' state is characterized by
\begin{align}
    E^{(0)}\,=\, \bigg|P^{(0)}-{1\over L}p_mp_w\bigg| +{1\over 2L}\left( {L^2\over 4\pi^2 }p_m^2 + {4\pi^2 \over L^2 }p_w^2\right)\ .
\end{align}
Note that for $\lambda=0$ we will have the 3D GS-like model with the usual light-cone target metric, and this is nothing but the familiar worldsheet BPS condition. Under the \ttb deformation, its deformed energy also follows the universal formula~\eqref{eq: deformed energy value}, and the deformed states also satisfy the BPS condition~\eqref{eq: bps condition 2} along the \ttb deformation,
\begin{align}
    E\,=\,&{L\over 2\lambda } \left[ \sqrt{1+ {4\lambda \over L}E^{(0)} +{4\lambda^2\over L^2}P^2 } - 1 \right]\cr
    =\,& {L\over 2\lambda}\left[\sqrt{1 + {4\lambda \over L }\bigg|P-{1\over L}p_mp_w\bigg| +{2\lambda \over L^2}\left( {L^2\over 4\pi^2 }p_m^2 + {4\pi^2 \over L^2 }p_w^2\right) + {4\lambda^2\over L^2}P^2 }-1\right]\ ,
\end{align}
where the momentum is invariant under the \ttb deformation (\ie $P=P^{(0)}$). In other words, the ``BPS'' condition~\eqref{eq: bps condition 2} already shows how it flows along the \ttb deformation. Note that this ``BPS'' state spontaneously breaks $\cN=(1,1)$ supersymmetry when it has non-vanishing zero-modes
\begin{align}
\bQ_\pm^1\, |\text{``BPS''}\rangle\,\ne\, 0 \hspace{7mm} \mbox{when}\quad p_m\,\ne\, 0\;\;\;\mbox{or}\;\;\; p_w\ne 0 \ .
\end{align}
From the point of view of \ttb deformation, this is interesting because ``BPS'' states will still be protected under the deformation even though it spontaneously breaks $\cN=(1,1)$ supersymmetry. On the other hand, this is natural from the point of view of $\cN=2$ Super Poincare, and the ``BPS'' state is annihilated by a linear combination of supercharges $\bQ_\pm^1$ and fermi global charges $\bQ_\pm^2$, respectively.

The \ttb deformation of $\cN=(0,1)$ SUSY model can easily be obtained by truncating the fermion $\psi_-$ from the $\cN=(1,1)$ case~\cite{Baggio:2018rpv}. Hence, we only have two fermi charges
\begin{align}
    Q^1_+\, =&\, \sqrt{2\pi \over L}\int d\sigma\; \psi_+(\spi+\phi')\ ,\\
    Q^2_+\, =&\, - 4 i \sqrt{2\pi \over L} \int d\sigma \; \fpi_+\ .
\end{align}
The first one is the supercharge of $\cN=(0,1)$ supersymmetry, and the second one generates a shift of the fermion $\psi_+$ by a Grassmannian-odd constant. This \ttb deformed $\cN=(0,1)$ SUSY model corresponds to the 3D $\cN=1$ GS-like model. $Q^a_+=\bQ_+^a$ $(a=1,2)$ are the supercharges of the $3D$ $\cN=1$ super Poincare algebra. As in $\cN=2$ GS-like model, the topological charge $\bW^-=-{L^2\over 4\pi^2} P$ appears in the supersymmetry algebra because of the WZ term,
\begin{align}
    [Q^\alpha, Q^\beta ]_+\,=\,2 (\Gamma^\mu C)^{\alpha \beta} P_\mu + {2\over 2\pi \ell_s^2} A^{\alpha \beta }\ ,
\end{align}
where $\Gamma^\mu C$ and $A^{\alpha\beta}$ are given in Eqs.~\eqref{eq: gamma C in algebra} and \eqref{eq: topological central charge}. As before, the ``BPS'' bound can be written as
\begin{align}
    &L(E+P)\left[ 1+ {\Lambda \over 4\pi^2} L (E-P)\right]-{1\over 2} \left( {L\over 2\pi }p_m+ {2\pi \over L }p_w\right)^2 \,\geqq\, 0 \ ,\label{eq: bps bound 3}\\
    &8\pi^2+2\Lambda L(E- P)+ L(E+ P) \,\geqq\, 0\ . \label{eq: bps bound 4}
\end{align}
From the two inequality, one can deduce that
\begin{align}
    L(E+P)\, \geqq\,  0\quad,\qquad L(E-P)\, \geqq\, -{4\pi^2\over \Lambda}\ .
\end{align}
Therefore, Eq.~\eqref{eq: bps bound 4} cannot be saturated. The ``BPS'' condition from Eq.~\eqref{eq: bps bound 3} is
\begin{align}
    E\,=\,{L\over 2\lambda}\left[\sqrt{1 - {4\lambda \over L }P+{2\lambda \over L^2}\left( {L\over 2\pi }p_m+ {2\pi \over L }p_w\right)^2+ {4\lambda^2\over L^2}P^2 }-1\right]\ .
\end{align}
This ``BPS'' state is also protected under the \ttb deformation as before.

\section{The Question of Which Energy-momentum Tensor}
\label{sec: instability}

So far, we have mostly studied the \ttb deformations constructed from the Noether energy-momentum tensor. It has been observed~\cite{Baggio:2018rpv,Frolov:2019nrr}, as we repeated here, that the \ttb deformed Lagrangian obtained from the worldsheet theory approach is consistent with this energy-momentum tensor from the Noether procedure. This is itself a little counter-intuitive as we are accustomed to the symmetric energy-momentum tensor, especially dealing with systems with dynamical gravity. Moreover, in the derivation of the \ttb deformed spectrum via the factorization formula~\cite{Zamolodchikov:2004ce}, one might think that the symmetric energy-momentum tensor is a more natural starting point, as obtained by variation of action with respect to the metric. This led us to wonder what would happen for \ttb deformations by the symmetric energy-momentum tensor. Do we obtain the same theory in the end, or if not, why did the worldsheet approaches above preferred the Noether energy-momentum? 
Let us start the investigation by listing some simplest examples of \ttb deformed theories where the symmetric energy-momentum tensor is used. 

A priori, deformation of fermionic theories by higher derivative interaction terms can easily suffer from superfluous degrees of freedom. Recall how, in ordinary fermion theories with at most a single time derivative, the system comes with a second-class constraint that relates the conjugate momenta of the fermion to the original fermion variables. The familiar canonical anti-commutator among the fermions is a result of the Dirac bracket based on this constraint. The latter results in, relative to the bosonic counterpart, halving of degrees of freedom. For instance, a complex Grassmannian field has the same number of canonical degrees of freedom as a single real scalar. 
\ttb deformations involve a higher-derivative operator and have the same potential danger of spoiling this counting. 

Therefore, a different kind of issue arises when the ``deformed" Lagrangian involves terms with more than one time derivatives on fermions; the usual second-class constraints that halve the degrees of freedom are no longer constraints, resulting in the degree of freedom being doubled. We find that, for the simplest examples, the \ttb based on the symmetric energy-momentum tensor leads to this phenomenon. Such higher derivative fermion theories are riddled with another problem, with unitarity. The naive canonical analysis 
leads to negative norm states, but we will find an alternate positive definite inner product on the Hilbert space on a closer look. We further show that this choice is actually the one consistent with the path integral, as can be seen from the partition function, and the unitarity can be intact. We will see below 
that this tends to happen quite easily in fermionic theories, especially with the symmetric energy-momentum. in particular,
the same happens if one deforms $\cN=(1,1)$ theory by a supersymmetric completion of ordinary \ttb. 

We illustrate the phenomenon by considering a $d=1$ toy model and a pair of $d=2$ pure fermion theories in the large $\lambda$ limit. We should warn ahead that in the latter examples, the Hamiltonian can be made Hermitian, hence the evolution operator is unitary, but only at the cost of the spatial momentum operator, a perfectly sensible physical quantity, becoming non-Hermitian. It is also unclear whether this recovery of the unitarity is an artifact of the $d=1$ toy model or the UV approximation we employed for the analysis in $d=2$.\footnote{A class of fermion theories with two time derivatives has been studied elsewhere with a similar conclusion, although their choice of the inner product is a little different \cite{LeClair:2007iy,Robinson:2009xm}.} 

Regardless of unitary or not, what does remain clear is that, for some \ttb deformations, the word ``deformation" is misleading. Even if such theories are well-defined, they lead to a rather different UV completion with an exponentially larger Hilbert space than naively anticipated.  In the IR, one must decouple half the degrees of freedom to match the free undeformed theory limit. The latter operation looks not too ill-motivated at least, thankfully, given how the relevant states are far removed from the physical ones due to the very large energy eigenvalues, divergent as $\lambda\rightarrow 0$. In retrospect, 
the very fact that we are yet to find such complications for the \ttb deformations with the Noether energy-momentum tensor should be a surprise.

\subsection{Examples of the Other \ttb Deformations} 
\label{sec: cov T example }

\subsubsection*{\ttb Deformation of Free Fermion by Symmetric $T_{\mu\nu}$}
Let us revisit the simplest model of a single Majorana fermion. 
%
%
%
%
The \ttb deformed Lagrangian of the free fermion by symmetric energy-momentum tensor is~\cite{Bonelli:2018kik}
\begin{align}
	\cL
	=\,& {i\over 2}\psi_+  \dot{\psi}_+ + {i\over 2} \psi_- \dot{\psi}_- - {i\over 2}\psi_+  \psi'_+ + {i\over 2} \psi_- \psi'_-  + {3\lambda\over 8}  \left( - \psi_+  \psi'_+ \psi_-\dot{\psi}_-  + \psi_+  \dot{\psi}_+ \psi_-\psi'_- \right)\cr
	& - {\lambda\over 8} \psi_+ \dot{\psi}_+\psi_-  \dot{\psi}_-  +  {\lambda\over 8} \psi_+ \psi'_+\psi_-  \psi'_-\ . \label{eq: lagrangian sym tt fermion}
\end{align}
From the Lagrangian, one can obtain the conjugate momentum $\fpi_\pm$ of the fermion $\psi_\pm$: 
\begin{align}
	\fpi_+\,=\,& {i\over 2} \psi_+ + {3\lambda \over 8 } \psi_+ \psi_- \psi'_- - {\lambda \over 8 } \psi_+ \psi_- \dot{\psi}_- \ ,\\
	\fpi_- \,=\,& {i\over 2} \psi_-  - {3\lambda \over 8 } \psi_+ \psi_- \psi'_- + {\lambda \over 8 } \psi_+ \psi_- \dot{\psi}_+\ .
\end{align}
Note that the right-hand sides contain $\dot{\psi}_\pm$, unlike the fermions in Section~\ref{sec: hamiltonian analysis of ttb deformation}. Therefore, they are not the second-class constraints for non-zero $\lambda$, although it is still not clear how to invert those equations to solve for $\dot{\psi}$ in general. Since we do not have the second-class constraints, we will have additional degrees of freedom, which would have been removed from the Hilbert space with the constraints. The term with more than one time derivatives is responsible for the emergence of the extra degrees of freedom for the case of free fermion. The higher time derivative term does not vanish either in the Hamiltonian, \ie
\begin{equation}
\begin{split}
	&\cH\,=\,  \fpi_+ \dot{\psi}_++\fpi_- \dot{\psi}_- -\cL \cr
	\,=\,& \left(\fpi^+ - {i\over 2} \psi_+ -{3\lambda \over 8}  \psi_+ \psi_- \psi'_- + {\lambda \over 8 } \psi_+ \psi_- \dot{\psi}_-  \right)\dot{\psi}_+ \cr
	& + \left( \fpi^- -  {i\over 2} \psi_-  + {3\lambda \over 8 } \psi_+ \psi_- \psi'_- -  {\lambda \over 8 } \psi_+ \psi_- \dot{\psi}_+\right) \dot{\psi}_- \cr
	& +{i\over 2} \psi_+ \psi'_+  - {i\over 2} \psi_- \psi'_- - {\lambda \over 8} \psi_+ \psi'_+ \psi_- \psi'_-  - {\lambda\over 8} \psi_+ \dot{\psi}_+\psi_-  \dot{\psi}_-\ .
\end{split}
\end{equation}
%

\subsubsection*{\ttb Deformation on Superspace}
\label{sec: superspace example }

There exist three types of \ttb deformed Lagrangians in the literature~\cite{Baggio:2018rpv,Chang:2018dge,Frolov:2019nrr,Coleman:2019dvf} of the simplest $\cN(1,1)$ model, whose free form Lagrangian is
\begin{align}
    \cL_0\,=\,&2\partial_\+\phi\partial_=\phi+i\psi_+\partial_=\psi_++i\psi_-\partial_\+\psi_-\ .
\end{align}
We have analyzed the canonical structure for one of them in Section~\ref{sec: hamiltonian analysis for susy model},
where the deformation involves the vanilla \ttb with no attempt at its supersymmetry completion. Here, 
we found no additional degrees of freedom appeared. 

On the other hand, the other Lagrangians were worked out in the superspace so that 
supersymmetry is manifestly preserved by construction~\cite{Chang:2018dge,Baggio:2018rpv}. 
In those two other Lagrangians, the conjugate momentum $\fpi_\pm$ of the fermions 
again includes the time derivatives of fermions. Denoting the Lagrangian generally 
as a Taylor expansion
\begin{equation}
	\mathcal L_\lambda\,=\,\mathcal L_0+\lambda\mathcal L_1+\lambda^2\mathcal L_2+\mathcal O(\lambda^3)\ ,
\end{equation}
the first-order correction $\cL_1$ of the  \ttb deformed Lagrangian in Ref.~\cite{Baggio:2018rpv} is found to be
\begin{align}
	\mathcal L_1\,=\,&-\left(2(\partial_\+\phi)^2+i\psi_+\partial_\+\psi_+\right)\left(2(\partial_=\phi)^2+i\psi_-\partial_=\psi_-\right)+\psi_+\partial_=\psi_+ \psi_-\partial_\+\psi_-\cr
	&-4i(\partial_\+\phi\partial_=\phi)\left(\psi_+\partial_=\psi_++\psi_-\partial_\+\psi_-\right)\ .
\end{align}
The conjugate momentum $\fpi_\pm$ of the fermion $\psi_\pm$ can be written as follows,
\begin{align}
\fpi_+\,=\,&\frac i2\bigg[1-\frac\lambda2\bigg(3 \spi^2-2\spi\phi'+2i\psi_-\dot\psi_--(\phi')^2\bigg)
\bigg]	\psi_++\mathcal O(\lambda^2)\ ,\\
\fpi_-\,=\,&\frac i2\bigg[1-\frac\lambda2\bigg(3 \spi^2+2\spi\phi'+2i\psi_+\dot\psi_+-(\phi')^2\bigg)
\bigg]\psi_-+\mathcal O(\lambda^2)	\ .
\end{align}
Since $\dot{\psi}_\pm$ appears on the right-hand side, these relationships are not constraints anymore and imply doubling of degrees of freedom warned above. 

The same phenomenon can be observed in the deformed Lagrangian in Ref.~\cite{Chang:2018dge} from the order $\cO(\lambda^2)$. After integrating out the auxiliary field, the  \ttb deformed Lagrangian in Ref.~\cite{Chang:2018dge} reads\footnote{Ref.~\cite{Chang:2018dge} used a different convention for light-cone from ours, \ie $\partial_{\pm\pm}=\frac1{\sqrt2}(\partial_t\pm\partial_x)$. Hence, we calculate Eqs.~\eqref{eq: chang lag1}$\sim$\eqref{eq: chang mom 2} following the convention in Ref.~\cite{Chang:2018dge}.} 
\begin{align}
\mathcal L_1\,=\,&-((\partial_\+\phi)^2+\psi_+\partial_\+\psi_+)((\partial_=\phi)^2+\psi_-\partial_=\psi_-)+\psi_+\partial_=\psi_+\psi_-\partial_\+\psi_-\cr
&-2(\partial_\+\phi\partial_=\phi)(\psi_+\partial_=\psi_++\psi_-\partial_\+\psi_-)\ ,\label{eq: chang lag1}\\[6pt]
\mathcal L_2\,=\,&3(\partial_\+\phi)^2\psi_+\partial_=\psi_+\psi_-\partial_=\psi_-+3(\partial_=\phi)^2\psi_+\partial_\+\psi_+\psi_-\partial_\+\psi_-	\cr
&+2(\partial_\+\phi\partial_=\phi)\psi_+\partial_=\psi_+\psi_-\partial_\+\psi_-+8(\partial_\+\phi\partial_=\phi)\psi_+\partial_\+\psi_+\psi_-\partial_=\psi_-\cr
&+6(\partial_\+\phi\partial_=\phi)^2(\psi_+\partial_=\psi_++\psi_-\partial_\+\psi_-)\cr
&+4(\partial_\+\phi\partial_=\phi)((\partial_\+\phi)^2\psi_-\partial_=\psi_-+(\partial_=\phi)^2\psi_+\partial_\+\psi_+)+2(\partial_\+\phi\partial_=\phi)^3\ ,
\end{align}
and the conjugate momentum $\fpi_\pm$ are found to be
\begin{align}
\fpi_+\,=\,&\frac1{\sqrt2}\bigg[1-\frac\lambda2\bigg(3\spi^2-2\spi\phi'+2\sqrt2\psi_-\psi'_--(\phi')^2\bigg)-\frac{\lambda^2}2\bigg(\spi^4+2\spi^3\phi'-17\sqrt2\spi^2\psi_-\dot\psi_-\cr
&+10\sqrt2\spi\phi'\psi_-\psi'_--2\spi(\phi')^3-4\sqrt2(\phi')^2\psi_-\psi'_-+3\sqrt2(\phi')^2\psi_-\dot\psi_--(\phi')^4\bigg)\bigg]\psi_++\mathcal O(\lambda^3)\ ,\\
\fpi_-=&\frac1{\sqrt2}\bigg[1-\frac\lambda2\bigg(3\spi^2+2\spi\phi'+2\sqrt2\psi_+\psi'_+-(\phi')^2\bigg)-\frac{\lambda^2}2\bigg(\spi^4-2\spi^3\phi'-17\sqrt2\spi^2\psi_+\dot\psi_+\cr
&+10\sqrt2\spi\phi'\psi_+\psi'_++2\spi(\phi')^3+4\sqrt2(\phi')^2\psi_+\psi'_++3\sqrt2(\phi')^2\psi_+\dot\psi_+-(\phi')^4\bigg)\bigg]\psi_-+\mathcal O(\lambda^3)\ .\label{eq: chang mom 2}
\end{align}
One can see again that $\dot{\psi}_\pm$ appears at the order $\cO(\lambda^2)$. 

Both Lagrangians are constructed from the superfield $\cT_{\mu\alpha}$ $(\mu=\+, =\;\;,\alpha=\pm)$ containing the supercurrent and energy-momentum tensor. The superfield $\cT_{\mu\alpha}$ is calculated as a supersymmetric Noether current. The deformation by this superfield induces the coupling to the extra degrees of freedom.

It was observed~\cite{Baggio:2018rpv,Chang:2018dge} that the superfield-squared deformation is equal to the deformation by the determinant of the energy-momentum tensor on-shell,
\begin{align}
    \int d^2\theta\; (\cT_{\++}\cT_{=-} + \cT_{\+-}\cT_{=+})\,=\, T_{\+\+ }T_{==}- T_{\+=}T_{=\+}\hspace{8mm}\mbox{(on-shell)}\ .\label{eq: susy tt deformation}
\end{align}
However, this does not mean that SUSY \ttb deformation is equivalent to non-SUSY \ttb deformation. Because the supercurrent does contribute to the deformation of Lagrangian off-shell, the energy-momentum tensor 
$T_{\mu\nu}$ on the right-hand side of Eq.~\eqref{eq: susy tt deformation} has a different form from 
that of non-SUSY \ttb deformed theory.

One can add further add other divergence-less terms to the energy-momentum tensor. Using the improved 
energy-momentum tensor, one can even get the higher-order Lagrangian of scalar fields in principle. This also leads to the coupling to the extra degrees of freedom.

\subsection{A Toy Model with $\cJ$-Hemiticity}
\label{sec: toy model}

To understand the emergence of the extra degrees of freedom, we will study a toy model of which Lagrangian is
\begin{align}
	L\,=\,{i\over 2} \bar{\psi} \dot{\psi} - {i\over 2} \dot{\bar{\psi}} \psi+m \bar{\psi} \psi - \lambda \dot{\bar{\psi}} \dot{\psi}\ .
\end{align}
Note that the last term characterizes the symplectic fermion in Ref.~\cite{LeClair:2007iy,Robinson:2009xm}. The conjugate momentum of $\psi$ and $\bar{\psi}$ can be obtained by 
\begin{align}
	\fpi\,=\,&{\overleftarrow{\delta} L\over \overleftarrow{\delta} \dot{\psi}}=  {i \over 2} \bar{\psi} - \lambda \dot{\bar{\psi}}\ ,\label{eq: def mom1}\\
	\bar{\fpi}\,=\,&{\overrightarrow{\delta} L\over \overrightarrow{\delta} \dot{\bar{\psi}} }= - {i \over 2} \psi - \lambda \dot{\psi}\ .\label{eq: def mom2}	
\end{align}
Note that we take the right and left derivative of the Lagrangian for $\pi$ and $\bar{\fpi}$ in order to demand that $\bar{\psi}$, $\bar{\fpi}$ be Hermitian conjugate to $\psi$ and $\fpi$, respectively. From this, we can express $\dot{\psi}$ and $\dot{\bar{\psi}}$ in terms of phase space variables,
\begin{align}
	\dot{\psi}\, =\,&-{1\over \lambda} \left( \bar{\fpi} + {i\over 2}\psi \right)\ ,\\
	\dot{\bar{\psi}}\, =\, &-{1\over \lambda} \left( \fpi - {i\over 2}\bar{\psi} \right)\ .
\end{align}
Now, we can obtain Hamiltonian,
\begin{align}
	H\,=\,& \fpi \dot{\psi} + \dot{\bar{\psi}} \bar{\fpi} - L
	\,=\, -m\bar{\psi} \psi - {1\over \lambda} \left( \fpi - {i\over 2} \bar{\psi}  \right)\left( \bar{\fpi} + {i \over 2} \psi \right)\ . \label{eq: hamiltonian of toy model}
\end{align}
Note that the order of fermi fields $\fpi \dot{\psi} + \dot{\bar{\psi}} \bar{\fpi}$ in Legendre transformation for Hamiltonian is also consistent with hermiticity condition for Hamiltonian. 
Moreover, this implies the anti-commutation relations of phase space variables is given by
\begin{align}
	\{\psi, \fpi\}\,=\, i\;\;,\qquad \{\bar{\psi}, \bar{\fpi} \}\,=\, -i\label{eq: psi fpi algebra toy model} \ .
\end{align}
Note that this anti-commutation is consistent with Hermitian conjugation. 

To construct Fock space, we will find the following transformation from $\psi\,,\,\bar{\psi}\,,\, \fpi\,,\, \bar{\fpi}$ to fermi oscillators $b, b^\dag, c, c^\dag $ parametrized by real constants $u_1\,,\,u_2\,,\,v_1\,,\,v_2$:
\begin{align}
	\bar{\fpi} +{i\over 2}\psi \,=\,& i (v_1 b+ v_2 c)\ ,\\
	\fpi -{i\over 2}\bar{\psi} \,=\,& -i (v_1 b^\dag+ v_2 c^\dag) \ , \\
	\bar{\fpi} -{i\over 2}\psi \,=\,& -i (u_1 b +u_2 c)\ ,\\
	\fpi +{i\over 2}\bar{\psi} \,=\,& i (u_1b^\dag+u_2 c^\dag)\ ,
\end{align}
where the oscillators obey the anti-commutation relations
\begin{align}
	\{b,b^\dag\}\,=\, 1\;\;,\qquad \{c,c^\dag\}\,=\,-1\ .\label{eq: b c algebra toy model}
\end{align}
Demanding that the algebras~\eqref{eq: psi fpi algebra toy model} and \eqref{eq: b c algebra toy model} are mapped by a Bogoliubov transformation, we obtain the conditions for the coefficients,
\begin{align}
	&u_1^2-u_2^2\,=\,1 \ ,\\
	&v_1^2-v_2^2\,=\,-1\ ,\\
	&u_1 v_1 - u_2v_2\,=\,0\ ,	
\end{align}
and they can be parameterized by $\theta$:
\begin{align}
	u_1\,=\,& \cosh\theta\quad,\qquad u_2 \,=\, \sinh \theta\ ,\cr
	v_1\,=\,& \sinh\theta\quad,\qquad v_2 \,=\,\cosh \theta\ .
\end{align}
Using this transformation, one can express the Hamiltonian in terms of the oscillators. 

We will choose the parameter $\theta$ in a way that cross terms such as $b^\dag c$ and $c^\dag b$ in the Hamiltonian vanish. The choice for vanishing cross terms will be clarified soon. From this requirement, the parameter $\theta$ is determined to be
\begin{align}
	\tanh 2\theta \,=\, - {2m\lambda \over 2m\lambda+ 1 }\ .
\end{align}
Because $\theta$ is real, this transformation is valid for
\begin{align}
	1+4m\lambda \,>\,0\ ,
\end{align}
and in this range of $\theta$, the Hamiltonian becomes
\begin{align}
	H\,=\, {1-\sqrt{1+4m\lambda} \over 2\lambda } b^\dag b - {1+\sqrt{1+4m\lambda} \over 2\lambda }c^\dag c \ .\label{eq: hamiltonian toy model cj hermitian}
\end{align}
Defining vacuum state $|0,0\rangle$ by
\begin{align}
    b|0,0\rangle\,=\, c|0,0\rangle\,=\,0\ ,
\end{align}
we construct Fock space by acting $b^\dag$ and $c^\dag$ on the vacuum. It is easy to see that these states diagonalize the Hamiltonian:
\begin{alignat}{4}
	&E_1\,=\,0\;\;,\qquad&&|\Psi_1\rangle\,=\,|0,0\rangle\ , \label{eq: eig1p}\\
	&E_2\,=\,{1-\sqrt{1+4m\lambda} \over 2\lambda }\;\;,\qquad&&|\Psi_2\rangle\,=\,|1,0\rangle \,\equiv\,b^\dag|0,0\rangle\ , \label{eq: eig2p}\\
	&E_3\,=\,{1+\sqrt{1+4m\lambda} \over 2\lambda }\;\;,\qquad&&|\Psi_3\rangle\,=\,|0,1\rangle\,\equiv\, c^\dag|0,0\rangle\ , \label{eq: eig3p}\\
	&E_4\,=\,{1\over \lambda}\;\;,\qquad&&|\Psi_4\rangle\,=\, |1,1\rangle \,\equiv\, b^\dag c^\dag|0,0\rangle \ .\label{eq: eig4p}
\end{alignat}
Due to the unusual anti-commutation relation of $c$ and $c^\dag$ in Eq.~\eqref{eq: b c algebra toy model}, this Fock space includes negative norm states. To see this, we define bra state $\langle \nu_1, \nu_2|$ by Hermitian conjugation of $|\nu_1,\nu_2\rangle$.  Using the anti-commutation relations, we have
\begin{align}
	\langle 0,0  | 0,0 \rangle\,=\,& \langle 0,0 | b^\dag b+ b b^\dag  | 0,0 \rangle\,=\, \langle 1,0  | 1,0 \rangle\ ,\\
	\langle 0,0  | 0,0 \rangle\,=\,&- \langle 0,0 | c^\dag c+ c c^\dag  | 0,0 \rangle\,=\, -\langle 0,1  | 0,1 \rangle\ ,\\
	\langle 0,1  | 0,1 \rangle\,=\,& \langle 0,0 | b^\dag b+ b b^\dag  | 0,0 \rangle\,=\, \langle 1,1 | 1,1 \rangle\ .
\end{align}
Assuming that the vacuum state $|0,0\rangle$ is normalized to be $1$, we can conclude that the negative norm state is inevitable:
\begin{align}
	\langle 0,0  | 0,0 \rangle\,=\,\langle 1,0  | 1,0 \rangle\,=\, - \langle 0,1  | 0,1 \rangle\,=\, - \langle 1,1  | 1,1 \rangle\,=\,1\ .\label{eq: norm of state}
\end{align}
Hence, the states constructed by $c^\dag$ have a negative norm, as had been observed in the symplectic fermion~\cite{Robinson:2009xm}. Therefore, although the Hamiltonian is Hermitian, the time evolution is not unitary because of the negative norm states. 

However, the unitarity depends on the definition of the inner product, and we will show that the unitarity can be recovered by an alternative definition of the inner product. For this, we introduce an operator $\cJ$~\cite{LeClair:2007iy,Robinson:2009xm} defined by
\begin{align}
	\cJ\,\equiv\, 1+ 2c^\dag c\ .
\end{align}
This operator is Hermitian and unitary,
\begin{align}
	\cJ^\dag \,=\, &\cJ\quad,\qquad \cJ^2\,=\, 1\ .
\end{align}
Moreover, the operator $\cJ$ anti-commutes with $c$ and $c^\dag$ while it commutes with $b$ and $b^\dag$,
\begin{align}
	&\cJ c \cJ\,=\, -c\;\;,\quad \cJ c^\dag \cJ\,=\,-c^\dag\;\;,\quad \cJ b \cJ\,=\, b\;\;,\quad \cJ b^\dag \cJ\,=\, b^\dag\ .\label{eq: action of j on b c}
\end{align}
Assuming that the vacuum is invariant under the action of operator $\cJ$
\begin{align}
	\cJ|0,0\rangle \,=\, |0,0\rangle\ ,
\end{align}
the operator $\cJ$ flips the sign of the state $|0,1\rangle$ and $|1,1\rangle$ which have negative norm:
\begin{align}
	\cJ|1,0\rangle \,=\, &|1,0\rangle\ ,\\
	\cJ|0,1\rangle \,=\, &-|0,1\rangle\ ,\\
	\cJ|1,1\rangle \,=\, &-|1,1\rangle\ .
\end{align}
Using the operator $\cJ$, the completeness relation can be written as
\begin{align}
	1
	\,=\,&|0,0\rangle \langle 0,0|\cJ + |1,0\rangle \langle 1,0|\cJ + |0,1\rangle \langle 0,1|\cJ  + |0,0\rangle \langle 0,0|\cJ \ .
\end{align}
Therefore, it is natural to define\footnote{In this paper, we employ the notation $\cJ$ instead of $\cC$ used in Refs.~\cite{LeClair:2007iy,Robinson:2009xm} to prevent any confusion. Especially, the definition of the inner product is different, \ie $\langle\; \rangle_{\text{\tiny here}}\,=\, \langle\; \rangle_{\cC\,,\, \text{\tiny there}}$ and $\langle \;\rangle_{\cJ\,,\, \text{\tiny here}}\,=\, \langle \;\rangle_{ \text{\tiny there}}$} $\cJ$-norm
\begin{align}
    \langle i |\cO| j \rangle_\cJ \,\equiv\, \langle i| \cJ  \cO | j \rangle\ .
\end{align}
Note that the $\cJ$-norm of the Fock space basis now becomes positive,
\begin{align}
    \langle i | j \rangle_\cJ\,=\, \delta_{ij}\ .
\end{align}
Also, the matrix representation for an operator, which is compatible with the above completeness relation, is defined in terms of the $\cJ$-norm
\begin{align}
    \cO_{ij}\,\equiv\, \langle i |\cO| j \rangle_\cJ \ .
\end{align}
For example, unlike the naive norm, the matrix representation of the Hamiltonian with $\cJ$-norm can reproduce the energy eigenvalues~\eqref{eq: eig1p}$\sim$\eqref{eq: eig4p}. When we use $\cJ$-norm, we need to introduce $\cJ$-conjugation $\cdag$ defined by
\begin{align}
	\cO^\cdag\,\equiv\, \cJ \cO^\dag \cJ\ .
\end{align}
This $\cJ$-conjugation is compatible with $\cJ$-norm, \ie
\begin{align}
	&\langle \cO \psi' | \psi  \rangle_\cJ \,=\, \langle \cO \psi' | \cJ | \psi  \rangle \,=\,\langle  \psi' |\cO^\dag \cJ | \psi  \rangle \,=\, \langle \psi' | \cJ \cO^\cdag |\psi\rangle \,=\, \langle  \psi' | \cO^\cdag  | \psi \rangle_\cJ \ .
\end{align}
Recall that we have chosen the parameter $\theta$ to eliminate the cross terms in the Hamiltonian. Then, the Hamiltonian~\eqref{eq: hamiltonian toy model cj hermitian} without the cross terms is $\cJ$-Hermitian,
\begin{align}
    H^\cdag \,= \, H\ .
\end{align}
Therefore, one can expect that the energy eigenstates (Fock space states) now enjoy the nice properties of the Hermitian matrices. 

Without the $\cJ$-Hermicity, the orthogonality of eigenstates would have been violated in spite of the positive-definite $\cJ$-norm. Now, the time evolution becomes unitary with respect to the $\cJ$-norm for $1+4m\lambda>0$. Recall that the transformation to $b$, $c$ is possible only for $1+4m\lambda>0$. For $1+4m\lambda>0$, this system exhibits nice physical properties such as real energy eigenvalues, orthonormal energy eigenstates, and unitary time evolution. However, for $1+4m\lambda\leqq0$, we lose all these sensible properties.

The above operator formalism with $\cJ$-norm is consistent with the path integral formalism,
in the sense that the latter naturally computes 
\begin{align}
	Z(\beta)\,=\,\int \cD \bar{\psi} \cD \psi\; \exp\left(-S_\beta[\bar{\psi},\psi]\right) \,=\,\Tr \left[ \cJ e^{-\beta H}\right] \ .
\end{align}
with the additional operator insertion of $\cJ$ on the canonical side. A formal proof of this is given in the  Appendix~\ref{app: path integral}, but we can easily convince ourselves of this relation by computing 
the (thermal) partition function by path integral directly. 

Imposing the usual anti-periodic boundary condition along the Euclidean time circle, the fermions can be expanded by
\begin{align}
	\psi(\tau)\,=\,&\sum_{n\in \mathbb{Z}+{1\over 2}}  \psi_n\; e^{{2\pi i n\tau \over \beta}}\ ,\\
	\bar{\psi}(\tau)\,=\,&\sum_{n\in \mathbb{Z}+{1\over 2}}  \bar{\psi}_n\; e^{-{2\pi i n\tau \over \beta}}\ ,
\end{align}
up to the normalization constant. The path integral can be evaluated as
\begin{align}
	Z\,=\,&\int \cD \bar{\psi} \cD \psi\; e^{-S_\beta[\bar{\psi},\psi]} = \prod_{n\in \mathbb{Z}+{1\over 2}} {\beta\over 2\pi} \left[ -\lambda \left({2\pi n \over \beta}\right)^2 + {2\pi i n \over \beta} - m  \right]\ ,\cr
	\,=\,&\prod_{n\in \mathbb{Z}_+\cup \{0\}+{1\over 2}} {4\pi^2\lambda^2 n^4\over \beta^2}  \left[  1+ {\beta^2 z^2\over 4\pi^2 n^2} \right]\left[  1+ {\beta^2 w^2\over 4\pi^2 n^2} \right]\ ,
\end{align}
where we defined
\begin{align}
	z\, \equiv\, \beta{1+\sqrt{1+4m\lambda}\over 2\lambda}\quad,\qquad w\, \equiv\,\beta {1-\sqrt{1+4m\lambda}\over 2\lambda}\ .
\end{align}
By regularizing the partition function by $L_0= -\lambda \dot{\bar{\psi}} \dot{\psi}$, we have
\begin{align}
	{Z\over Z_0}\,=\,& \cosh { z\over 2}\cosh {w\over 2}
	\,=\,{1\over 4} e^{-\beta{1\over 2\lambda}}\left[ 1+ e^{\beta\left(1- \sqrt{1+4m\lambda}\over 2\lambda\right)}+e^{\beta\left(1+ \sqrt{1+4m\lambda}\over 2\lambda\right)}+ e^{\beta {1\over \lambda}} \right]\ .\label{eq: partition function path integral}
\end{align}
On the other hand, in operator formalism, the partition function can be found to be
\begin{align}
	Z\,=\,& \Tr\left[ \cJ e^{-\beta H}\right]
	\,=\,1+ e^{\beta\left(1- \sqrt{1+4m\lambda}\over 2\lambda\right)}+e^{\beta\left(1+ \sqrt{1+4m\lambda}\over 2\lambda\right)}+ e^{\beta {1\over \lambda}} \ ,
\end{align}
which reproduces the result~\eqref{eq: partition function path integral} from the path integral up to 
a normal ordering, of which we were not careful, of the Hamiltonian. The path integral consists of the
four terms, all with positive sign, and the exponents match energy eigenvalues we obtained on the 
canonical side. Clearly, the end result is consistent with the $\cJ$-norm and with the above 
Hamiltonian, modulo a normal ordering issue.

In the small $\lambda$ limit, the energy eigenstates cluster pairwise. $E_{1,2}$ and  $E_{3,4}$ each
cluster together, relatively separated by a large gap $\sim 1/\lambda$. Concentrating on
one pair, say, $E_{1,2}$, we find $E_1-E_2\simeq m \ll \lambda$. 
%
%
The two sectors are separated by a divergent energy gap as $\lambda\rightarrow 0$, so
a consistent decoupling of state $|\Psi_3\rangle$ and $|\Psi_4\rangle$ from the other two
would be possible in the $\lambda \rightarrow + 0$ limit. In this limit, we can restrict our 
attention to $\{|0,0\rangle, |1,0\rangle\}$, disregarding $c\,, \, c^\dag$ (\ie $c\,=\,c^\dag\,=\,0$). 
With this truncation, the fermion $\psi$ and $\bar{\psi}$ reduce to
\begin{align}
	\psi\,=\,b\;\;,\qquad \bar{\psi}\,=\,b^\dag\ ,
\end{align}
with the usual anti-commutation relation of the usual free fermion 
\begin{align}
	\{\psi,\bar{\psi}\}\,=\, \{b,b^\dag\}\,=\,1 \ .
\end{align}
One can even say that, at least in this toy model, the system reduces to 
that of the ordinary fermion oscillator defined at $\lambda=0$ by decoupling 
the highly gapped states.

So far, we have analyzed the toy model where the higher time derivative term is still quadratic; this allowed an easy and explicit analysis and a relatively simple understanding of the extra emergent states for all the values of $\lambda$. What would happen if the higher derivative term appears at the quartic order? As we have seen before, some of $d=2$ \ttb deformation such as those in Section~\ref{sec: cov T example } come with a quartic interaction of type $\psi_+\dot{\psi}\psi_-\dot{\psi}$. Our quadratic toy model would be insufficient even as a qualitative model. As such,  let us consider, instead, the following $d=1$ toy model of $N$ complex fermions. The large $N$ limit can emulate some essential features of $d=2$ higher derivative theories by a dimensional reduction,  
\begin{align}
    S\;=\;\sum_{j=1}^N \big(\bar{\psi}_j\dot{\psi}_j + m_j \bar{\psi}_j \psi_j \big) + \lambda \sum_{j<k} \bar{\psi}_j \dot{\psi}_j \bar{\psi}_k\dot{\psi}_k\ .
\end{align}
Due to the double time derivatives in the quartic terms, we again lose the second-class constraint so that the dimension of the phase space is $2N$, instead of the usual $N$. Thus, there will be $2^{2N}$ states instead of $2^N$.

Now one immediate question is whether these extra states decouple in the small $\lambda$ limit, at least, the same as with the above quadratic model?
Consider the thermal partition function, schematically written as,
\begin{align}
    Z(\beta,\lambda)\;=\; \sum_{j=1}^{2^{2N}} e^{-\beta E_j(\lambda)}\ .
\end{align}
The partition function should be again analytic in $\lambda$. Taking the limit $\lambda\rightarrow 0$, one thus should recover the partition function of $N$ free complex fermion,
\begin{align}
    \lim_{\lambda \rightarrow 0} Z(\beta,\lambda)\; = \; Z_{\text{\tiny free}}(\beta) =\sum_{j=1}^{2^N} e^{-\beta E_j^{\text{\tiny free}}} \ .
\end{align}
The implication is that in the limit $\lambda\rightarrow 0$ only $2^N$ states have finite energies while the energies of the other $2^{2N}-2^N$ states diverge. This implies, in turn, that these $N$ degrees of freedom must all be infinitely gapped and decouple in the small $\lambda$ limit, just as in the quadratic toy model.

Note that in both classes above, one cannot really consider the finite $\lambda$ theories to be a ``deformation" of the $\lambda=0$ theory in any strict sense. Instead, the $\lambda=0$ theory is embedded into and emerges from, in the infrared limit, a one-parameter family of higher derivative theories with a much larger Hilbert space. We will presently see that the same behavior occurs for some \ttb deformations and other similarly irrelevant deformations when the theory contains fermions, to begin with.

\subsection{Two-dimensional Models with $\cJ$-Hermiticity near UV}
\label{sec: negative mode nsrlike model}

In the simplest of $d=1$ toy model above with a single complex fermion but with double time derivatives, we managed to reduce the Hamiltonian to a pair of free oscillators and find a positive-definite norm, twisted by $\cJ$, under which the Hamiltonian 
is Hermitian. Without such twists, one would have found negative norm states, but
fortunately, we showed that this twisted $\cJ$-norm is actually the one demanded by the path integral. 

In the small $\lambda $ limit, one set of oscillators become highly gapped, of order $\sim 1/\lambda$, leaving behind the ordinary pair that would have emerged at $\lambda=0$ after the second-class constraint is used. In the later, more general quantum mechanics that had quartic fermion interaction terms again with double time derivatives, we could not carry out such precise analysis but did see that the degrees of freedom again split into two classes in the small $\lambda$ limit; one is composed of would-be harmonic oscillators at $\lambda=0$ while the other invokes divergent energy gaps and decouples as $\lambda\rightarrow 0$. In this limited sense, the finite $\lambda$ theory with double the degrees of freedom reduces to the ordinary first-order theory sitting at $\lambda=0$.

Here, we lift the same set of questions to $d=2$ theories with higher time derivatives and find these
exhibit some common behaviors with the $d=1$ toy models. The quantum mechanics with the quartic fermion term can be considered as a very rough image of various momentum sectors of such $d=2$ theories, so we expect that, for small $\lambda$, $d=2$ theory would again split into two sectors, one with light degrees of freedom and the other with heavy degrees of freedom which incur large energy gap, relative to the light ones, scaling inversely with small $\lambda$. Unfortunately, the full analysis of this limit with all degrees of freedom kept is enormously complicated. All we can say is that as $\lambda\rightarrow 0$, the heavy part of the theory would again decouple from the ordinary light fields relevant for strict $\lambda=0$. 

Also unclear is whether the $\cJ$-Hermiticity of the simplest toy model above can be extended here for $d=2$ theories with such higher derivative interaction terms. In the most general setting, it is widely believed that ghost sectors are generic once such higher time-derivative terms lift the second-class constraint for the fermion \cite{Henneaux:1992ig}. However, we will take up a pair of the simplest theories deformed by quartic fermion interactions with double time-derivatives and show that one can define a positive definite pairing in the Hilbert at least for the UV limit space, with respect to which the evolution is unitary. Unfortunately, not all of the usual observables can be made $\cJ$-Hermitian simultaneously, adding further uncertainties on such high derivative theories in $d=2$. 

The first example is the truncated Lagrangian obtained from the NSR-like action in Section~\ref{sec: rns superstring}. Although it is not clear if this is a \ttb deformation based on some version of the energy-momentum tensor, its spectra share the common eigenvalue flows and have a rather simple structure of the irrelevant perturbation. It serves as a useful starting point for $d=2$ investigation of 
the above issue. The action is 
\begin{align}
	S^{\text{\tiny tr}}_{\text{\tiny NSR}} \,=\,& \int d\tau d\sigma \left[ {i\over 2} ( \psi_+\dot{\psi}_+ +\psi_-\dot{\psi}_- - \psi_+\psi'_+  + \psi_-\psi'_-  ) \right]\cr
	- &\Lambda\int d\tau d\sigma \left[ \psi_+ \dot{\psi}_+\psi_- \dot{\psi}_-  - \psi_+ \dot{\psi}_+\psi_- \psi'_- + \psi_+ \psi'_+\psi_- \dot{\psi}_- - \psi_+ \psi'_+\psi_- \psi'_- \right]\ , \label{eq: truncated nsr action section 5}
\end{align}
where we rescaled\footnote{Or, one may rescale the fermion $\psi_\pm$ and the worldsheet coordinates $(\tau,\sigma)$ to make contact with the usual \ttb deformation.} $\lambda$ by ${L^2\over 4\pi^2}$ to define dimensionless parameter $\Lambda$.
\begin{align}
    \Lambda \,\equiv\, {4\pi^2 \lambda \over L^2}\ .
\end{align}
Recall that we integrated out the worldsheet metric from the NSR-like action with the static gauge, and we truncated the bosonic degrees of freedom. This action has a term that is quadratic in the time derivatives of fermions, which leads to negative norm states. 

The conjugate momentum of the fermion $\psi_\pm$ is 
\begin{align}
	\pi_+ \,=\,& {i\over 2} \psi_+ - \Lambda  \psi_+ \psi_- \dot{\psi}_- + \Lambda \psi_+ \psi_- \psi'_-\ , \label{eq: nsrlike fermion momentum 1}\\
	\pi_- \,=\,& {i\over 2} \psi_- + \Lambda  \psi_+ \psi_- \dot{\psi}_+ + \Lambda  \psi_+ \psi_- \psi'_+\ .\label{eq: nsrlike fermion momentum 2}
\end{align}
The relations~\eqref{eq: nsrlike fermion momentum 1} and \eqref{eq: nsrlike fermion momentum 2} do not form the second-class constraint any more for $\Lambda\,\ne\,0$. Moreover, the problematic term $\psi_+ \dot{\psi}_+\psi_- \dot{\psi}_-$ still appears in the Hamiltonian
\begin{align}
	H\,=\, \int d\sigma\;\bigg[  {i\over 2} \psi_+ \psi'_+- {i\over 2} \psi_-\psi'_- -  \Lambda\psi_+ \dot{\psi}_+ \psi_- \dot{\psi}_- -\Lambda \psi_+ \psi'_+ \psi_-\psi'_- \bigg]\label{eq: app ham nsr like ham}
\end{align}

To see what happens in the large $\Lambda$ limit more concretely, let us take such a limit while keeping the phase space variables being of order $\cO(\Lambda^0)$, \ie
\begin{align}
    \psi_\pm\;,\;\; \fpi_\pm \;\; \sim\;\; \cO(\Lambda^0)\ .
\end{align}
To analyze the Hamiltonian we need to express $\dot{\psi}_\pm$ in terms of other phase space variables. For this, we rewrite the relations~\eqref{eq: nsrlike fermion momentum 1} and \eqref{eq: nsrlike fermion momentum 2} as follows,
\begin{align}
	\Lambda \psi_+ \psi_- ( \dot{\psi}_- - \psi'_-)\,=\,& -\fpi_+ +{i\over 2}\psi_+\ ,\\
	\Lambda \psi_+ \psi_- ( \dot{\psi}_+ + \psi'_+)\,=\,& \fpi_- - {i\over 2}\psi_-\ .
\end{align}
In a large $\Lambda$ limit, these can be inverted perturbatively,
\begin{align}
	\dot{\psi}_+\,=\,& - \psi'_+ +{1\over \Lambda } \eta_+\ , \\
	\dot{\psi}_-\,=\,&  \psi'_- +{1\over \Lambda } \eta_-\ ,
\end{align}
where $\eta_\pm$ satisfies
\begin{align}
	 \psi_+ \psi_- \eta_- \,=\,& -\fpi_+  + {i\over 2}\psi_+\ ,\\
	 \psi_+ \psi_-\eta_+ \,=\,& \fpi_- - {i\over 2}\psi_-\ .
\end{align}
Note that $\eta_\pm$ can, in principle, have any terms which vanish when we act $\psi_+\psi_-$ on it. Those terms may have an arbitrarily higher order in $\lambda$. But, since $\eta_\pm$ appears together with $\psi_+\psi_-$ in the Hamiltonian, they will not contribute to the Hamiltonian. Using them, one can expand the Hamiltonian~\eqref{eq: app ham nsr like ham}
\begin{align}
	H
	\,=\,&  \int d\sigma\; \big[\fpi_- \psi'_- - i\psi_- \psi'_- - \fpi_+ \psi'_+ + i\psi_+ \psi'_+ \big] + \cO(\Lambda^{-1})\ .
\end{align}
The anti-commutation relation of $\psi_\pm$ and $\fpi_\pm$ is given by
\begin{align}
	\{\psi_\pm(\sigma_1),\fpi_\pm(\sigma_2)\}=i\delta(\sigma_1-\sigma_2)\ .
\end{align}
As in the toy model, we consider a linear transformation of $\psi_+$ and $\fpi_+$ such that their Fourier modes $b_p\,,\; c_p\,,\; \bar{b}_p\,,\; \bar{c}_p$ satisfies the following anti-commutation relation:
\begin{align}
    \{b_p,b_q\}=\{\bar{b}_p,\bar{b}_q\}\,=\,\delta_{p+q,0}\;\;,\quad \{c_p,c_q\}=\{\bar{c}_p,\bar{c}_q\}\,=\,-\delta_{p+q,0}\ .
\end{align}
Such a linear transformation is found to be
\begin{align}
	u_+ \psi_+(\sigma) - {1\over 2u_+} i\fpi_+ (\sigma)\,=\,& {1\over \sqrt{2\pi}} \sum_p b_p e^{ip\sigma}\ ,\\
	{1\over 2u_+} \psi_+(\sigma) + u_+ i \fpi_+(\sigma)\,=\,& {1\over \sqrt{2\pi}} \sum_p c_p e^{ip\sigma}\ ,\\
	u_- \psi_-(\sigma) - {1\over 2u_-} i\fpi_- (\sigma)\,=\,& {1\over \sqrt{2\pi}} \sum_p \bar{b}_p e^{ip\sigma}\ ,\\
	{1\over 2u_-} \psi_-(\sigma) + u_- i \fpi_-(\sigma)\,=\,& {1\over \sqrt{2\pi}} \sum_p \bar{c}_p e^{ip\sigma}\ ,
\end{align}
where $u_\pm$ are real constants. Under this transformation, the Hamiltonian can be written as
\begin{align}
	&H
	\,=\, \kappa\sum_{p>0} p \left[ (1 - 2u_+^2)b_{-p}b_p  - ( 1 +{1\over 2 u_+^2})c_{-p}c_p  - \left( u_+^2+1 -{1\over 4u_+^2} \right)(b_{-p}c_p+  c_{-p}b_p)   \right]\cr
	&-\kappa \sum_{p>0} p \left[ (1 - 2u_-^2)\bar{b}_{-p}\bar{b}_p  - ( 1 +{1\over 2 u_-^2})\bar{c}_{-p}\bar{c}_p - \left( u_-^2+1 -{1\over 4u_-^2} \right)(\bar{b}_{-p}\bar{c}_p+  \bar{c}_{-p}\bar{b}_p)   \right] \ ,
\end{align}
where we defined $\kappa \,\equiv\, {16u^4\over (1+4u^4)^2} $. We determine the parameter $u_\pm$ by demanding that the cross term such as $b_{-p}c_p+  c_{-p}b_p$ vanish to make Hamiltonian $\cJ$-Hermitian. We obtain 
\begin{align}
    u^2_\pm\,=\,{\sqrt{2}-1\over 2} \ ,\label{eq: upm value nsrlike}
\end{align}
and the Hamiltonian becomes
\begin{align}
    H\,=\, \sum_{p>0} p \left[ (2-\sqrt{2})b_{-p}b_p  - (2+\sqrt{2})c_{-p}c_p  - (2-\sqrt{2})\bar{b}_{-p}\bar{b}_p  + (2+\sqrt{2})\bar{c}_{-p}\bar{c}_p  \right] \ .
\end{align}
Since the Hamiltonian is $\cJ$-Hermitian, the time evolution is unitary with $\cJ$-norm. 

One can also express the momentum operator $P$ obtained by the Noether procedure in terms of the above oscillators.  
%
%
However, at the value of $u_\pm$ in Eq.~\eqref{eq: upm value nsrlike}, the cross term does not vanish, which implies that the momentum operator is not $\cJ$-Hermitian although it is Hermitian in the usual sense. Or, one can find another value $\tilde{u}_\pm$ such that the cross term in the momentum operator vanishes. While the momentum operator becomes $\widetilde{\cJ}$-Hermitian\footnote{For each value of $u$, $\cJ$ operator is different. Hence, $\cJ$ inner product and $\cJ$-Hermiticity depend on the value of $u$.}, the Hamiltonian is not $\widetilde{\cJ}$-Hermitian anymore. 

The second example is the \ttb deformation of free fermion by the symmetric energy-momentum tensor discussed at the beginning of Section~\ref{sec: cov T example }. Recall that the Lagrangian is given by
\begin{align}
	\cL\,=\,& {i\over 2}\psi_+  \dot{\psi}_+ + {i\over 2} \psi_- \dot{\psi}_- - {i\over 2}\psi_+  \psi'_+ + {i\over 2} \psi_- \psi'_-  + {3\lambda\over 8}  \left( - \psi_+  \psi'_+ \psi_-\dot{\psi}_-  + \psi_+  \dot{\psi}_+ \psi_-\psi'_- \right)\cr
	& - {\lambda\over 8} \psi_+ \dot{\psi}_+\psi_-  \dot{\psi}_-  +  {\lambda\over 8} \psi_+ \psi'_+\psi_-  \psi'_-\ , \label{eq: lagrangian sym tt fermion 2}
\end{align}
and this gives the relation for the conjugate momentum $\fpi_\pm$ as follows, 
\begin{align}
	-{\lambda \over 8}\psi_+ \psi_-( \dot{\psi}_- - 3 \psi'_-) \,\equiv\,& \fpi_+ -{i\over 2}\psi_+ \ , \label{eq: mom fermion cov t 1}\\
	{\lambda \over 8}\psi_+\psi_- (\dot{\psi}_+ +3 \psi'_+)\,\equiv\,& \fpi_-  -{i\over 2}\psi_- \ . \label{eq: mom fermion cov t 2}
\end{align}
%
%
%
%
%
%
The Hamiltonian can be written as
\begin{align}
	&H\,=\, \int dx\; \bigg[  {i\over 2} \psi_+ \psi'_+  - {i\over 2} \psi_- \psi'_- - {\lambda\over 8} \psi_+ \dot{\psi}_+\psi_-  \dot{\psi}_- - {\lambda \over 8} \psi_+ \psi'_+ \psi_- \psi'_- \bigg] \ .
\end{align}
In large $\lambda$ limit (with $\psi_\pm,\fpi_\pm\sim \cO(\lambda^0)$, one can invert Eqs.~\eqref{eq: mom fermion cov t 1} and \eqref{eq: mom fermion cov t 2} perturbatively to have
\begin{align}
	\dot{\psi}_+\,=\,-3\psi'_+ + {1\over \lambda} \eta_+ \;\;,\qquad \dot{\psi}_-\,=\,3\psi'_-+ {1\over \lambda} \eta_- \ ,
\end{align}
where the ${1\over \lambda}$ correction $\eta_\pm$ satisfies
\begin{align}
	\psi_+\psi_-\eta_+\,=\,8\left[ \fpi_- -{i\over 2} \psi_- \right]\;\;,\qquad \psi_+\psi_-\eta_- \,=\,-8\left[ \fpi_+ -{i\over 2} \psi_+ \right]\ .
\end{align}
Therefore, the ${1\over \lambda}$ expansion of the Hamiltonian is 
\begin{align}
	H
	\,=\,&\int dx\; \bigg[ \lambda \psi_+\psi'_+\psi_-\psi'_- + 3(\fpi_--{i\over 2} \psi_-)\psi'_- - 3 (\fpi_+ -{i\over 2}\psi_+) \psi'_+  + {i\over 2}\psi_+  \psi'_+ - {i\over 2} \psi_- \psi'_- \bigg] \cr
	&+ \cO(\lambda^{-1})\ .
\end{align}
When the canonical variables are of order $\cO(\lambda^0)$, the energy would be of order $\cO(\lambda)$ in general, which is seemingly not consistent with the large $\lambda$ limit of the \ttb deformed spectrum~\eqref{eq: deformed energy value} with the unperturbed energy and momentum of order $\cO(\lambda^0)$. First, due to the existence of the additional degrees of freedom with the negative norm, it is not guaranteed that the deformation of the energy~\eqref{eq: deformed energy value} still holds. Furthermore, it is not clear that keeping the canonical variable $\psi_\pm$ and $\fpi_\pm$ being of order $\cO(\lambda^0)$ is equivalent to keeping the unperturbed energy and momentum being of order $\cO(\lambda^0)$.

Since the leading term is quartic, it is still difficult to analyze the spectrum. Hence, we will consider a subsector by taking a constraint,
\begin{align}
    \psi'_-\,=\,0\ .
\end{align}
For consistent truncation, we need to check the secondary constraint:
\begin{align}
	[H,\psi'_-]\,=\, -3i\psi''_-\ .
\end{align}
Therefore, $\psi'_-\,=\,0$ is the first-class constraint and (by taking a gauge condition $\fpi_-\,=\,0$), we have
\begin{align}
    H\,=\, \int dx\; \big[ - 3 \fpi_+\psi'_+  + 2i \psi_+  \psi'_+ \big]+ \cO(\lambda^{-1})\ .
\end{align}
%
%
%
As in the NSR-like model, the Hamiltonian can be expressed in terms of fermi oscillators $b_p\,,\,c_p$ for each $u$: 
\begin{align}
	H
	\,=\,&{16u^4\over (1+4u^4)^2} \sum_{p>0} p \left[ (3 -4u^2)b_{-p}b_p  -(3+{1\over u^2})c_{-p}c_p \right.\cr
	&\hspace{50mm}\left.- \left( 3u^2+2 -{3\over 4u^2} \right)(b_{-p}c_p+  c_{-p}b_p)   \right]+E_0\ .
\end{align}
where $E_0$ is a constant from the operator ordering. Demanding that the cross term $b_{-p}c_p+  c_{-p}b_p$ vanish to make Hamiltonian $\cJ$-Hermitian, we get
\begin{align}
	u\,=\,\sqrt{\sqrt{13}-2 \over 6}\;\;:\;\; \mbox{real}\ .\label{eq: app u value}
\end{align}
And the Hamiltonian becomes
\begin{align}
	H\,=\,&{9\over 13} \sum_{p>0}{p\over 3}\left[ (13-2\sqrt{13})b_{-p}b_p -(13+2\sqrt{13})c_{-p}c_p \right]\  ,
\end{align}
which is $\cJ$-Hermitian explicitly. Hence, the time evolution recovers unitarity. However, at this value of $u$, the momentum has the cross term, and it is not $\cJ$-Hermitian. As in the NSR-like case, one cannot make both Hamiltonian and momentum $\cJ$-Hermitian at the same time. 
Therefore, the eigenstates of the momentum operator lose the properties of the eigenstate of the Hermitian operators. (\eg orthogonality and unitarity of $e^{-iP x}$ \etc). Without the $\cJ$-Hermiticity, one cannot use the orthogonality of energy and momentum eigenstates, an important step in the factorization formula of the \ttb operator~\cite{Zamolodchikov:2004ce}.

In these two examples, we find a common pattern in the large $\lambda$ limit. The naive inner product yields negative norm states, yet there does exist a modified pairing such that the norm is positive definite and at the same time the Hamiltonian is Hermitian. This guarantees that the evolution operator is unitary. On the other hand, the spatial momentum is not $\cJ$-Hermitian, suggesting that there are certainly not ordinary quantum field theories. We should also emphasize that the exercise was done in the large $\lambda$ limit where the Hamiltonian is greatly simplified; it is not clear whether both the positive-definiteness of $\cJ$-norm and the unitary evolution extends to finite nonzero $\lambda$.

Either way, it is clear that the degrees of freedom are doubled compared to $\lambda=0$ limit, so one cannot consider this one-parameter family of theories as a ``deformation" $\lambda=0$ theory. Rather, a more sensible interpretation would be that in the infrared limit $\lambda\rightarrow 0$ of this much bigger theory, one can find a very small subset of the Hilbert space, separated from the rest by a divergent energy gap, which happens to match the sensible $\lambda=0$ theory. 

In this final part of the note, we gave two examples of ``deformation" that incur doubling of the degrees of freedom with a potential unitarity issue. Interestingly, for those \ttb deformation based on Noether energy-momentum, we are yet to find a problem of this kind. It is unclear whether the latter is a general feature of the Noether energy-momentum or simply due to the small fermion content of models we relied on. 

\acknowledgments

We would like to thank Ioannis Papadimitriou, Byungmin Kang, Ryo Suzuki, Jong-Hyun Baek and Changrim Ahn for discussions and Andre LeClair for private communications. 
KL and JY thank Asia Pacific Center for Theoretical Physics (APCTP) for the hospitality, and
JY is grateful to the Galileo Galilei Institute for Theoretical Physics (GGI) for the hospitality.
PY and JY were supported in part by KIAS Individual Grants (PG005714 and PG070101) at Korea Institute for Advanced Study, 
and JY also by the National Research Foundation of Korea~(NRF) grant funded by the Korea government~(MSIT) (2019R1F1A1045971). JY is supported by an appointment to the JRG Program at the APCTP through the Science and Technology Promotion Fund and Lottery Fund of the Korean Government. This is also supported by the Korean Local Governments - Gyeongsangbuk-do Province and Pohang City. KL was supported by Basic Science Research Program through the National Research Foundation of Korea(NRF) funded by the Ministry of Science, ICT \& Future Planning (NRF-2021R1A2C1006791) and (NRF-2020R1I1A2054376).

\appendix

\section{Convention}
\label{app: convention}

\subsection*{GS-like and NSR-like Models}

\begin{itemize}
    \item Gamma matrices
    \begin{equation}
    	\Gamma^0\,=\,i\sigma_2\;,\qquad \Gamma^1\,=\,\sigma_1\;,\qquad\Gamma^2=\sigma_3\ .
    \end{equation}
    \item $\Psi$ : 3D space-time Majorana spinor. $\bar{\Psi}\,=\,\Psi^T C$ where $C\,=\,\Gamma^0\,=\,i\sigma_2$.
    \item $(\tau,\sigma)$ : (dimensionless) worldsheet coordinates. $\sigma \;\sim\; \sigma+2\pi$
    \item $\Lambda\,=\,{ 4\pi^2\lambda\over L^2}$ : (dimensionless) deformation parameter.
    \item $\momp_\mu\,\equiv \, {\delta S \over \delta \dot{X}^\mu}$ :  the conjugate momentum of $X^\mu$.
    \item $\pz_\mu$ : zero-mode of the conjugate momentum $\momp_\mu$, $\pz_\mu\,\equiv \, {1\over 2\pi} \oint d\sigma\; \momp_\mu\ .$
    \item $\bP_\mu$ : momentum operator generating the translation along $X^\mu$, 
    
    $\bP_\mu\,\equiv\, \oint d\sigma\;  \momp_\mu\,=\, 2\pi \pz_\mu\ .$
    \item $\mpi_\mu\,\equiv\,\frac{\delta \mathcal S}{\delta\bPi^\mu_0}$ : Auxiliary momentum-like variable.
    \item $\bW^\mu$ : winding number operator of $X^\mu$, $\bW^\mu\,\equiv \, \oint d\sigma\; \partial_\sigma X^\mu\ .$
    \item $\sigma^\pm$ : Worldsheet light-cone coordinates.
    \begin{align}
        	\sigma^\pm\,\equiv\, \tau \pm \sigma\ .
    \end{align}
    \begin{align}
	\partial_\+\,=\,{1\over 2}(\partial_0 + \partial_1)\;\;,\qquad \partial_= \,=\,{1\over 2}(\partial_0 - \partial_1)\ .
    \end{align}
\end{itemize}

\subsection*{Target Light-cone Coordinates in Section~\ref{sec: string theory and ttbar deformed hamiltonian}}

\begin{itemize}
    \item Target light-cone coordinates
    \begin{equation}
        X^+\,\equiv\,{1\over 2} X^1+ {1\over 2} X^0\;\;,\qquad X^-\,\equiv\,X^1-X^0 \ .
    \end{equation}

    \item Target space metric with light-cone
    \begin{align}
        ds^2\,=\,2dX^+ dX^-+(dX^2)^2\ .
    \end{align}
    \item Gamma matrices in the light-cone
    \begin{equation}
        \Gamma^+\,=\,{1\over 2} \Gamma^1+{1\over 2} \Gamma^0 \,=\,\begin{pmatrix}
        0 & 1\\
        0 & 0\\
        \end{pmatrix}\;\;,\qquad \Gamma^-\,=\,\Gamma^1-\Gamma^0\,=\,\begin{pmatrix}
        0 & 0\\
        2 & 0\\
        \end{pmatrix} \ .
    \end{equation}
\end{itemize}

\subsection*{Target Shifted Light-cone Coordinates in Section~\ref{sec: ttbar deformation and gs shifted background}}

\begin{itemize}
    \item Target Shifted light-cone coordinates.
    \begin{equation}
        X^+\,\equiv\,\bigg({1\over 2}-\Lambda\bigg) X^1+ \bigg({1\over 2}+\Lambda\bigg) X^0\;\;,\qquad X^-\,\equiv\,X^1-X^0 \ .
    \end{equation}
    \item Target space flat metric in terms of shifted light-cone coordinates.
    \begin{align}
        ds^2\,=\,2dX^+ dX^-+2\Lambda(dX^-)^2+(dX^2)^2\ .
    \end{align}
    \item Gamma matrices in the shifted light-cone.
    \begin{equation}
        \Gamma^+\,=\,\left({1\over 2}-\Lambda\right) \Gamma^1+\left({1\over 2}+\Lambda\right)\Gamma^0\,=\,\begin{pmatrix}
        0 & 1\\
        -2\Lambda & 0\\
        \end{pmatrix}\;\;,\quad \Gamma^-\,=\,\Gamma^1-\Gamma^0\,=\,\begin{pmatrix}
        0 & 0\\
        2 & 0\\
        \end{pmatrix}\ .
    \end{equation}
    \item Projector
    \begin{align}
        \Upsilon^+\,\equiv\, {1\over 2}(\Gamma^1+\Gamma^0)\,=\, \begin{pmatrix}
        0 & 1 \\
        0 & 0 \\
        \end{pmatrix}\;\;,\qquad \Upsilon^-\,\equiv\, {1\over 2}(\Gamma^1-\Gamma^0)\,=\, \begin{pmatrix}
        0 & 0 \\
        1 & 0 \\
        \end{pmatrix}\ .
    \end{align}
\end{itemize}

\subsection*{\ttb Deformation}

\begin{itemize}
    \item $(t,x)$ : (dimensionful) coordinates for 2D space-time 
    \item Light-cone coordinates
    \begin{align}
        	x^\pm\,\equiv\, t\pm x\ .
    \end{align}
    \begin{align}
	\partial_\+\,=\,{1\over 2}(\partial_0 + \partial_1)\;\;,\qquad \partial_= \,=\,{1\over 2}(\partial_0 - \partial_1)\ .
    \end{align}
    \item $L$ : Circumference of the coordinates $x$
    \item $\lambda$ : \ttb deformation parameter of dimension length-squared.
    \item $\phi,\psi_\pm$ and $\spi,\fpi_\pm$ : scalar field, fermion and its conjugate momentum.
    \item $\cH, \cP$ : Hamiltonian and momentum density of \ttb deformed model
    \item $\cH_{b,0}, \cP_{b,0}, \cH_{f,0}, \cP_{f,0}$ : Hamiltonian and momentum density of bosonic and fermi part of the undeformed model, respectively.
    \item $H, P$ : Hamiltonian and momentum operator.
\end{itemize}

\section{Details of Perturbative Calculations}
\label{app: perturbation}

\subsection{\ttb Deformation of Free Scalar Field}
\label{app: perturbation scalar field}

In this appendix, we will provide the detailed calculations in Section~\ref{sec: perturbation scalar}. Let us continue with the equation~\eqref{eq: ham eq scalar sec 2}
\begin{align}
	H[A,\bar{A}]\,=\,\wH [\alpha,\bar{\alpha}]\ . \label{eq: ham eq scalar field app}
\end{align}
The Hamiltonian $H[A,\bar{A}]$ is evaluated from the Hamiltonian density
\begin{align}
	\cH\,=\,&{1\over 2 \lambda} \left[ \sqrt{1+ 4\lambda \left({1\over 2} \spi^2 + {1\over 2} \phi'^2\right) + 4\lambda^2 (\spi \phi')^2 } -1 \right]\ ,
\end{align}
in terms of the Fourier modes $A_k$ and $\bar{A}_k$ of $\phi$ and $\spi$ while $\wH [\alpha,\bar{\alpha}]$ is defined in terms of the free harmonic oscillators $\alpha$'s and $\bar{\alpha}$'s by
\begin{align}
    \wH [\alpha_k,\bar{\alpha}_k]\,=\,& {L \over 2\lambda } \left[ \sqrt{1+ {4\lambda\over  L } (H_++H_-) + { 4\lambda^2 \over L^2} ( H_+ - H_-)^2 } - 1 \right]\ ,\\
    H_+\,\equiv \,& {\pi \over L }\sum_k  \alpha_{-k}\alpha_k \;\;,\qquad H_-\, \equiv\, {\pi \over L } \sum_k  \bar{\alpha}_{-k} \bar{\alpha}_k \ .
\end{align}
which explicitly gives rise to the \ttb deformed spectrum.

We might be able to solve directly this equation~\eqref{eq: ham eq scalar field app} together with the condition for canonical transformation. However, we will develop a practical procedure to find the transformation. If two Hamiltonians are equal under the transformation \ie $H[A,\bar{A}]\,=\,\wH [\alpha,\bar{\alpha}]$, the Poisson brackets of any fields with them should be identical,
\begin{equation}
	\{ A_k, H \}_\pb\,=\, \{ A_K, \wH\}_\pb\ .\label{eq: pb of a and h}
\end{equation}
Noting that the transformation from $A_k$ to $\alpha$ is canonical, we will evaluate the left-hand side of Eq.~\eqref{eq: pb of a and h} with respect to $A_k$ and $\bar{A}_k$ and compute the Poisson bracket on the right-hand side of Eq.~\eqref{eq: pb of a and h} with respect to $\alpha_k$ and $\bar{\alpha}_k$: 
\begin{equation}
	\{ A_k, H \}_{\pb, A,\bar{A}}\,=\, \{ A_k, \wH\}_{\pb,\alpha,\bar{\alpha}}\ .\label{eq: pb of a and h 2 app}
\end{equation}
Using the expansion of $A_k$ and $\bar{A}_k$
\begin{align}
	A_k\,=\,\alpha_k + {\lambda\over L^2} A^{(1)}_k +\cdots\;\;,\qquad \bar{A}_k\,=\,\bar{\alpha}_k + {\lambda\over L^2} \bar{A}^{(1)}_k +\cdots\ ,
\end{align}
the left-hand side of Eq.~\eqref{eq: pb of a and h 2 app} is found to be
\begin{align}
	\{ A_k, H \}_{\pb, A,\bar{A}}\,=\, &-{2\pi i k\over L} A_k + {4 \pi \lambda \over L^2}  {2\pi i k \over L} \sum_{r,s}A_{k-r-s} \bar{A}_{-r} \bar{A}_{-s} +\cO(\lambda^2)\\
	\,=\,&-{2\pi i k\over L} \alpha_k - {2\pi i k\lambda \over L^3} A_k^{(1)} + {4 \pi \lambda \over L^2}  {2\pi i k \over L}\sum_{r,s} \alpha_{k-r-s} \bar{\alpha}_{-r} \bar{\alpha}_{-s} +\cO(\lambda^2) \ .
\end{align}
On the other hand, the right-hand side of Eq.~\eqref{eq: pb of a and h 2 app} becomes
\begin{align}
	&\left\{ \alpha_k+ {\lambda \over L^2} A_k^{(1)}+ \cO(\lambda^2), \wH \right\}_{\pb,\alpha,\bar{\alpha}}\cr
	\,=\,& -{2\pi i k \over L}\alpha_k  + {8\pi i k \lambda\over L^2}\alpha_k H_- - {\lambda\over L^2}\sum_u\left( {2\pi i u\over L}  \alpha_u {\delta A_k^{(1)} \over \delta \alpha_u}+  {2\pi i u\over L}  \bar{\alpha}_u {\delta A_k^{(1)} \over \delta \bar{\alpha}_u} \right)+\cO(\lambda^2)\ ,
\end{align}
where we used the Poisson bracket
\begin{align}
	 \{ F, G\}_{\pb,\alpha,\bar{\alpha}}\,=\,  \sum_u \left[-i u {\delta F \over \delta \alpha_u}{\delta G \over \delta \alpha_{-u}} - i u {\delta F \over \delta \bar{\alpha}_u}{\delta G \over \delta \bar{\alpha}_{-u}}\right]\ .
\end{align}
At order $\cO(\lambda)$ we have a functional differential equation for $A^{(1)}_k$ (and similarly for $\bar{A}_k$)
%
%
%
\begin{align}
	\sum_u\left(   u  \alpha_u {\delta A_k^{(1)} \over \delta \alpha_u}+    u  \bar{\alpha}_u {\delta A_k^{(1)} \over \delta \bar{\alpha}_u} \right)  -   k  A_k^{(1)}  \,=\,&-  4\pi  k \sum_{\substack{r,s \\ r+s\ne 0}} \alpha_{k-r-s} \bar{\alpha}_{-r} \bar{\alpha}_{-s} \ ,\\
	\sum_u\left(   u  \alpha_u {\delta \bar{A}_k^{(1)} \over \delta \alpha_u}+    u  \bar{\alpha}_u {\delta \bar{A}_k^{(1)} \over \delta \bar{\alpha}_u} \right)  -   k  A_k^{(1)}  \,=\,&-  4\pi  k \sum_{\substack{r,s \\ r+s\ne 0}} \alpha_{-r} \alpha_{-s} \bar{\alpha}_{k-r-s}  \ .
\end{align}
These are inhomogeneous first-order differential equations, and a solution is found to be
\begin{align}
	A^{(1)}_k\,=\,&  2 \pi  \sum_{\substack{r,s\\ r+s\ne 0}}  {k\over r+s} \alpha_{k-r-s}\bar{\alpha}_{-r} \bar{\alpha}_{-s} + \sum_r f_{k,r}  \alpha_{k}\bar{\alpha}_{-r} \bar{\alpha}_{r} \ , \\
	\bar{A}^{(1)}_k\,=\,&  2 \pi \sum_{\substack{r,s\\ r+s\ne 0}} {k\over r+s} \alpha_{-r} \alpha_{-s} \bar{\alpha}_{k-r-s} +  \sum_{r} \bar{f}_{k,r}  \alpha_{-r} \alpha_{r} \bar{\alpha}_{k} \ ,
\end{align}
where the second terms are solutions of the homogeneous parts. By construction, we have
\begin{equation}
	f_{k,r}\,=\, f_{k,-r}\quad,\quad \bar{f}_{k,r}\,=\, \bar{f}_{k,-r}\ .
\end{equation}
The Poisson relations of $A_k$ and $\bar{A}_k$ 
\begin{align}
	[A_k,A_q]\,=\,& k \delta_{k+q,0}\ ,\label{eq: com relation of a1 app}\\
	[\bar{A}_k,\bar{A}_q]\,=\,&k\delta_{k+q,0}\ ,\label{eq: com relation of a2 app}\\
	[A_k,\bar{A}_q]\,=\,&0\ ,\label{eq: com relation of a3 app}
\end{align}
leads to constraints on $f_{k,r}$ and $\bar{f}_{k,r}$: 
\begin{align}
	i\{A^{(1)}_k, \alpha_q\}_\pb+ i\{\alpha_k , A^{(1)}_q\}_\pb
	\,=\,& \sum_r  k (f_{k,r} + f_{-k,r} ) \delta_{k+q,0}\bar{\alpha}_{-r} \bar{\alpha}_{r}=0\ ,\\
	\{\bar{A}^{(1)}_k, \bar{\alpha}_q\}_\pb+ \{\bar{\alpha}_k , \bar{A}^{(1)}_q\}_\pb 
	\,=\,&  \sum_r  k (\bar{f}_{k,r} + \bar{f}_{-k,r} ) \delta_{k+q,0} \alpha_{-r}  \alpha_{r}=0\ ,
\end{align}
and
\begin{align}
	i\{A^{(1)}_k, \bar{\alpha}_q\}_\pb+i\{\alpha_k, \bar{A}^{(1)}_q\}_\pb
	\,=\,& -2( qf_{k,q}- k \bar{f}_{q,k} )\alpha_k \bar{\alpha}_q \,=\,0 \ .
\end{align}
In sum, we have
\begin{align}
	&f_{k,r}\,=\,f_{k,-r}\ ,\\
	&\bar{f}_{k,r}\,=\,\bar{f}_{k,-r}\ ,\\
	&f_{k,r}\,=\,- f_{-k,r}\ ,\\
	&\bar{f}_{k,r}\,=\,- \bar{f}_{-k,r}\ ,\\
	&qf_{k,q}\,=\,k\bar{f}_{q,k}\ .
\end{align}
Because $f_{k,q}$ and $\bar{f}_{k,q}$ are dimensionless, one can take the following ansatz:
\begin{equation}
	f_{k,q}\,=\,f\bigg({k\over q}\bigg)\ .
\end{equation}
But, the above conditions lead to
\begin{equation}
	f\bigg({k\over q}\bigg)\,=\, f\bigg({k\over -q}\bigg)\,=\,-f\bigg({-k\over q}\bigg)\ ,
\end{equation}
and we can conclude that
\begin{equation}
	f\bigg({k\over q}\bigg)\,=\,0 \ .
\end{equation}
Hence, the first-order solution is found to be
\begin{align}
	A^{(1)}_k\,=\,&  2 \pi  \sum_{\substack{r,s\\ r+s\ne 0}}  {k\over r+s} \alpha_{k-r-s}\bar{\alpha}_{-r} \bar{\alpha}_{-s} \ , \label{eq: first order sol scalar 1 app} \\
	\bar{A}^{(1)}_k\,=\,&  2 \pi \sum_{\substack{r,s\\ r+s\ne 0}} {k\over r+s} \alpha_{-r} \alpha_{-s} \bar{\alpha}_{k-r-s} \ . \label{eq: first order sol scalar 2 app}
\end{align}

The small $\lambda$ expansion of the Hamiltonian $H[A,\bar{A}]$ is given by
\begin{align}
	H
	\,=\,&{2\pi \over 2L } \left( A_{-k}A_k + \bar{A}_{-k} \bar{A}_k \right) - {4\pi^2 \lambda\over L^3 } \sum_{k,q,r} A_k \bar{A}_{-q} A_{-k-q-r} \bar{A}_{-r} + \cO(\lambda^2)\ .
\end{align}
Using the solution, one can express the Hamiltonian in terms of the free oscillators $\alpha$'s, and one can expand it with respect to $\lambda$ again,
\begin{align}
    	H\,=\,&H^{(0)}[\alpha,\bar{\alpha}] + {\lambda \over L^2} H^{(1)}[\alpha,\bar{\alpha}] + \cO(\lambda^2)\ .
\end{align}
Then, the leading and the first-order contribution are 
\begin{align}
	H^{(0)}\,=\,&{2\pi \over 2L } \left( \alpha_{-k}\alpha_k + \bar{\alpha}_{-k} \bar{\alpha}_k \right)\equiv H_+ +H_-\ ,\\
	H^{(1)}\,=\,& {2\pi \over L} \sum_k \left( \alpha_{-k} A^{(1)}_k+ \bar{\alpha}_{-k} \bar{A}^{(1)}_k  \right) - {4\pi^2\over L}  \sum_{k,q,r} \alpha_k \bar{\alpha}_{-q} \alpha_{-k-q-r} \bar{\alpha}_{-r}\cr
	\,=\,&- {4\pi^2 \over L} \sum_{ k,q } \alpha_{-k} \alpha_{k} \bar{\alpha}_{-q}\bar{\alpha}_{q} \,=\,-4L H_+H_-\ .
\end{align}
Therefore, we have
\begin{align}
	H\,=\, H_+ + H_- -{4\lambda \over L} H_+ H_- + \cO(\lambda^2)\ .
\end{align}
One can repeat the same calculation for momentum $P$. The momentum density of the \ttb deformed scalar field is given by
\begin{align}
	{T^0}_1\,=\, {\delta \cL \over  \delta \partial_0 \phi} \partial_1\phi = {\phi' \dot{\phi}\over \sqrt{1+\lambda(-\dot{\phi}^2+\phi'^2)}}\,=\, \spi \phi'\ ,
\end{align}
and the momentum operator is 
\begin{align}
	P\,=\,\int dx \; \spi \phi'\,=\, - {\pi \over L}\sum_k [A_{-k}A_{k}- \bar{A}_{-k}\bar{A}_k]\ .
\end{align}
The leading momentum operator is the same as that of the free scalar field,
\begin{align}
	P^{(0)}\,=\,- {\pi \over L} \sum_k [\alpha_{-k}\alpha_{k}- \bar{\alpha}_{-k}\bar{\alpha}_k]\,=\,-H_+ + H_-\ .
\end{align}
One can confirm that the $\cO(\lambda)$ correction of momentum operator vanishes,
\begin{align}
	P^{(1)}\,=\,- {2\pi  \over L} \sum_k [\alpha_{-k}A^{(1)}_{k}- \bar{\alpha}_{-k}\bar{A}^{(1)}_k]\,=\,0 \ .
\end{align}

One can repeat the same procedure to obtain the transformation at order $\cO(\lambda^2)$ at the classical level. Expanding $A_k$ with respect to ${\lambda\over L^2}$
\begin{align}
	A_k\,=\,&\alpha_k+ {\lambda\over L^2} A^{(1)}_k[\alpha,\bar{\alpha}]+ {\lambda^2 \over L^4} A^{(2)}_k[\alpha,\bar{\alpha}]+\cO(\lambda^3)\ ,
\end{align}
we get inhomogeneous differential equations for $A_k^{(2)}$ from the equation~\eqref{eq: pb of a and h} at order $\cO(\lambda^2)$:
\begin{align}
	& \sum_u\left(   u \alpha_u {\delta A_k^{(2)} \over \delta \alpha_u}+   u   \bar{\alpha}_u {\delta A_k^{(2)} \over \delta \bar{\alpha}_u} \right) -  k  A_k^{(2)} \cr
	=\,& - 4\pi^2  k \sum_{\substack{r,s,u,v\\u+v\ne 0, r+s\ne 0}} {(r+s+u+v)(k-r-s-u-v)\over (u+v)(r+s)} \alpha_{k-r-s -u -v} \bar{\alpha}_{-u} \bar{\alpha}_{-v} \bar{\alpha}_{-r}\bar{\alpha}_{-s} \cr
	&+ 8 \pi  k  \sum_{\substack{u,v\\u+v\ne 0}} \alpha_{k-u-v} \bar{ \alpha}_{-u}\bar{ \alpha}_{-v}  L(H_++H_-) - 8\pi^2  k \sum_{\substack{r,s,u,v\\r+s\ne 0}} {u+v \over r+s}  \alpha_{k-r-s-u-v} \alpha_{r}\alpha_{s}\bar{\alpha}_{-u}  \bar{\alpha}_{-v}\ .\cr
\end{align}
A solution for this equation can be written as
\begin{align}
	A_k^{(2)}\,=\,& 2\pi^2  k \sum_{\substack{r,s,u,v\\u+v\ne 0,r+s\ne 0}} {k-r-s-u-v\over (u+v)(r+s)}\alpha_{k-r-s -u -v} \bar{\alpha}_{-u} \bar{\alpha}_{-v} \bar{\alpha}_{-r}\bar{\alpha}_{-s} \cr
	&- 4 \pi  k  \sum_{\substack{u,v\\u+v\ne 0}} {1\over u+v} \alpha_{k-u-v} \bar{ \alpha}_{-u}\bar{ \alpha}_{-v} L(H_++H_-)  + 4\pi^2  k \sum_{\substack{r,s,u,v\\r+s\ne 0}} {1 \over r+s}  \alpha_{k-r-s-u-v} \alpha_{r}\alpha_{s}\bar{\alpha}_{-u}  \bar{\alpha}_{-v}\cr
	& + \alpha_k \sum_{r,u,v} c_{k;r,u,v} \alpha_{-r} \alpha_{r-u-v} \bar{\alpha}_{-u} \bar{\alpha}_{-v} + \alpha_k \sum_{r,u,v} d_{k;r,u,v} \bar{\alpha}_{-r-u-v} \bar{\alpha}_{r} \bar{\alpha}_{u} \bar{\alpha}_{v}\ ,
\end{align}
and similar for $\bar{A}_k^{(2)}$. Furthermore, for the canonical transformation, we demand
\begin{align}
	&i\{A^{(1)}_k, A^{(1)}_q\}_\pb+i\{\alpha_k, A^{(2)}_q\}_\pb+i\{ A^{(2)}_k, \alpha_q\}_\pb\,=\,0\ ,\\
	&i\{\bar{A}^{(1)}_k, \bar{A}^{(1)}_q\}_\pb+i\{\bar{\alpha}_k, \bar{A}^{(2)}_q\}_\pb+i\{ \bar{A}^{(2)}_k, \bar{\alpha}_q\}_\pb\,=\,0\ ,\\
	&i\{A^{(1)}_k, \bar{A}^{(1)}_q\}_\pb+i\{\alpha_k, \bar{A}^{(2)}_q\}_\pb+i\{ A^{(2)}_k, \bar{\alpha}_q\}_\pb\,=\,0\ .
\end{align}
These equations are satisfied by
\begin{align}
	c_{k,r,u,v}\,=\,d_{k,r,u,v}\,=\,\bar{c}_{k,r,u,v}\,=\,\bar{d}_{k,r,u,v}\,=\,0\ .
\end{align}
In sum, we find that the canonical transformation from $A_k, \bar{A}_k$ to $\alpha_k, \bar{\alpha}_k$ is
\begin{align}
	A_k\,=\,&\alpha_k+ {\lambda\over L^2} A^{(1)}_k[\alpha,\bar{\alpha}]+ {\lambda^2 \over L^4} A^{(2)}_k[\alpha,\bar{\alpha}]+\cO(\lambda^3)\ ,\\
	\bar{A}_k\,=\,&\bar{\alpha}_k+ {\lambda\over L^2} \bar{A}^{(1)}_k[\alpha,\bar{\alpha}]+ {\lambda^2 \over L^4} \bar{A}^{(2)}_k[\alpha,\bar{\alpha}]+\cO(\lambda^3)\ ,
\end{align}
where
%
%
%
\begin{align}
	&A_k^{(2)}\,=\, 2\pi^2  k \sum_{\substack{r,s,u,v\\u+v\ne 0,r+s\ne 0}} {k-r-s-u-v\over (u+v)(r+s)}\alpha_{k-r-s -u -v} \bar{\alpha}_{-u} \bar{\alpha}_{-v} \bar{\alpha}_{-r}\bar{\alpha}_{-s} \cr
	&- 4 \pi  k  \sum_{\substack{u,v\\u+v\ne 0}} {1\over u+v} \alpha_{k-u-v} \bar{ \alpha}_{-u}\bar{ \alpha}_{-v} L(H_++H_-)  + 4\pi^2  k \sum_{\substack{r,s,u,v\\r+s\ne 0}} {1 \over r+s}  \alpha_{k-r-s-u-v} \alpha_{r}\alpha_{s}\bar{\alpha}_{-u}  \bar{\alpha}_{-v}\ ,\cr\\
	&\bar{A}_k^{(2)}\,=\, 2\pi^2  k \sum_{\substack{r,s,u,v\\u+v\ne 0,r+s\ne 0}} {k-r-s-u-v\over (u+v)(r+s)}\bar{\alpha}_{k-r-s -u -v}  \alpha_{-u}  \alpha_{-v} \alpha_{-r} \alpha_{-s} \cr
	&- 4 \pi  k  \sum_{\substack{u,v\\u+v\ne 0}} {1\over u+v} \bar{\alpha}_{k-u-v}  \alpha_{-u}  \alpha_{-v} L(H_+ + H_-)  + 4\pi^2  k \sum_{\substack{r,s,u,v\\r+s\ne 0}} {1 \over r+s}  \bar{\alpha}_{k-r-s-u-v} \bar{\alpha}_{r}\bar{\alpha}_{s}  \alpha_{-u}   \alpha_{-v}\ .
\end{align}
Under the transformation, the Hamiltonian and the momentum become
\begin{align}
	H\,=\,&H_++ H_- -4{\lambda\over L} H_+ H_- + {8\lambda^2\over L^2} H_+H_-(H_++H_-)+\cO(\lambda^3)\ ,\cr
	\,=\,&{1\over 2\lambda} \left[\sqrt{1+ 4\lambda(H_++ H_-)^2 + 4\lambda^2(H_+ - H_-)^2 }-1\right]+ \cO(\lambda^3)\ ,\\
	P\,=\,&-H_+ + H_- + \cO(\lambda^3)\ .
\end{align}
%

\subsection{\ttb Deformation of Free Fermion}
\label{app: perturbation fermion}

This appendix will present the details of the perturbative calculations for the fermion in Section~\ref{app: perturbation fermion}. We proceed with the equation~\eqref{eq: fermion ttb hamiltonian equation sec 2}
\begin{align}
    H[\psi_+,\psi_-]\,=\, \widetilde{H}[b,\bar{b}]\ . \label{eq: ham eq fermion app}
\end{align}
The Hamiltonian $H[\psi_+,\psi_-]$ is obtained from the Hamiltonian density~\eqref{eq: ham density of free Majorana fermion} of the \ttb deformed fermion, and it can be expressed in terms of the Fourier modes $\psi_{\pm, k}$ of the fermion
\begin{align}
    H[\psi_+,\psi_-]\,=\,{\pi \over L}\sum_{k} (-k\psi_{+,-k}\psi_{+,k} + k\psi_{-,-k} \psi_{-,k})\ .
\end{align}
On the other hand, $\widetilde{H}[b,\bar{b}]$ on the right-hand side is defined in terms of the free fermi oscillators $b_k$ and $\bar{b}_k$ by
\begin{align}
	\wH [b,\bar{b}]\,=\, &{L \over 2\lambda } \left[ \sqrt{1+ {4\lambda\over  L } (H_++H_-) + { 4\lambda^2 \over L^2} ( H_+ - H_-)^2 } - 1 \right]\ ,\\
	H_+\,\equiv\,& -{\pi \over L} \sum_k k b_{-k} b_k\;\;,\qquad H_- \,\equiv\,  {\pi \over L} \sum_k k \bar{b}_{-k} \bar{b}_k\ .
\end{align}
And \ttb deformed spectrum immediately follows from $\widetilde{H}[b,\bar{b}]$.

Although the idea here differs a little from the bosonic case, since there the transformation preserves the canonical commutator, a key identity that helps us to solve the problem is again, 
\begin{equation}
	\{ \psi_{+,k}, H \}_{\dirac}= \{ \psi_{+,k}, \wH\}_{\dirac} \ .\label{eq: pb of a and h 1 fermi app}
\end{equation}
In addition, we demand that the algebra of $\psi_{+,k}\,,\; \psi_{-,k}$ 
\begin{align}
	i\{\psi_{+, k},\psi_{+,q}\}_\dirac\,=\,& \delta_{k+q,0} + { \lambda\over L^2} S_{-,k+q} +  {2\lambda^2 \over L^4}(S_+S_-)_{k+q}\ , \label{eq: dirac bracket mode 1 app}\\
	i\{\psi_{-, k},\psi_{-,q}\}_\dirac\,=\,& \delta_{k+q,0} - {  \lambda\over L^2} S_{+,k+q} + {2 \lambda^2 \over L^4}(S_+S_-)_{k+q} \ , \label{eq: dirac bracket mode 2 app}\\
	i\{\psi_{+, k},\psi_{-,q}\}_\dirac\,=\,&-{  \lambda \over L^2} K_{k+q} \ .\label{eq: dirac bracket mode 3 app}
\end{align}
is realized by $b_k\,,\;\bar{b}_k$ which obey the following Dirac brackets:
\begin{align}
	i\{b_k,b_q\}_\dirac\,=\,\delta_{k+q,0}\;\;,\;\; i \{\bar{b}_k,\bar{b}_q\}_\dirac \, =\,  \delta_{k+q,0}\;\;,\;\; i\{b_k,\bar{b}_q\}_\dirac\,=\,0\ .\label{eq: dirac bracket of b app}
\end{align}
Hence, we may evaluate the Dirac bracket on the left-hand side of Eq.~\eqref{eq: pb of a and h 1 fermi app} with respect to $\psi_{\pm,k}$ whereas we may calculate the Dirac bracket on the right-hand side of Eq.~\eqref{eq: pb of a and h 1 fermi app} with respect to $b_k$ and $\bar{b}_k$, 
\begin{equation}
	\{ \psi_{+,k}, H \}_{\dirac, \psi_+, \psi_-}= \{ \psi_{+,k}, \wH\}_{\dirac,b,\bar{b}} \ .\label{eq: pb of a and h 2 fermi app}
\end{equation}
To avoid the ordering ambiguity, we will find the map classically. The left-hand side of Eq.~\eqref{eq: pb of a and h 2 fermi app} is found to be
\begin{align}
	\{ \psi_{+,k} ,H\}_\dirac\,=\,&{2\pi i k \over L } \psi_{+,k} - {8\pi^2 i \lambda \over L^3}\sum_{r,s} (k-r-s)s \psi_{+,k-r-s} \psi_{-,r}\psi_{-,s}\ , \label{eq: eq lhs fermion 1 app} \\
	\{ \psi_{-,k} ,H\}_\dirac\,=\,&-{2\pi i k \over L } \psi_{-,k} - {8\pi^2 i \lambda \over L^3}\sum_{r,s} (k-r-s)s \psi_{+,r} \psi_{+,s}\psi_{-,k-r-s}\ . \label{eq: eq lhs fermion 2 app} 
\end{align}
Using the expansion of $\psi_{+,k}$ and $\psi_{-,k}$
\begin{align}
	\psi_{+,k}\,=\, b_k + {\lambda\over L^2} \psi^{(1)}_{+,k} +\cdots\;\;,\qquad \psi_{-,k}\,=\,\bar{b}_k + {\lambda\over L^2} \bar{\psi}^{(1)}_{-,k} +\cdots\ , \label{eq: expansion of psi app}
\end{align}
one can write Eqs.~\eqref{eq: eq lhs fermion 1 app} and \eqref{eq: eq lhs fermion 2 app} as
\begin{align}
	\{ \psi_{+,k} ,H\}_\dirac\,=\,&{2\pi i k \over L } b_{k}  + {\lambda \over L^2} \left({2\pi i k \over L } \psi^{(1)}_{+,k} - {8\pi^2 i  \over L} \sum_{r,s} (k-r-s)s b_{k-r-s} \bar{b}_{r}\bar{b}_{s}\right)+\cO(\lambda^2)\ , \\
	\{\psi_{-,k},H\}_\dirac\,=\,&- {2\pi i k \over L } \bar{b}_{k} + {\lambda \over L^2}\left( -{2\pi i k \over L } \psi^{(1)}_{-,k} - {8\pi^2 i  \over L} \sum_{r,s} (k-r-s)s b_{r} b_{s}\bar{b}_{k-r-s}\right)+\cO(\lambda^2)\ .
\end{align}
On the other hand, using the Dirac brackets of the free fermi oscillators $b_k$ and $\bar{b}_k$, one can calculate the right-hand side of Eq.~\eqref{eq: pb of a and h 2 fermi app},
\begin{align}
	\{\psi_{+,k}, \wH\}_\dirac\,=\,& {2\pi i k\over L}b_k+ {\lambda\over L^2}\left[    \sum_u \left(  {2\pi i u \over L}{\overleftarrow{\delta}  \psi^{(1)}_{+,k} \over \overleftarrow{\delta} b_{u} } b_u - {2\pi i u \over L} {\overleftarrow{\delta} \psi^{(1)}_{+,k} \over \overleftarrow{\delta} \bar{b}_{u} }\bar{b}_u \right)  - {8\pi^2 i  \over L}k r b_k \bar{b}_{-r}\bar{b}_r \right] +\cO(\lambda^2)\ ,\\
	\{\psi_{-,k}, \wH\}_\dirac\,=\,& -{2\pi i k \over L } \bar{b}_{k} + {\lambda \over L^2} \left[ \sum_u \left(  {2\pi i u \over L}{\overleftarrow{\delta}  \psi^{(1)}_{-,k} \over \overleftarrow{\delta} b_{u} } b_u - {2\pi i u \over L} {\overleftarrow{\delta} \psi^{(1)}_{-,k} \over \overleftarrow{\delta} \bar{b}_{u} }\bar{b}_u \right) - {8\pi^2 i  \over L}  k r  b_{-r} b_{r}  \bar{b}_{k}  \right] + \cO(\lambda^2)\ ,
\end{align}
where we represented the Dirac bracket as follows:
\begin{align}
	\{A,B\}_\dirac\,\equiv \, -i \sum_u \left( {\overleftarrow{\delta}  A \over \overleftarrow{\delta} b_{-u} } { \overrightarrow{\delta} B \over \overrightarrow{\delta}  b_u } + {\overleftarrow{\delta} A \over \overleftarrow{\delta} \bar{b}_{-u} } { \overrightarrow{\delta} B \over \overrightarrow{\delta}  \bar{b}_u }  \right)\ .
\end{align}
Putting all together, Eq.~\eqref{eq: pb of a and h 2 fermi app} becomes inhomogeneous first-order differential equations,
\begin{align}
	  \sum_u \left(   u {\overleftarrow{\delta}  \psi^{(1)}_{+,k} \over \overleftarrow{\delta} b_{u} } b_u -   u  {\overleftarrow{\delta} \psi^{(1)}_{+,k} \over \overleftarrow{\delta} \bar{b}_{u} }\bar{b}_u \right)  -  k  \psi^{(1)}_{+,k} = - 4\pi   \sum_{\substack{r,s\\r+s\ne 0}} (k-r-s)s b_{k-r-s} \bar{b}_{r}\bar{b}_{s}\ ,\label{eq: 1st order eq fermion 1 app}\\
	  \sum_u \left(   u  {\overleftarrow{\delta}  \psi^{(1)}_{-,k} \over \overleftarrow{\delta} b_{u} } b_u -    u  {\overleftarrow{\delta} \psi^{(1)}_{-,k} \over \overleftarrow{\delta} \bar{b}_{u} }\bar{b}_u \right)  +  k  \psi^{(1)}_{-,k} = - 4\pi  \sum_{\substack{r,s\\r+s\ne 0}} (k-r-s)s b_{r} b_{s}\bar{b}_{k-r-s} \ .\label{eq: 1st order eq fermion 2 app}
\end{align}
We take the following ansatz for a particular solution $\psi_{\pm,k}^{(1)}$ of the equations~\eqref{eq: 1st order eq fermion 1 app} and \eqref{eq: 1st order eq fermion 2 app},
\begin{align}
	\psi_{+,k}^{(1)}\,=\,&\sum_{r,s} f_{k;r,s} b_{k-r-s} \bar{b}_r \bar{b}_s \ ,\\
	\psi_{-,k}^{(1)}\,=\,&\sum_{r,s} \bar{f}_{k;r,s}  b_r b_s \bar{b}_{k-r-s}\ ,  
\end{align}
where $f_{k;r,s}$ and $\bar{f}_{k;r,s}$ is a function of $k, r, s$. Then, Eqs.~\eqref{eq: 1st order eq fermion 1 app} and \eqref{eq: 1st order eq fermion 2 app} become
\begin{align}
	-  2 \sum_{r,s}(r+s) f_{k;r,s}  b_{k-r-s}\bar{b}_r \bar{b}_s\,=\,&- 4\pi \sum_{\substack{r,s\\r+s\ne 0}}   (k-r-s)s \; b_{k-r-s} \bar{b}_r \bar{b}_s\ ,\label{eq: fermi sol first order 1 app}\\
	2\sum_{r,s} (r+s) \bar{f}_{k;r,s} b_r b_s \bar{b}_{k-r-s}\,=\,& - 4\pi \sum_{\substack{r,s\\r+s\ne 0}}  (k-r-s)s \; b_r b_s \bar{b}_{k-r-s}\ , \label{eq: fermi sol first order 2 app}
\end{align}
and we have
\begin{align}
	f_{k;r,s}\,=\,& 2\pi {(k-r-s)s\over r+s}\hspace{8mm} ( r+s\ne 0)\ ,\\
	\bar{f}_{k;r,s}= &-2\pi {(k-r-s)s\over r+s}\hspace{8mm} ( r+s\ne 0)\ .
\end{align}
Adding a solution of the homogeneous part of the equations, $\psi_{\pm,k}^{(1)}$ can be written as
\begin{align}
	\psi_{+,k}^{(1)}\,=\,&2\pi \sum_{\substack{r,s\\r+s\ne 0}} {(k-r-s)s\over r+s}b_{k-r-s} \bar{b}_r \bar{b}_s + \sum_r g_{k,r} b_k \bar{b}_{-r} \bar{b}_r \ , \label{eq: fermi map sol 1 app}\\
	\psi_{-,k}^{(1)}\,=\,&- 2\pi \sum_{\substack{r,s\\r+s\ne 0}}{(k-r-s)s\over r+s} b_r b_s \bar{b}_{k-r-s} + \sum_r \bar{g}_{k,r}   b_{-r} b_r \bar{b}_k\ .\label{eq: fermi map sol 2 app}
\end{align}
By construction, $g_{k,r}$ and $\bar{g}_{k,r}$ satisfy
\begin{equation}
	g_{k,r}\,=\,- g_{k,-r}\quad,\quad \bar{g}_{k,r}\,=\,-\bar{g}_{k,-r}\ .
\end{equation}
Now, we determine $g_{k,r}$ and $\bar{g}_{k,r}$ from the algebras. Note that in terms of $b$ and $\bar{b}$, $S_{\pm,p}$ can be written as
\begin{align}
	S_{+,p}\,=\,& - 2\pi \sum_k k  b_{p-k}b_k +\cO(\lambda)\ ,\\
	S_{-,p}\,=\, &- 2\pi \sum_k k  \bar{b}_{p-k} \bar{b}_k +\cO(\lambda)\ .
\end{align}
It is useful to define $\tS_{p}, \bar{\tS}_{p}$ and $\tK_p$ by
\begin{align}
	\tS_{p}\,\equiv \,&- 2\pi \sum_{k} k b_{p-k} b_k\ ,\\
	\bar{\tS}_{p}\,\equiv \,&- 2\pi \sum_{k} k \bar{b}_{p-k} \bar{b}_k \ , \\
	\tK_p\, \equiv\,&- 2\pi  p \sum_{k} b_k \bar{b}_{p-k}\ .
\end{align}
In order to demand that the algebra of $A_k\,,\; \bar{A}_k$ in \eqref{eq: dirac bracket mode 1 app}$\sim$\eqref{eq: dirac bracket mode 3 app} is realized by $\alpha_k\,,\;\bar{\alpha}_k$ via the solution~\eqref{eq: fermi map sol 1 app} and \eqref{eq: fermi map sol 2 app}, we have at order $\cO(\lambda)$
\begin{align}
	i\{b_k, \psi^{(1)}_{+,q}\}_\dirac +i \{ \psi^{(1)}_{+,k}, b_q\}_\dirac\,=\, &  \bar{\tS}_{k+q}\ ,\label{eq: 1st order fermion algebra 1 app}\\
	i\{\bar{b}_k, \psi^{(1)}_{-,q}\}_\dirac + i\{ \psi^{(1)}_{-,k}, \bar{b}_q\}_\dirac\,=\,& -\tS_{k+q}\ ,\label{eq: 1st order fermion algebra 2 app}\\
	i\{ b_k, \psi^{(1)}_{-,q}\}_\dirac +i \{ \psi^{(1)}_{+,k}, \bar{b}_q\}_\dirac \,=\,& - \tK_{k+q}\ .\label{eq: 1st order fermion algebra 3 app}
\end{align}
First, Eq.~\eqref{eq: 1st order fermion algebra 1 app} gives
\begin{align}
	- 2\pi \sum_r  s \bar{b}_{k+q-r} \bar{b}_r(1-\delta_{k+q,0}) + \sum_r (g_{k,r} + g_{q,r}) \delta_{k+q,0}  \bar{b}_{-r}\bar{b}_r\,=\,& \bar{\tS}_{k+q}\ ,
\end{align}
and we obtain
\begin{equation}
	g_{k,r}+g_{-k,r}\,=\, - 2\pi r \ .\label{eq: 1st order eq homogeneous sol 1 app}
\end{equation}
In the same way with Eqs.~\eqref{eq: 1st order fermion algebra 2 app} and \eqref{eq: 1st order fermion algebra 3 app}, one can get
%
%
%
%
\begin{align}
	\bar{g}_{k,r}+\bar{g}_{-k,r}\,=\,&2\pi r\ ,\label{eq: 1st order eq homogeneous sol 2 app}\\
    \bar{g}_{q,k}  - g_{k,q} \,=\,&\pi (k+q)\ .\label{eq: 1st order eq homogeneous sol 3 app}
\end{align}
%
%
%
%
%
A simple solution of Eqs.~\eqref{eq: 1st order eq homogeneous sol 1 app}, \eqref{eq: 1st order eq homogeneous sol 2 app} and \eqref{eq: 1st order eq homogeneous sol 3 app} is found to be
\begin{equation}
	g_{k,r}\,=\, -\pi r \quad,\quad \bar{g}_{k,r}\,=\, \pi r\ .
\end{equation}
Then, the map from $\psi_{\pm,k}$ to $b_k$ and $\bar{b}_k$ is written as
\begin{align}
	\psi_{+,k}^{(1)}\,=\,&2\pi \sum_{\substack{r,s\\r+s\ne 0}} {(k-r-s)s\over r+s}b_{k-r-s}  \bar{b}_r \bar{b}_s - \pi    b_k \sum_r  r   \bar{b}_{-r} \bar{b}_r \ , \\
	\psi_{-,k}^{(1)}\,=\,&-2\pi \sum_{\substack{r,s\\r+s\ne 0}}{(k-r-s)s\over r+s} b_r b_s \bar{b}_{k-r-s} +  \pi  \sum_r r  b_{-r} b_r \bar{b}_k\ .
\end{align}
%
%
%
It is easy to confirm that this map indeed gives the conjectured form of Hamiltonian and the momentum up to order $\cO(\lambda)$:
\begin{align}
	H
	\,=\,&{\pi \over L } \sum_k ( -k  b_{-k}b_k + k \bar{b}_{-k} \bar{b}_k )+{2\pi \over L } {\lambda \over L^2} \sum_k  \left[ - k  b_{-k}\psi^{(1)}_{+,k}  + k  \bar{b}_{-k}\psi^{(1)}_{-,k} \right]+\cO(\lambda^2)\ ,\cr
	\,=\,& H_+ + H_- - {4 \lambda \over L^2}H_+ H_- + \cO(\lambda^2)\ ,\\
	P
	\,=\,&\int dx\; {i\over 2}\left( \psi_+ \psi'_+ + \psi_-\psi'_-   \right)= H_+ - H_- + \cO(\lambda^2)\ .
\end{align}
%

\section{Path Integral and $\cJ$-Norm}
\label{app: path integral}

In this appendix, we will show how the path integral of the quantum mechanical toy model in Section~\ref{sec: toy model} 
implies the operator formalism with $\cJ$-norm. 

Before discussing the  toy model, let us first recall how the canonical formulation and the path integral are
related using a simple example of the fermi oscillator, with the free Hamiltonian,
\begin{align}
    H\,=\, - m b^\dag b\ ,
\end{align}
where the fermi oscillator $b$ and $b^\dag$ obeys the anti-commutation relation
\begin{align}
	\{b,b^\dag\}\,=\, 1 \ .
\end{align}
For the relation between path integral and operator formalism, we introduce the coherent state defined by
\begin{align}
	| \eta \rangle\, \equiv\, e^{-\eta b^\dag }|0,0\rangle\,=\, (1-\eta b^\dag)|0,0\rangle\ .
\end{align}
where $\eta$ is a complex Grassmannian variable. With the coherent state, the completeness relation can be written as
\begin{align}
	1\,=\, \int  d \bar{\eta} d\eta \; e^{ -  \bar{\eta} \eta } |\eta\rangle \langle \bar{\eta}|\ .\label{eq: b completeness relation app}
\end{align}
We will find the path integral representation of the transition amplitude $\langle \bar{\eta}_{\text{\tiny out}}| e^{-i T H} | \eta_{\text{\tiny in}} \rangle$. For this, we discretize the time, and we insert the completeness relations~\eqref{eq: b completeness relation app} of $\eta_j$ ($j=1,2,\cdots, N$) into the transition amplitude at each time slice. Then, we have
\begin{align}
    \langle \bar{\eta}_{\text{\tiny out}} | e^{-i T H}  | \eta_{\text{\tiny in}} \rangle\,=\,& \int \prod_{j=1}^{N} d\bar{\eta}_jd\eta_j\; \exp\bigg[ {1\over 2} \bar{\eta}_{N+1} \eta_N - {1\over 2}\sum_{j=1}^{N} \bar{\eta}_j(\eta_j-\eta_{j-1})  \cr
    &\hspace{16mm} + {1\over 2} \bar{\eta}_{1} \eta_0 + {1\over 2}\sum_{j=1}^{N} (\bar{\eta}_{j+1} -\bar{\eta}_j )\eta_j + i \epsilon \sum_{j=1}^N m \bar{\eta}_j\eta_{j-1}  \bigg]\ ,
\end{align}
where $\epsilon$ denotes the interval of the discrete time $T\,=\, N\epsilon$. Also, we defined $\bar{\eta}_{N+1}$ and $\eta_0$ by
\begin{align}
    \bar{\eta}_{N+1}\,\equiv\, \eta_{\text{\tiny out}}\;\;,\quad \eta_0\,\equiv \, \eta_{\text{\tiny in}} \ .
\end{align}
In the continuum limit ($N\,\rightarrow \,\infty$), one can express the transition amplitude by the path integral as follows,
\begin{align}
    \langle \bar{\eta}_{\text{\tiny out}} | e^{-iT H} | \eta_{\text{\tiny in}} \rangle\, =\, \int_{\eta(0)=\eta_{\text{\tiny in}}}^{\bar{\eta}(T)=\bar{\eta}_{\text{\tiny out}}} \cD \bar{\eta} \cD \eta\; e^{i \int dt \; L + {1\over 2}\bar{\eta}_{\text{\tiny out}} \eta(T)+ {1\over 2}\bar{\eta}(0) \eta_{\text{\tiny in}}}\ ,\label{eq: path integral represtion of transition amplitude example app}
\end{align}
where the first order Lagrangian $L$ is given by
\begin{align}
    L\,=\, {i\over 2} \bar{\eta} \dot{\eta} - {i\over 2} \dot{\bar{\eta}} \eta + m \bar{\eta}\eta= {i\over 2}\bar{\eta} \dot{\eta}- {i\over 2} \dot{\bar{\eta}}\eta - H\ . \label{eq: 1st order lag c app}
\end{align}
Note that the first order Lagrangian~\eqref{eq: 1st order lag c app} is expressed in terms of $\eta(t)$ and $\bar{\eta}(t)$ because of the second-class constraint of this system. The extra term ${1\over 2}\bar{\eta}_{\text{\tiny out}} \eta(T)+ {1\over 2}\bar{\eta}(0) \eta_{\text{\tiny in}}$ in Eq.~\eqref{eq: path integral represtion of transition amplitude example app} plays a crucial role in imposing the anti-periodic boundary condition of the path integral representation of the thermal partition function. Namely, after the Wick rotation, we have
\begin{align}
    &\tr (e^{-\beta H})\,=\, \int d\bar{\eta}_{\text{\tiny out}}d\eta_{\text{\tiny in}}\; e^{\bar{\eta}_{\text{\tiny out}}\eta_{\text{\tiny in}} }  \langle \bar{\eta}_{\text{\tiny out}} | e^{-\beta H} | \eta_{\text{\tiny in}} \rangle\ , \cr
    =\,& \int d\bar{\eta}_{\text{\tiny out}}d\eta_{\text{\tiny in}} \int_{\eta(0)=\eta_{\text{\tiny in}}}^{\bar{\eta}(T)=\bar{\eta}_{\text{\tiny out}}} \cD \bar{\eta} \cD \eta \; e^{{1\over2}\bar{\eta}_{\text{\tiny out}}\big[\eta_{\text{\tiny in}} + \eta(T) \big]+ {1\over2}\big[\bar{\eta}_{\text{\tiny out}} + \bar{\eta}(0) \big] \eta_{\text{\tiny in}} } e^{- S_\beta }\ ,\cr
    \,=\,&\int_{\eta(0)=-\eta(T)\,,\, \bar{\eta}(0)=-\bar{\eta}(T)} \cD \bar{\eta} \cD \eta \; e^{- S_\beta }\ ,
\end{align}
where $S_\beta$ denotes the Euclidean action given by
\begin{align}
    S_\beta\,=\,\int_0^\beta d\tau \; \bigg[{1\over 2} \bar{\eta} \partial_\tau \eta- {1\over 2} \partial_\tau \bar{\eta} \eta -m \bar{\eta}\eta \bigg]\ .
\end{align}

Now, we will find the path integral representation of the transition amplitude of our toy model in Section~\ref{sec: toy model}. In this appendix, we will not demand that Hamiltonian is $\cJ$-Hermitian. That is, for the arbitrary real value of $\theta$ we define the fermi oscillators by
\begin{align}
	\bar{\fpi} +{i\over 2}\psi \,=\,& i (\sinh \theta b + \cosh\theta c )\ ,\label{eq: transf app 1}\\
	\fpi -{i\over 2}\bar{\psi} \,=\,& -i (\sinh \theta b^\dag + \cosh\theta c^\dag) \ , \\
	\bar{\fpi} -{i\over 2}\psi \,=\,& -i (\cosh\theta b + \sinh \theta c)\ ,\\
	\fpi +{i\over 2}\bar{\psi} \,=\,& i (\cosh\theta b^\dag + \sinh \theta c^\dag)\ , \label{eq: transf app 4}
\end{align}
Recall that by construction the fermi oscillators $b\,,\,b^\dag\,,\,c$ and $c^\dag$ still obey the anti-commutation relations 
\begin{align}
	\{b,b^\dag\}\,=\, 1\;\;,\qquad \{c,c^\dag\}\,=\,-1\ .
\end{align}
For the arbitrary value of $\theta$, we can still define the unitary and Hermitian operator $\cJ$ as before,
\begin{align}
    \cJ\,\equiv \,1+ 2c^\dag c\ .
\end{align}
and it has the same properties 
\begin{align}
	&\cJ c \cJ\,=\, -c\;\;,\quad \cJ c^\dag \cJ\,=\,-c^\dag\;\;,\quad \cJ b \cJ\,=\, b\;\;,\quad \cJ b^\dag \cJ\,=\, b^\dag\ .
\end{align}
Note that the operator $\cJ$ and the oscillators depend on the value of $\theta$ (\eg $\cJ=\cJ(\theta)$), but we omit $\theta$ for simplicity. Now, the Hamiltonian is not $\cJ$-Hermitian except for the special value of $\theta$ (\ie $\tanh \theta\,=\,-{ 2m\lambda \over 2m\lambda+1}$.)

Using the fermi oscillators, we define a coherent state by 
\begin{align}
	| \eta,\zeta \rangle\, =\, e^{-\eta b^\dag -\zeta c^\dag}|0,0\rangle\,=\, (1-\eta b^\dag)(1-\zeta c^\dag)|0,0\rangle\ .
\end{align}
One can evaluate the eigenvalue of the oscillators with respect to the coherent state,
\begin{align}
    b|\eta,\zeta\rangle\,=\,&\eta |\eta,\zeta\rangle\ ,\\
    c|\eta,\zeta\rangle\,=\,&-\zeta |\eta,\zeta\rangle\ ,\\
    \langle\bar{\eta},\bar{\zeta}| b^\dag\,=\,& \langle\bar{\eta},\bar{\zeta}| \bar{\eta}\ ,\\
    \langle\bar{\eta},\bar{\zeta}| c^\dag\,=\, &-\langle\bar{\eta},\bar{\zeta}| \bar{\zeta}\ .
\end{align}
Due to the anti-commutation of $c$ and $c^\dag$, the inner product of the coherent state is also different from that of usual fermi oscillators. Namely,
\begin{align}
	\langle \bar{\eta},\bar{\zeta} | \eta' ,\zeta' \rangle\,=\, e^{\bar{\eta} \eta' - \bar{\zeta} \zeta' }\ .
\end{align}
As a result, the completeness relation is given by
\begin{align}
	1\,=\, \int d\bar{\eta} d\eta d \bar{\zeta} d\zeta \; e^{ - \bar{\eta} \eta' + \bar{\zeta} \zeta' } |\eta,\zeta\rangle \langle \bar{\eta},\bar{\zeta}|\ .
\end{align}
Note that the term $\bar{\zeta}\zeta'$ in the measure of the completeness relation, which will become symplectic one-form of the first order Lagrangian, has an opposite sign to the usual fermi oscillators. Since we need to use the $\cJ$-norm in the operator formalism, it is more convenient to rewrite the completeness relation with the $\cJ$ operator as follows,
\begin{align}
	1\,=\,\int d\bar{\eta} d\eta d \bar{\zeta} d\zeta \; e^{ - \bar{\eta} \eta' - \bar{\zeta} \zeta' } |\eta,\zeta\rangle \langle \bar{\eta},\bar{\zeta}| \cJ\ .\label{eq: completeness relation with j}
\end{align}
Then, one can treat all $\zeta$'s uniformly.\footnote{Otherwise, one needs to take care of $\zeta$ on the boundary (\eg $\zeta_1$ or $\zeta_{N}$) separately.} Also note that the $\cJ$-inner product of the coherent state is given by
\begin{align}
	\langle \eta,\zeta | \eta' ,\zeta' \rangle_{\cJ}\,=\, e^{\bar{\eta} \eta' + \bar{\zeta} \zeta' }\ .
\end{align}

Now, we will find the path integral representation of the transition amplitude defined with $\cJ$-norm\footnote{Since the Hamiltonian is not $\cJ$-Hermitian, the transition amplitude depends on the position of the insertion of $\cJ$ operator. Here, we insert $\cJ$ in front of $e^{-i TH}$ because it will be easier to use the completeness relation~\eqref{eq: completeness relation with j}. If we insert $\cJ$ behind of $e^{-iHT}$, we can put $\cJ$ operator in front of the ket in the completeness relation in Eq.~\eqref{eq: completeness relation with j}. Although one should carefully distinguish $\langle \bar{\eta}_{\text{\tiny out}},\bar{\zeta}_{\text{\tiny out}}| \cJ e^{-iTH} | \eta_{\text{\tiny in}},\zeta_{\text{\tiny in}}\rangle$ from $\langle\bar{\eta}_{\text{\tiny out}},\bar{\zeta}_{\text{\tiny out}}| e^{-iTH}\cJ |\eta_{\text{\tiny in}},\zeta_{\text{\tiny in}}\rangle$ in general, their traces (\eg thermal partition function) give identical results.}
\begin{align}
    \langle \bar{\eta}_{\text{\tiny out}},\bar{\zeta}_{\text{\tiny out}} | e^{-iT H} | \eta_{\text{\tiny in}},\zeta_{\text{\tiny in}} \rangle_\cJ\,=\, \langle \bar{\eta}_{\text{\tiny out}},\bar{\zeta}_{\text{\tiny out}} | \,\cJ\, e^{-iT H} | \eta_{\text{\tiny in}},\zeta_{\text{\tiny in}} \rangle \ .
\end{align}
After discretizing the time as before, we insert the completeness relation~\eqref{eq: completeness relation with j} to the transition amplitude at each discrete time,
\begin{align}
    &\langle \bar{\eta}_{\text{\tiny out}},\bar{\zeta}_{\text{\tiny out}}  |\, e^{-iTH}\,| \eta_{\text{\tiny in}},\zeta_{\text{\tiny in}} \rangle_\cJ \cr
    \,=\,& \int \prod_{j=1}^N d\bar{\eta}_j d\eta_jd\bar{\zeta}_jd\zeta_j \; \exp\bigg( {1\over 2}\bar{\eta}_{N+1}\eta_N + {1\over 2}\bar{\zeta}_{N+1}\zeta_N -{1\over 2}\sum_{j=1}^N\big[ \bar{\eta}_j(\eta_j-\eta_{j-1}) + \bar{\zeta}_j(\zeta_j-\zeta_{j-1})\big]   \bigg) \cr
    &\hspace{12mm} \times \exp\bigg( {1\over 2}\bar{\eta}_{1}\eta_0 + {1\over 2}\bar{\zeta}_{1}\zeta_0 +{1\over 2}\sum_{j=1}^N\big[ (\bar{\eta}_{j+1}-\bar{\eta}_j )\eta_j + (\bar{\zeta}_{j+1}-\bar{\eta}_j)\zeta_j \big]  \bigg) \cr
    &\hspace{12mm}\times \exp\bigg(- i \epsilon \sum_{j=1}^N H[\eta_{j-1},\bar{\eta}_j,\zeta_{j-1},-\bar{\zeta}_j  ] \bigg)\ .
\end{align}
Here, we defined $\bar{\eta}_{N+1}\,,\, \bar{\eta}_0\,,\, \bar{\zeta}_{N+1}$ and $\zeta_0$ by
\begin{align}
    \bar{\eta}_{N+1}\,\equiv\, \eta_{\text{\tiny out}}\;\;,\quad \eta_0\,\equiv \, \eta_{\text{\tiny in}}\;\;,\quad \bar{\zeta}_{N+1}\,\equiv\, \zeta_{\text{\tiny out}}\;\;,\quad \zeta_0\,\equiv \, \zeta_{\text{\tiny in}} \ .
\end{align}
Note that $c^\dag$ in the Hamiltonian is replaced by $-\bar{\zeta}$'s because of the $\cJ$ operator insertion in front of $e^{-i \epsilon H}$. In the continuum limit $(N\,\rightarrow \,\infty)$, we obtain the path integral representation of the transition amplitude,
\begin{align}
    &\langle \bar{\eta}_{\text{\tiny out}},\bar{\zeta}_{\text{\tiny out}}  |\, e^{-iTH}\,| \eta_{\text{\tiny in}},\zeta_{\text{\tiny in}} \rangle_\cJ \cr
    \,=\,& \int_{\eta(0)=\eta_{\text{\tiny in}}\,,\, \zeta(0)=\zeta_{\text{\tiny in}}}^{\bar{\eta}(T)=\eta_{\text{\tiny out}}\,,\,\bar{\zeta}(T)=\zeta_{\text{\tiny out}}} \cD \bar{\eta} \cD \eta \cD\bar{\zeta} \cD \zeta\; e^{i \int dt \; L + {1\over 2}\bar{\eta}_{\text{\tiny out}}\eta(T)+{1\over 2}\bar{\zeta}_{\text{\tiny out}}\zeta(T) + {1\over 2} \bar{\eta}(0)\eta_{\text{\tiny in}}+ {1\over 2} \bar{\zeta}(0)\zeta_{\text{\tiny in}} }\ ,\label{eq: path integral representation toy model 1 app}
\end{align}
where $L$ is found to be
\begin{align}
    L\,=\, {i\over 2} \bar{\eta} \dot{\eta}- {i\over 2} \dot{\bar{\eta} } \eta+ {i\over 2} \bar{\zeta} \dot{\zeta} - {i\over 2} \dot{\bar{\zeta}} \zeta -  H[\eta,\bar{\eta},\zeta,-\bar{\zeta}]\ .
\end{align}
Using the transformation similar to \eqref{eq: transf app 1}$\sim$\eqref{eq: transf app 4} (\ie $b\,,\,b^\dag \,,\, c\,,\, c^\dag \;\;\longrightarrow \;\;\eta\,,\, \bar{\eta}\,,\, \zeta\,,\, -\bar{\zeta} $),
\begin{align}
	\bar{\fpi} +{i\over 2}\psi \,=\,& i (\sinh \theta \eta + \cosh\theta \zeta )\ ,\\
	\fpi -{i\over 2}\bar{\psi} \,=\,& -i (\sinh \theta \bar{\eta} - \cosh\theta \bar{\zeta}) \ , \\
	\bar{\fpi} -{i\over 2}\psi \,=\,& -i (\cosh\theta \eta + \sinh \theta \zeta)\ ,\\
	\fpi +{i\over 2}\bar{\psi} \,=\,& i (\cosh\theta \bar{\eta} - \sinh \theta \bar{\zeta})\ , 
\end{align}
we can express $L$ in terms of $\psi\,,\,\pi\,,\, \bar{\psi}$ and $\bar{\pi}$, and it exactly agrees with the first order Lagrangian of the toy model,
\begin{align}
    L\,=\, \pi \dot{\psi} + \dot{\bar{\psi}} \bar{\pi} - \bigg[-m\bar{\psi} \psi -{1\over \lambda}\bigg(\pi-{i\over 2} \bar{\psi}\bigg)\bigg(\bar{\pi}+{i\over 2} \psi\bigg) \bigg]\ .
\end{align}
Compared to the transition amplitude~\eqref{eq: path integral representation of transition amplitude example app} of the ordinary fermion, the degrees of freedom in the path integral representation~\eqref{eq: path integral representation toy model 1 app} of the toy model is doubled, which is due to the absence of the second-class constraints.

The additional terms ${1\over 2}\bar{\eta}_{\text{\tiny out}}\eta(T)+\cdots$ in the path integral representation~\eqref{eq: path integral representation toy model 1 app} will impose the anti-periodic boundary condition of the thermal partition function. To see this, after the Wick rotation, we take the trace of the transition amplitude~\eqref{eq: path integral representation toy model 1 app}:
\begin{align}
    &\tr (\cJ e^{-\beta H})\,=\, \int d\bar{\eta}_{\text{\tiny out}} d\eta_{\text{\tiny in}} d\bar{\zeta}_{\text{\tiny out}} d\zeta_{\text{\tiny in}}\; e^{\bar{\eta}_{\text{\tiny out}}\eta_{\text{\tiny in}} + \bar{\zeta}_{\text{\tiny out}}\zeta_{\text{\tiny in}} }  \langle \bar{\eta}_{\text{\tiny out}}, \bar{\zeta}_{\text{\tiny out}} | e^{-\beta H} | \eta_{\text{\tiny in}},  \zeta_{\text{\tiny in}}  \rangle\ , \cr
    =\,& \int  d\bar{\eta}_{\text{\tiny out}} d\eta_{\text{\tiny in}} d\bar{\zeta}_{\text{\tiny out}} d\zeta_{\text{\tiny in}} \int_{\eta(0)=\eta_{\text{\tiny in}}\,,\, \zeta(0)=\zeta_{\text{\tiny in}}}^{\bar{\eta}(\beta)=\eta_{\text{\tiny out}}\,,\,\bar{\zeta}(\beta)=\zeta_{\text{\tiny out}}} \cD \bar{\eta} \cD \eta \cD\bar{\zeta} \cD \zeta \; e^{{1\over2}\bar{\eta}_{\text{\tiny out}}\big[\eta_{\text{\tiny in}} + \eta(\beta) \big]+ {1\over2}\big[\bar{\eta}_{\text{\tiny out}} + \bar{\eta}(0) \big] \eta_{\text{\tiny in}} }\cr
    &\hspace{72mm}\times e^{{1\over2}\bar{\zeta}_{\text{\tiny out}}\big[\zeta_{\text{\tiny in}} + \eta(\beta) \big]+ {1\over2}\big[\bar{\zeta}_{\text{\tiny out}} + \bar{\zeta}(0) \big] \zeta_{\text{\tiny in}} }  e^{- S_\beta }\ ,\cr
    \,=\,&\int^{\pi(0)=-\pi(\beta)\,,\, \bar{\pi}(0)=-\bar{\pi}(\beta)}_{\psi(0)=-\psi(\beta)\,,\, \bar{\psi}(0)=-\bar{\psi}(\beta)}\cD \bar{\pi} \cD\bar{\psi} \cD \pi \cD \psi \;  e^{- S_\beta[\psi,\bar{\psi},\pi,\bar{\pi}] }\ ,\label{eq: path integral representation toy model 2 app}
\end{align}
where the Euclidean action $S_\beta[\psi,\bar{\psi},\pi,\bar{\pi}]$ is found to be
\begin{align}
    S_\beta\,=\,\int_0^\beta d\tau \;\bigg( \pi \dot{\psi} + \dot{\bar{\psi}} \bar{\pi} - \bigg[-m\bar{\psi} \psi -{1\over \lambda}\bigg(\pi-{i\over 2} \bar{\psi}\bigg)\bigg(\bar{\pi}+{i\over 2} \psi\bigg) \bigg] \bigg)\ .
\end{align}
We can integrate out the conjugate momentum $\pi$ and $\bar{\pi}$ in the path integral representation~\eqref{eq: path integral representation toy model 2 app},
\begin{align}
    &\tr (\cJ e^{-\beta H})\,=\,\int_{\psi(0)=-\psi(T)\,,\, \bar{\psi}(0)=-\bar{\psi}(T)} \cD\bar{\psi}  \cD \psi \;  e^{- S_\beta }\ ,
\end{align}
and we recover the Euclidean action of our toy model:
\begin{align}
    S_\beta\,=\,\int_0^\beta d\tau \;\bigg( {1\over 2} \bar{\psi}\partial_\tau \psi -{1\over 2} \partial_\tau \bar{\psi} \psi -\lambda \partial_\tau \bar{\psi} \partial_\tau \psi -m \bar{\psi} \psi  \bigg)\ .
\end{align}
This proves the equivalence of the operator formalism with $\cJ$-norm and the path integral formalism for the thermal partition function, as was glimpsed at in Section~\ref{sec: toy model}.

\bibliographystyle{JHEP}
\bibliography{ttbar}

\end{document}